\documentclass[10pt,preprint]{aastex}
\usepackage{color}
\pdfoutput=1

\newcommand{\av}{$A_V$}

\newcommand{\etal}{et~al.}

\newcommand{\teff}{$T_{\rm eff}$}

\newcommand{\mum}{$\mu$m}

\begin{document}

\title{YSOVAR: Mid-Infrared Variability in NGC 1333}

\slugcomment{Version from \today}

\author{L.~M.~Rebull\altaffilmark{1,2}, 
J.~R.~Stauffer\altaffilmark{2},
A.~M.~Cody\altaffilmark{2,3},
H.~M.~G\"unther\altaffilmark{4,5}
L.~A.~Hillenbrand\altaffilmark{6} ,
K.~Poppenhaeger\altaffilmark{7,4}
S.~J.~Wolk\altaffilmark{4},
J.~Hora\altaffilmark{4},
J.~Hernandez\altaffilmark{8},
A.~Bayo\altaffilmark{9,10},          
K.~Covey\altaffilmark{11,12},
J.~Forbrich\altaffilmark{13,4},
R.~Gutermuth\altaffilmark{14},
M.~Morales-Calder\'on\altaffilmark{15},
P.~Plavchan\altaffilmark{16},
I.~Song\altaffilmark{17}, 
H.~Bouy\altaffilmark{15}, 
S.~Terebey\altaffilmark{18}
J.~C.~Cuillandre\altaffilmark{19}, 
L.~E.~Allen\altaffilmark{20} 
}

\altaffiltext{1}{Infrared Science Archive (IRSA), 
Infrared Processing and Analysis Center (IPAC), 1200 E.\
California Blvd., California Institute of Technology, Pasadena, CA
91125; rebull@ipac.caltech.edu}
\altaffiltext{2}{Spitzer Science Center (SSC), 1200 E.\ California
Blvd., California Institute of Technology, Pasadena, CA 91125}
\altaffiltext{3}{NASA Ames Research Center, Moffett Field, CA 94035}
\altaffiltext{4}{Harvard-Smithsonian Center for Astrophysics, 60 Garden St.,
Cambridge, 02138 MA, USA}
\altaffiltext{5}{Massachusetts Institute of Technology, Kavli Institute
for Astrophysics and Space Research, 77 Massachusetts Avenue,  NE83-557
Cambridge, MA 02139, USA}
\altaffiltext{6}{Department of Astronomy, California Institute of
Technology, Pasadena, CA 91125}
\altaffiltext{7}{NASA Sagan Fellow}
\altaffiltext{8}{Centro de Investigaciones de Astronom\'ia, Apdo.
Postal 264, M\'erida 5101-A, Venezuela}
\altaffiltext{9}{Max Planck Institut f\"ur Astronomie, K\"onigstuhl
17, 69117, Heidelberg, Germany}
\altaffiltext{10}{Departamento de F\'isica y
Astronom\'ia, Facultad de Ciencias, Universidad de Valpara\'iso, Av.
Gran Breta\~na 1111, 5030 Casilla, Valpara\'iso, Chile} 
\altaffiltext{11}{Lowell Observatory, 1400 West Mars Hill Road, Flagstaff,
AZ 86001 USA}
\altaffiltext{12}{Department of Physics and Astronomy, Western
Washington University, 516 High Street, Bellingham WA 98225 USA}
\altaffiltext{13}{University of Vienna, Department of Astrophysics,
T\"urkenschanzstr. 17, 1180 Vienna, Austria}
\altaffiltext{14}{Dept.\ of Astronomy, University of Massachusetts, Amherst, MA  01003 USA}
\altaffiltext{15}{Depto. Astrof\'{\i}sica, Centro de Astrobiolog\'{\i}a
(INTA-CSIC), ESAC campus, P.O. Box 78, E-28691 Villanueva de la Ca\~nada, Spain}
\altaffiltext{16}{NASA Exoplanet Science Institute (NExScI), Infrared Processing and
Analysis Center (IPAC), 1200 E.\ California Blvd., California
Institute of Technology, Pasadena, CA 91125 USA, and Missouri State
University, 901 S National Ave, Springfield, MO 65897 USA}
\altaffiltext{17}{Physics and Astronomy Department, University of
Georgia, Athens, GA 30602-2451 USA}
\altaffiltext{18}{Department of Physics and Astronomy, 5151
State University Drive, California State University at Los Angeles,
Los Angeles, CA 90032 USA}
\altaffiltext{19}{Canada-France-Hawaii Telescope Corporation, 65-1238 Mamalahoa
Highway, Kamuela, HI96743, USA}
\altaffiltext{20}{NOAO, 950 N. Cherry Ave., Tucson, AZ USA}

\begin{abstract}

As part of the Young Stellar Object VARiability (YSOVAR) program, we
monitored NGC~1333 for $\sim$35 days at 3.6 and 4.5 \mum\ using the
Spitzer Space Telescope.  We report here on the mid-infrared
variability of the point sources in the
$\sim10\arcmin\times\sim20\arcmin$ area centered on 03:29:06,
+31:19:30 (J2000). Out of 701 light curves in either channel, we find
78 variables over the YSOVAR campaign.  About half of the members are
variable. The variable fraction for the most embedded SEDs (Class~I,
flat) is higher than that for less embedded SEDs (Class~II), which is
in turn higher than the star-like SEDs (Class~III).  A few objects
have amplitudes (10-90th percentile brightness) in [3.6] or
[4.5]$>$0.2 mag; a more typical amplitude is 0.1-0.15 mag.  The
largest color change is $>$0.2 mag.  There are 24 periodic objects,
with 40\% of them being flat SED class. This may mean that the periodic
signal is primarily from the disk, not the photosphere, in those
cases. We find 9 variables likely to be `dippers', where texture in
the disk occults the central star, and 11 likely to be `bursters',
where accretion instabilities create brightness bursts. There are 39
objects that have significant trends in [3.6]$-$[4.5] color over the
campaign, about evenly divided between redder-when-fainter (consistent
with extinction variations) and bluer-when-fainter. About a third of
the 17 Class 0 and/or jet-driving sources from the literature are
variable over the YSOVAR campaign, and a larger fraction ($\sim$half)
are variable between the YSOVAR campaign and the cryogenic-era Spitzer
observations (6-7 years), perhaps because it takes time for the
envelope to respond to changes in the central source. The NGC~1333
brown dwarfs do not stand out from the stellar light curves in any way
except there is a much larger fraction of periodic objects ($\sim$60\%
of variable brown dwarfs are periodic, compared to $\sim$30\% of the
variables overall).

\end{abstract}


\section{Introduction}
\label{sec:intro}

Located on the western edge of the Perseus molecular cloud, NGC 1333
is only $\sim$235 pc (Hirota \etal\ 2008, 2011) away from us. Its
stars are thought to have an average age of 1-2 Myr (e.g., Bally
\etal\ 2008), but there are also several Class 0 objects, which are
objects in the earliest stages of star formation (see, e.g., Sadavoy
\etal\ 2014 or Sandell \& Knee 2001). The average extinction towards
NGC 1333 could be as high as \av$\sim$6-7 mag (Ridge \etal\ 2006),
with condensations of higher extinction. Few of the cluster members
are easily visible in optical bands. 

NGC 1333 was first mapped in the infrared (IR) in 1976 by Strom, Vrba,
\& Strom, who found 25 candidate young members. The region is host to
numerous Herbig-Haro (HH) objects (see, e.g., Strom \etal\ 1974); in
fact, protostellar outflows may be important in the evolution of this
cluster (see, e.g., Walawender \etal\ 2008). The Spitzer Space
Telescope (Werner \etal\ 2004) 4.5 \mum\ image of this region is
riddled with outflows from young stars (e.g., Plunkett \etal\ 2013).

Because NGC 1333 is very young and relatively nearby, and because the
extinction towards this region is high enough that optical monitoring
of cluster members is very difficult, we selected NGC 1333 for
inclusion in the YSOVAR project.  YSOVAR (Young Stellar Object
VARiability) is the name of a coordinated effort to probe the mid-IR
variability of young stars. This project monitored a dozen
star-forming regions with the Spitzer Infrared Array Camera (IRAC;
Fazio \etal\ 2004) in the post-cryogen era, at 3.6 and 4.5 \mum. At
these wavelengths, we can penetrate high extinction, and detect both
photospheres and dust.  Since Spitzer is space-based, there is no
day-night aliasing, and monitoring campaigns can be conducted over
weeks to months. Rebull \etal\ (2014; hereafter R14) provided an
overall introduction to the YSOVAR project, including the project's
goals and data reduction. One of the project's primary goals is to
understand the mid-IR variability characteristics of the most embedded
young stars. IR variability in young stars can contain within it
signatures of a variety of processes including accretion, structure in
the disk rotating into and out of view, disk scale height structure or
changes, and geometric effects of companions (or protoplanets). NGC
1333 has many very embedded young stars, making it an ideal laboratory
for studying the IR variability of such objects.   

Walawender \etal\ (2008) provide a recent review of star formation in
NGC 1333. Rebull (2015; hereafter R15) collects more than 25 studies
and catalogs into a point source catalog of objects in the direction
of NGC 1333; this catalog forms the basis on which we proceed in this
paper. There are a few papers specifically reporting on variability of
specific objects in NGC 1333, which we now summarize. 

SVS 13, sometimes called SSV 13\footnote{SVS 13 is known by a wide
variety of synonyms, including V512 Per; in R15 it is R15-NGC1333
J032903.75+311603.9.}, is one of the objects discovered by Strom
\etal\ (1976); it is very bright in the IR.  Liseau, Lorenzetti, \&
Molinari (1992) reported that SVS 13 brightened by about a magnitude
at $K$ sometime between 1989 January and 1990 February. They
postulated that the variations originated from thermal instabilities
in the disk caused by variable accretion. Aspin \& Sandell (1994)
monitored this object over three years and found additional
significant variability in the near-IR. Harvey \etal\ (1998) observed
this object in the far-IR with the Kuiper Airborne Observatory,
finding that the object brightened by a factor of 1.5-2 between the
1980s and the early 1990s, and concluding that the changes were a
result of a true change in the YSO's luminosity, as opposed to changes
in the foreground and/or the YSO's intrinsic extinction. (See also
Hodapp 2015 and references therein.) Although this famous object is in
the region we monitored as part of YSOVAR, it is very bright. It is
certainly saturated in our 4.5 \mum\ data, and just above saturation
in 3.6 \mum. Because this source is of particular interest, we
manually extracted photometry for it, assuming it has not quite
saturated. Its light curve was not included in the rest of the
analysis here, but it appears in the Appendix, and it has faded by
$\sim$0.15 mag in [3.6] from 2004 to 2011 (cryo-to-YSOVAR), which
appears consistent with the $K$-band trend in Hodapp (2015).

Herbst \etal\ (2006) monitored this region in the optical, but  report
on variations from only a single object, HBC 338. They find evidence
for differential rotation in this star, based on the large changes in
the measured period. Unfortunately, this object is well away from the
region we monitored as part of YSOVAR.

More recently, Forbrich \etal\ (2011) monitored this region
simultaneously in X-rays and in radio. Very few YSOs were detected in
both bands, and the authors did not find a close correlation between
X-rays and radio luminosities, which was surprising because such a
correlation is found in older active stars.  

Scholz (2012) placed constraints on the frequency of large variations
in the near-IR for a variety of young clusters, including NGC 1333, on
timescales up to $\sim$2000 d. Of the clusters studied, the largest
amplitudes of near-IR variability were found in NGC 1333, and are
attributed to this cluster's youth. Scholz identifies three objects
(LAL 166, LAL 189, LAL 296) as NIR variables, all of which are in our
monitored fields (SSTYSV J032858.41+312217.6, J032903.13+312238.1, and
J032920.04+312407.6, respectively). All are independently identified
by us as variable, but not with particularly large amplitudes at 3.6
and 4.5 \mum, and only one source is recovered as variable over a 6-7
year timescale. 

There is also literature specifically on the variability found in the
jets and clumps in NGC 1333 (see, e.g., Choi \etal\ 2006, Khanzadyan
\etal\ 2003, and references therein). 
Changes in the outflows seen in
Spitzer data are beyond the scope of this paper; Raga \etal\ (2013)
report on proper motions of the outflows in this region using the
YSOVAR data. We will discuss in this paper the mid-IR variability
properties of the sources thought to be driving jets (see R15 for
details on how the jet-drivers were identified). 

We report here on the mid-IR variability identified in NGC 1333 during
the YSOVAR (R14) campaign.  Section \ref{sec:obsdatared} covers the
observations and data reduction, and Section \ref{sec:largearchival}
the archival data sets included in our analysis.
Section~\ref{sec:importantdatasubsets} defines  members, variables,
and other terms, and Section~\ref{sec:countingthings} calculates
mid-IR variability fractions for various subsets of the data. 
Section~\ref{sec:amplitudes} examines the amplitudes of brightness and
color changes, and Section~\ref{sec:timescales} analyzes the
distributions of timescales in NGC 1333, including the periodic
objects. Section~\ref{sec:dippers} describes dippers and bursters,
objects in our data that have light curve structure similar to objects
identified in Spitzer monitoring of NGC 2264 (e.g., Cody \etal\ 2014)
and Orion (e.g., Morales-Calder\'on \etal\ 2011).
Section~\ref{sec:colortrends} identifies objects that have significant
trends in the color-magnitude diagrams over the YSOVAR campaign.
Finally, Section~\ref{sec:specialsources} calls out some special
sources in NGC 1333, and Section~\ref{sec:concl} summarizes the paper.

\section{Observations and Data Reduction}
\label{sec:obsdatared}

All YSOVAR Spitzer light curve tabular data are available via the
YSOVAR data delivery to the Infrared Science Archive (IRSA). Plots of
all the light curves are provided as part of that delivery; only
certain light curves are shown in the present paper.

\begin{figure}[h]
\epsscale{0.5}
\plotone{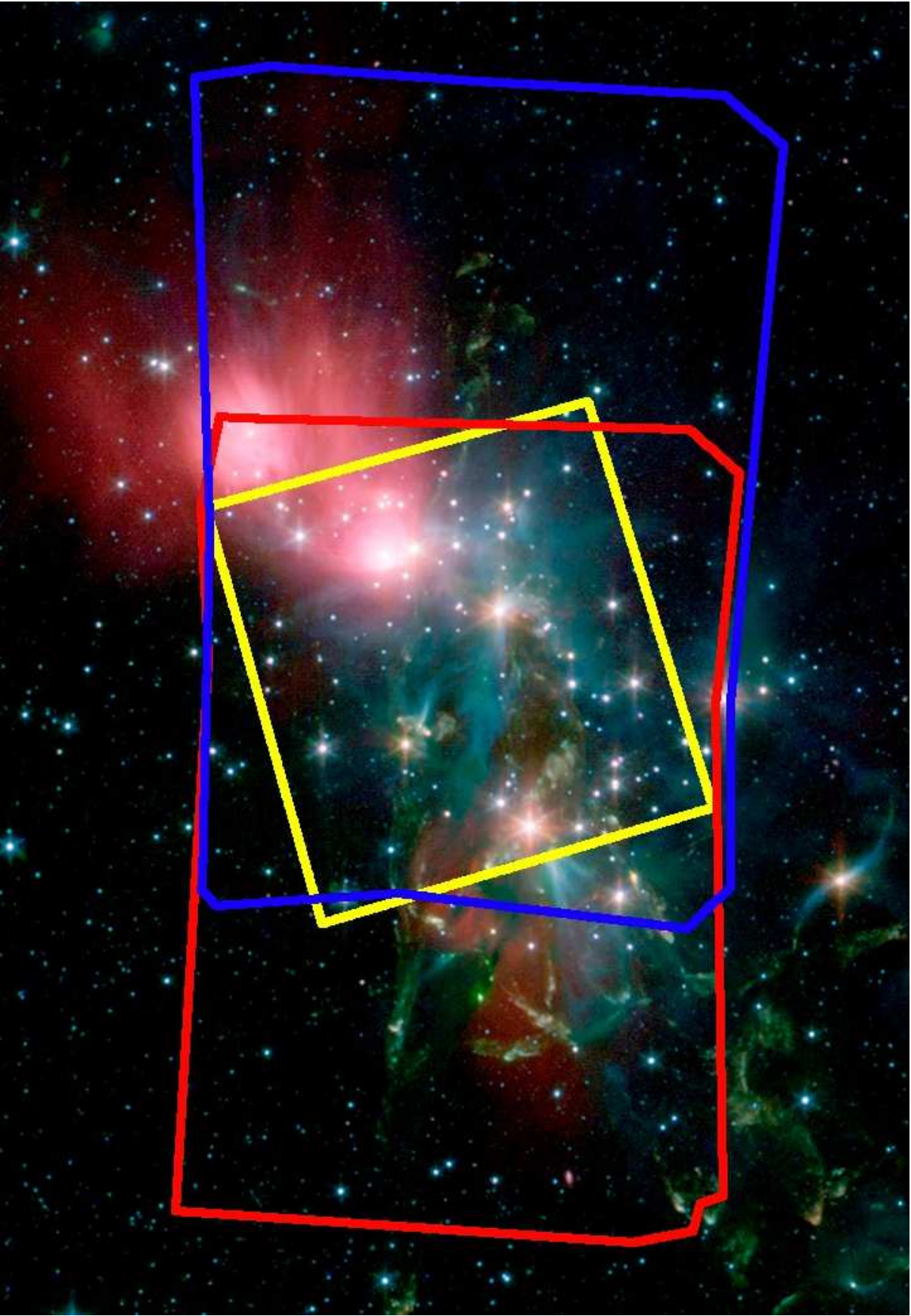}
\caption{Solid red and blue lines indicate the approximate sky
coverage (footprint) for a summed image of
of YSOVAR NGC 1333 observations, superimposed on a 3-color image of
NGC 1333  obtained during the cryogenic mission (image: SSC press
release 2005-24a, NASA/JPL-Caltech/R. A. Gutermuth,
Harvard-Smithsonian CfA). The blue line is 3.6 \mum, the red line is
4.5 \mum\ (the region where they overlap is where we have 2-band
light curves). The yellow square indicates archival Chandra data
coverage. The image is about 25$\arcmin$ top-to-bottom; the center of
the YSOVAR coverage is  3h29m05.75s,+31d19m30.0s (J2000).}
\label{fig:n1333footprints}
\end{figure}

\subsection{YSOVAR Spitzer Data}
\label{sec:ysovarspitzer}

The observation design, operational considerations, cadence, and data
reduction procedure for YSOVAR Spitzer monitoring data are discussed
in detail in R14. We summarize here briefly the most salient points,
but refer the reader to R14 for more details.

Figure~\ref{fig:n1333footprints} shows the sky coverage of the YSOVAR
monitoring observations of NGC 1333. The observations were 2$\times$2
IRAC FOVs in a mapping mode Astronomical Observation Request (AOR),
with a 5-point small Gaussian dither pattern of 12-s
high-dynamic-range (HDR) frames at each map position, resulting in a
median depth of 54 s per epoch.
The sky observed by the two IRAC channels (3.6 and 4.5 \mum, or IRAC-1
and -2) fields of view (FOVs) is slightly offset with the central
$\sim10\arcmin\times10\arcmin$ region, centered on 03:29:06 +31:19:30,
covered by both channels. Although the observations in the central
region were not obtained at exactly the same time in both bands, they
were obtained within minutes, and we take them to be functionally
simultaneous. These observations were obtained under program ID 61026,
between 2011 Oct 10 and 2011 Nov 14.  This campaign was conducted
entirely in the fast cadence mode described in R14 (8 observations
made every 3.5 days with non-uniform spacing ranging from 4 to 16 hrs
to reduce aliasing). NGC 1333's location near the ecliptic provides
for minimal field rotation during the Spitzer monitoring program, such
that the majority of objects detected in this region were able to be
monitored for the full campaign, producing light curves with 72 epochs
over 35 days for most of the standard set of members (selected as
described in R14 and below in Sec.~\ref{sec:members}). 

Because NGC 1333 is not exactly on
the ecliptic, there was some field rotation, and because of the very
IR-bright objects in the middle of the field, the rotation of the
diffraction spikes was enough to affect some photometry of some nearby
faint objects in the region surrounding the bright sources. All
individual objects identified as variable in the rest of this paper
were specifically investigated for this effect, and bad points were
removed. 

We note that our photometry (regardless of the nature of
the source) becomes noisy and incomplete fainter than
[3.6]$\sim$[4.5]$\sim$16 mags (Fig.~\ref{fig:magdist} below);
sources this faint receive special scrutiny before being identified as
variable. 

\clearpage

\begin{figure}[h]
\epsscale{0.8}
\plotone{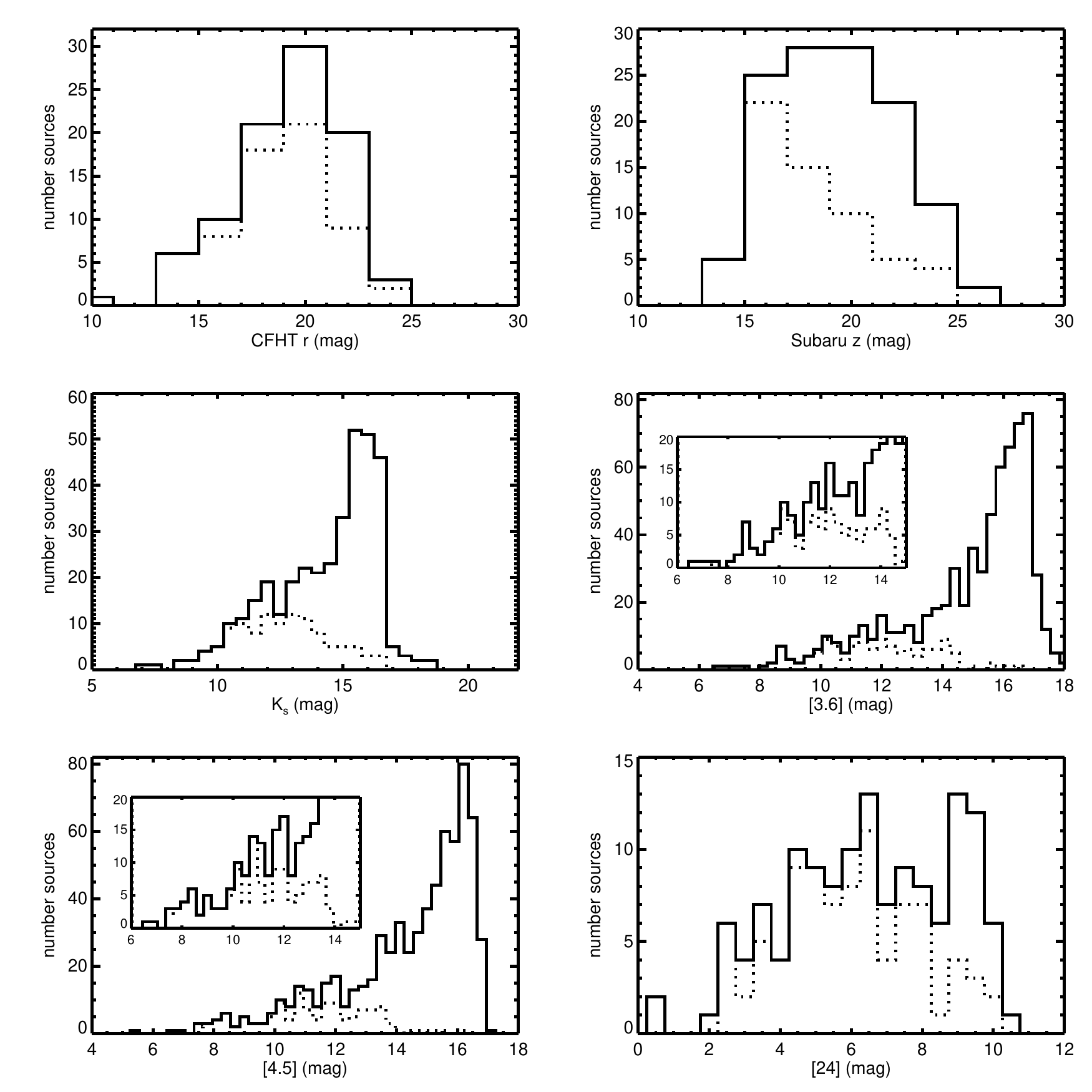}
\caption{Histograms of some of the single-epoch optical, NIR,
and MIR measurements used to assemble SEDs in NGC 1333, just for
objects with light curves. Solid line is everything in the database
(with a light curve); dotted line is standard set of cluster members 
(selected as described in R14 and below in Sec.~\ref{sec:members}). 
The CFHT $r$ and Subaru $z$ bands are representative of the optical
bands; the set of objects with optical counterparts is strongly biased
towards members (the standard set) and variables because a complete
catalog over the whole NGC 1333 field was not obtained. $K_s$ is
representative of the NIR bands; the available data reach far deeper
than the standard set of members, and the standard set of members
represent most if not all of the brightest objects. The [3.6] and
[4.5] histograms show the range over which there are YSOVAR light
curves (down to [3.6]$\sim$[4.5]$\sim$16), and the relatively small,
bright fraction that are members. (The inset plots for both of these
channels enlarge the transition where the fraction of members is
falling off.) By [24], typically, only the members are bright enough
to be detected, and the sample of all detections is again dominated by
members. The dataset spans very different brightnesses as a function
of wavelength, but the standard set of members always dominates the
bright end of the distribution. 
\label{fig:magdist}}
\end{figure}

\begin{deluxetable}{ccccp{7cm}}
\tabletypesize{\scriptsize}
\tablecaption{Contents of Single-Value Object
Catalog\tablenotemark{a}\label{tab:singlevaluedata}}
\tablewidth{0pt}
\tablehead{\colhead{Number} & \colhead{Format}
&\colhead{Units} & \colhead{Label} & \colhead{Explanations}}
\startdata
    1  & A18   &  --- &    cat       & SSTYSV catalog name (HHMMSS.ss+DDMMSS.s; J2000)\\
    2  & F10.6 & deg & RA & RA in decimal degrees, J2000\\
    3  & F10.6 & deg & Dec & Dec in decimal degrees, J2000\\
    4  &F7.2   & AB mag & cfht\_umag & $u$-band AB magnitude from CFHT \\
    5  &F7.2   & AB mag & e\_cfht\_umag & error on $u$-band AB magnitude from CFHT \\
    6  &F7.2   & AB mag & cfht\_gmag & $g$-band AB magnitude from CFHT \\
    7  &F7.2   & AB mag & e\_cfht\_gmag & error on $g$-band AB magnitude from CFHT \\
    8  &F7.2   & AB mag & cfht\_rmag & $r$-band AB magnitude from CFHT \\
    9  &F7.2   & AB mag & e\_cfht\_rmag & error on $r$-band AB magnitude from CFHT \\
   10  &F7.2   & AB mag & cfht\_imag & $i$-band AB magnitude from CFHT \\
   11  &F7.2   & AB mag & e\_cfht\_imag & error on $i$-band AB magnitude from CFHT \\
   12  &F7.2   & AB mag & subaru\_imag & $i^{\prime}$-band AB magnitude from Subaru \\
   13  &F7.2   & AB mag & e\_subaru\_imag & error on $i^{\prime}$-band AB magnitude from Subaru \\
   14  &F7.2   & AB mag & subaru\_zmag & $z^{\prime}$-band AB magnitude from Subaru \\
   15  &F7.2   & AB mag & e\_subaru\_zmag & error on $z^{\prime}$-band AB magnitude from Subaru \\
   16  &A4     &--- &    l\_J       &Limit flag on J                                    \\ 
   17  &F6.2   &mag &    J         &J band magnitude                                   \\ 
   18  &F6.2   &mag &    e\_J       &Uncertainty in J                                   \\ 
   19  &A4     &--- &    l\_H       &Limit flag on H                                    \\ 
   20  &F6.2   &mag &    H         &H band magnitude                                   \\ 
   21  &F6.2   &mag &    e\_H       &Uncertainty in H                                   \\ 
   22  &A4     &--- &    l\_Ks      &Limit flag on Ks                                   \\ 
   23  &F6.2   &mag &    Ks        &Ks band magnitude                                  \\ 
   24  &F6.2   &mag &    e\_Ks      &Uncertainty in Ks                                  \\ 
   25  &A4     &--- &    l\_[3.6]   &Limit flag on [3.6]                                \\ 
   26  &F6.2   &mag &    [3.6]     &Spitzer/IRAC 3.6 \mum\ band magnitude               \\ 
   27  &F6.2   &mag &    e\_[3.6]   &Uncertainty in [3.6]                               \\ 
   28  &A4     &--- &    l\_[4.5]   &Limit flag on [4.5]                                \\ 
   29  &F6.2   &mag &    [4.5]     &Spitzer/IRAC 4.5 \mum\ band magnitude               \\ 
   30  &F6.2   &mag &    e\_[4.5]   &Uncertainty in [4.5]                               \\ 
   31  &A4     &--- &    l\_[5.8]   &Limit flag on [5.8]                                \\ 
   32  &F6.2   &mag &    [5.8]     & Spitzer/IRAC 5.8 \mum\ band magnitude              \\ 
   33  &F6.2   &mag &    e\_[5.8]   &Uncertainty in [5.8]                               \\ 
   34  &A4     &--- &    l\_[8.0]   &Limit flag on [8.0]                                \\ 
   35  &F6.2   &mag &    [8.0]     &Spitzer/IRAC 8.0 \mum\ band magnitude               \\ 
   36  &F6.2   &mag &    e\_[8.0]   &Uncertainty in [8]                                 \\ 
   37  &A4     &--- &    l\_[24]    &Limit flag on [24]                                 \\ 
   38  &F6.2   &mag &    [24]      &Spitzer/MIPS 24 \mum\ band magnitude               \\ 
   39  &F6.2   &mag &    e\_[24]    &Uncertainty in [24]                                \\ 
   40  &A4     &--- &    l\_[70]    &Limit flag on [70]                                 \\ 
   41  &F6.2   &mag &    [70]      & Spitzer/MIPS 70 \mum\ band magnitude              \\ 
   42  &F6.2   &mag &    e\_[70]    &Uncertainty in [70]                                \\ 
   43  &F6.2   & --- & logfx & log of $F_x$, where $F_x$ is in units
   of ergs cm$^{-2}$ s$^{-1}$\\
   44  &F6.2   & --- & logfxerr & error on log of $F_x$, where $F_x$ is in units
   of ergs cm$^{-2}$ s$^{-1}$\\
   45  &F6.2   & --- & loglx & log $L_x$, where $L_x$ is in units
   of ergs s$^{-1}$\\
   46  & I3    &     & glvary\_index & Gregory-Loredo index (see R14);
   the larger the number, the greater the X-ray variability \\
   47  &A12    &--- &    SpTy      & Spectral type                                     \\ 
   48  &A10    &--- &    SpTySrc   & Origin of spectral type (literature)              \\ 
   49  &I4     &K   &    Teff      & \teff\ from Foster \etal\ (2015)                    \\ 
   50  &F6.2   &---& oursedslope24 & SED slope (from 2 to 24 \mum; see text)\\
   51  &A5     &---& oursedclass24 & SED class (from 2 to 24 \mum; see text)\\
   52  &I2     &---&   YSOlit & Is this a literature YSO (see text)? 1=yes, 0=no \\
   53  &I2     &---&   YSOstd & Is this a part of the standard set of YSOs (see text)? 1=yes, 0=no \\
   54  &I2     &---&   YSOx & Was this selected for the standard set of YSOs based on X-rays (see text)? 1=yes, 0=no \\
   55  &I2     &---&   YSOaug & Is this a part of the augmented set of YSOs (see text)? 1=yes, 0=no \\
   56  &I2     &---&   CYvar     & Is this CY variable (see text)? 1=yes, 0=no \\
   57  &I2     &---&   Var    & Is this variable over the YSOVAR campaign? 1=yes, 0=no \\
   58  &I2     &---&   VarMem    & Is this a newly identified member based on variability? 1=yes, 0=no \\
   59  &I4     &---& Npts\_i1 & Number of points in [3.6] for YSOVAR campaign\\
   60  &F6.2   & mag & Mean\_i1 & Mean [3.6]  \\
   61  &F6.2   & mag & Median\_i1 & Median [3.6] \\
   62  &F6.2   & mag & Sdev\_i1 & Standard deviation [3.6]\\
   63  &F6.2   & mag & Max\_i1 & Maximium [3.6] \\
   64  &F6.2   & mag & Min\_i1 & Minimum [3.6] \\
   65  &F6.2   & days & Range\_JD\_i1 & Length of [3.6] light curve\\
   66  &F6.2   & mag & Ampl1090\_i1 & Amplitude of [3.6] light curve, defined as 10-90\% range.\\
   67  &F6.2   & --- & Chisq\_i1 & $\chi^2([3.6])$\\
   68  &F6.2   & days & Timescale\_i1 & Timescale for [3.6] \\
   69  &F6.2   & days & Period\_i1 & Period for [3.6], if applicable\\
   70  &F6.2   & --- & M\_i1 & $M$ metric for [3.6] as per Cody \etal\ (2014)\\
   71  &F6.2   & --- & Ms\_i1 & $M$ metric for smoothed [3.6] as per Cody \etal\ (2014)\\
   72  &F6.2   & --- & Q\_i1 & $Q$ metric for [3.6] as per Cody \etal\ (2014)\\
   73  &F6.2   & --- & Qlt\_i1 & $Q$ metric for long-term-trend-subtracted [3.6] as per Cody \etal\ (2014)\\
   74  &I4     &---& Npts\_i2 & Number of points in [4.5] for YSOVAR campaign\\
   75  &F6.2   & mag & Mean\_i2 & Mean [4.5]  \\
   76  &F6.2   & mag & Median\_i2 & Median [4.5] \\
   77  &F6.2   & mag & Sdev\_i2 & Standard deviation [4.5] \\
   78  &F6.2   & mag & Max\_i2 & Maximium [4.5] \\
   79  &F6.2   & mag & Min\_i2 & Minimum [4.5] \\
   80  &F6.2   & days & Range\_JD\_i2 & Length of [4.5] light curve\\
   81  &F6.2   & mag & Ampl1090\_i2 & Amplitude of [4.5] light curve, defined as 10-90\% range.\\
   82  &F6.2   & --- & Chisq\_i2 & $\chi^2([4.5])$\\
   83  &F6.2   & days & Timescale\_i2 & Timescale for [4.5]\\
   84  &F6.2   & days & Period\_i2 & Period for [4.5], if applicable\\
   85  &F6.2   & --- & M\_i2 & $M$ metric for [4.5] as per Cody \etal\ (2014)\\
   86  &F6.2   & --- & Ms\_i2 & $M$ metric for smoothed [4.5] as per Cody \etal\ (2014)\\
   87  &F6.2   & --- & Q\_i2 & $Q$ metric for [4.5] as per Cody \etal\ (2014)\\
   88  &F6.2   & --- & Qlt\_i2 & $Q$ metric for long-term-trend-subtracted [4.5] as per Cody \etal\ (2014)\\
   89  &F6.2   & mag & Mean\_i1i2 & Mean [3.6]$-$[4.5]  \\
   90  &F6.2   & mag & Sdev\_i1i2 & Standard deviation [3.6]$-$[4.5] \\
   91  &F6.2   & days & Period\_i1i2 & Period for [3.6]$-$[4.5], if applicable\\
   92  &F6.2   & --- & Stetson\_i1i2 & Stetson index for [3.6], [4.5]\\
   93  &F6.2   & --- & Correl\_i1i1i2 & Correlation coefficient for [3.6] vs.~[3.6]$-$[4.5] CMD.\\
   94  &F6.2   & --- & CorrelProb\_i1i1i2 & Probability that correlation is real in [3.6] vs.~[3.6]$-$[4.5] CMD.\\
   95  &F6.2   & --- & Correl\_i2i1i2 & Correlation coefficient for [4.5] vs.~[3.6]$-$[4.5] CMD.\\
   96  &F6.2   & --- & CorrelProb\_i2i1i2 & Probability that correlation is real in [4.5] vs.~[3.6]$-$[4.5] CMD.\\
\enddata
\tablenotetext{a}{Entire data table available online at the journal.
This table provided as a guide to the contents of the table.}
\end{deluxetable}
\clearpage

\subsection{YSOVAR Data from PAIRITEL}

Near-IR observations were obtained using the Peters Automated Infrared
Imaging Telescope (PAIRITEL; Bloom \etal\ 2006), which is an automated
1.3 m telescope. PAIRITEL was located at the Whipple Observatory at Mt.
Hopkins, Arizona;  the same telescope and camera used for the northern
part of 2MASS, PAIRITEL acquired simultaneous {\it J, H,} and {\it K$_s$}
images. The observations of NGC 1333 were obtained over 24 epochs from
7 October 2010 to 19 November 2012. The PAIRITEL pipeline version 3
was used, which linearizes the data and combines individual exposures
to produce mosaics and ``weight'' map products for each epoch and
filter. The mosaic products from this pipeline have a scale of
1\farcs0 pixel$^{-1}$ and the world coordinate system (WCS) has a
typical pointing error of less than an arcsec.

Photometry from the PAIRITEL images was extracted using a data
reduction pipeline developed at SAO. We examined each epoch's mosaics
to exclude data with insufficient coadds due to poor weather and other
artifacts in the data. We then checked each weight frame to find an
appropriate cutoff level (in terms of numbers of coadds) for the
epoch, then masked the unreliable data so that it would not impact the
derived photometry.  The SAO pipeline uses the IRAF\footnote{IRAF is
distributed by the National Optical Astronomy Observatory, which is
operated by the Association of Universities for Research in Astronomy
(AURA) under cooperative agreement with the National Science
Foundation.} package {\em daofind} to detect all point sources
4$\sigma$ above the background noise level with a full-width
half-maximum of 2.2$\arcsec$. Photometry was performed with the IRAF
{\em phot} task, using an aperture radius of 5 pixels and a sky
annulus radius of 10 pixels. We combined the data from the three bands
into a common catalog using a script to match objects by position.
This process is iterated, applying small shifts to the individual
catalogs, to correct their relative offsets and minimize the mean
position error between the bands. After obtaining the 3-band catalog,
another script matches the objects to 2MASS positions, refines the
PAIRITEL astrometry, generates object IDs and associates the PAIRITEL
detections with their corresponding 2MASS objects where possible. The
photometry is then calibrated using an iterative routine that
minimizes the offset between the 2MASS and PAIRITEL photometry
measurements, using all objects with a signal to noise $>$30. This
process was repeated for each epoch, and the photometry combined into
a master catalog that contains the photometric measurements of all
bands and sources. The typical standard deviation of sources not
detected as variable is on the order of 0.05 mag for {\it J} and {\it
H} band, and 0.08 mag for {\it K$_s$} band.

The PAIRITEL data were taken over a span of $\sim$2.5 years, far
longer than the YSOVAR campaign. Due largely to bad weather, there are
at most 5-8 epochs during the YSOVAR campaign, and two immediately
following it. We can find PAIRITEL counterpart light curves for 122
objects with YSOVAR light curves (out of about 300 in the region that
was monitored in both $JHK_s$ and IRAC). Out of the $\sim$65 light
curves determined below to be {\em variable} over the YSOVAR campaign,
there are 53 with a $J$, $H$, and/or $K_s$ light curve. For about 30\%
of these light curves, the $JHK_s$ data light curves have very similar
structure (and often amplitude) as for the IRAC light curves.
Morales-Calder\'on \etal\ (2011) found many such cases in Orion;
similar light curve structure from 1 to 5 \mum\ is expected for stars
without disks (arising in stellar phenomena such as rotation of
surface inhomogeneities) and can also occur for some physical
mechanisms for stars with disks (such as accretion bursts,
particularly when our view angle is from well above the disk). 
However, many NGC 1333 members (standard or augmented set of members)
show near and mid-IR light curves with quite disparate structure,
indicative of different physical mechanisms at work.

\clearpage

\section{Large Archival Datasets}
\label{sec:largearchival}

We compiled additional photometry from several different sources in
order to assemble spectral energy distributions (SEDs) for our targets
and to identify members; this process is reported in R15. While many
of our targets vary significantly, single-epoch archival data can help
define the SED such that, in some cases, the assembled SED can reveal
the underlying nature of the source, or at least help narrow the
possibilities for the nature of the source.  We also use the slope of
the SED between 2 and 24 \mum\ to define SED classes into which our
objects fall; see R14 and R15. 

To summarize the most salient points from R15, all of the catalogs
described here were merged by position with a catalog-dependent search
radius (usually $\sim$1$\arcsec$). Many sources, especially those in
regions where many sources are close together on the sky, or those
that had particularly strange SEDs, were also inspected and matched by
hand. Data from the Widefield Infrared Survey Explorer (WISE; Wright
\etal\ 2010) at 3.4, 4.5, 12, and 22 \mum\ were included in a few
specific cases after inspection of the WISE images.

In order to give a sense of the brightness, faintness, and
completeness limits of these data sets, representative histograms of
these single-epoch measurements for the objects with light curves
appear in Figure~\ref{fig:magdist}. The baseline catalog from R15 was
assembled initially without regard to membership, variability, or the
existence of light curves; however, in this paper, we restrict
ourselves to those objects with light curves. The histograms in
Fig.~\ref{fig:magdist} are shown for both the entire set of objects
with light curves, and for those in the standard set of members
(selected as described in R14 and below in Sec.~\ref{sec:members},
where issues of contamination are also raised).  

Single-valued measurements (such as the single-epoch optical
photometry) or metrics (such as the SED slope and class, or mean from
the light curve, etc.) for all of the objects with light curves are
collected in Table~\ref{tab:singlevaluedata}. The $JHK$ and cryo-era
Spitzer data (as well as the SED slopes and classes) are reported in
R15, but are repeated in Table~\ref{tab:singlevaluedata} for
convenience. The multi-epoch data tables will appear in the delivery
of these data to IRSA.

\subsection{Cryogenic-era Spitzer Archival Data}

Early in the Spitzer mission, NGC 1333 was observed by both the
guaranteed time observations (GTO) and the Cores-to-Disks
(c2d) Legacy program (Evans \etal\ 2003, 2009), with both IRAC and the
Multiband Imaging Photometer for Spitzer (MIPS; Rieke \etal\ 2004). 
Gutermuth \etal\ (2008, 2009, 2010) present methodology for
identifying YSOs from the cryogenic catalog. The details of the
selection process appear in those papers, but in summary, multiple
cuts in multiple color-color and color-magnitude diagrams are used to
identify YSO candidates, as distinct from, e.g., extragalactic and
nebular contamination. We compute this classification as part of the
YSOVAR processing, and we have adopted this YSO selection mechanism as
part of one of the primary YSOVAR sample definitions (R14), which we
apply here to the NGC 1333 data set.

Spitzer data are also available from the c2d program data deliveries,
served by IRSA. The data used for these deliveries are typically the
same BCDs as were used in the cryogenic data that we re-reduced above.
As such, then, they are not independent measurements, and these data
were only used to supplement our cryogenic-era catalog if a band was
missing (which may happen due to low signal-to-noise). The c2d catalog
includes measurements at 70 \mum\ and band-filled upper
limits\footnote{That is, if a source was known to be at a given
location because of information from other bands, but it was not
detected independently in a given band, then an aperture was placed at
the location expected for the source and an upper limit determined for
that band.} between 3.6 and 24 \mum, which we incorporated.

Histograms of some of these cryo-era measurements for objects with
YSOVAR light curves appear in Figure~\ref{fig:magdist}. For the
objects with light curves, there are 3.6 \mum\ (IRAC-1) measurements
complete down to about 16.5 mags, and there are 4.5 \mum\ (IRAC-2)
measurements complete down to about 16 mags. The histograms are not
shown, but there are  5.8 \mum\ (IRAC-3) measurements complete down to
about 15 mags, and there are  8 \mum\ (IRAC-4) measurements complete
down to about 14 mags. Completeness is harder to assess for the MIPS
bands. There are 24 \mum\ (MIPS-1) measurements as faint as 10th mag
(Fig.~\ref{fig:magdist}). There are just five objects with light
curves and 70 \mum\ (MIPS-2) measurements; they range from $-$3.7 to
1.1 mags. Overall, the objects identified as part of the standard set
of members (defined in R14 and below in Sec.\ref{sec:members}) in
Figure~\ref{fig:magdist} are distinctly brighter than the rest of the
sources. While the overall peak [3.6] in Fig.~\ref{fig:magdist} is
fainter than 16th mag, the histogram of the standard set of members is
much flatter, and peaks at 14th mag. Similar behavior can be seen in
the other three IRAC bands. The sources bright enough to be seen at 24
\mum, however, are strongly biased towards those sources that are also
in the standard set of members, though these members are on average
brighter. 

Cryogenic data between 3.6 and 8 \mum\ are available for typically
90-100\% of all the objects with light curves, and essentially all of
the members (standard or augmented) or variables. Data at 24 \mum\
are rarer, with 62\% of all the light curves having a counterpart at
[24], though 80-90\% of the members (standard or augmented) or
variables have counterparts. Less than 1\% of the light curves have a
match at [70], and $\sim$4\% of the standard members or variables have
a match at [70]. 

\subsection{2MASS and 2MASS 6$\times$}

NGC 1333 was covered in the Two-Micron All Sky Survey (2MASS;
Skrutskie \etal\ 2006) and was also located in a field targeted by the
long exposure 6$\times$ 2MASS program. As described in R15, we
included these main 2MASS catalog and deeper 6$\times$ catalog NIR
$JHK_s$ data into our database.  We also include the $\sim$30 deep
$JH$ space-based measurements from Greissl \etal\ (2007) for the
targets discussed there. These data were merged into the rest of the
catalog by position with a 1$\arcsec$ search radius.

Histograms of the $K_s$ measurements for those objects with light
curves appear in Figure~\ref{fig:magdist}, and the $JHK_s$ values
themselves appear in R15 but are repeated in
Table~\ref{tab:singlevaluedata} for reference.  The $J$ histogram (not
shown) peaks at about 17.5 mag, the $H$ histogram (not shown) peaks at
about 17th mag, and the $K_s$ histogram (Fig.~\ref{fig:magdist}) peaks
at about 16th mag.  As at IRAC wavelengths, the NGC 1333 standard set
of members is distinctly brighter in $JHK_s$ than the rest of the
catalog. In the $K_s$ histogram, the distribution of standard members
peaks at $\sim$13, 3 magnitudes brighter than the peak of the rest of
the catalog.   Similar behavior can be seen at $J$ and $H$.

About half of all the objects with light curves have a counterpart at
$JHK_s$, and $\sim$90\% of the the members (standard or augmented) and
variables have such counterparts.

\subsection{Chandra ACIS}

Chandra X-ray Observatory Advanced CCD Imaging Spectrometer for
wide-field imaging (ACIS-I) observations of NGC 1333 were first
reported in Getman (2002) and then Winston \etal\ (2009, 2010). There
are three pointings in NGC 1333, with obsids 642, 6436, and 6437, with
a total exposure time of 119.3 ks. 

As we described in R14, we re-reduced the Chandra data in a
self-consistent way across most of the YSOVAR clusters. Source
detection was performed using CIAO (Chandra Interactive Analysis of
Observations; Fruscione \etal\ 2006). Sources, even faint ones, were
retained if they had a counterpart in the cryogenic IRAC catalog. We
also tested the X-ray light curves for variability using the
Gregory-Loredo method (GL-vary; Gregory \& Loredo 1992). This method
uses maximum-likelihood statistics and evaluates a large number of
possible break points from the prediction of constancy. This method
assigns an index to each lightcurve -- the higher the value of the
index, the greater the variability. Index values greater than 7
indicate $>99$\% variability probability. Values of the GL-vary index
$>$9 usually indicate flares. The GL-vary index is not reliable,
however, for sources with less than about 30 raw counts.  Values for
$F_x$, $L_x$, and GL-vary for sources in our region with X-ray
detections appear in Table~\ref{tab:singlevaluedata}. Sources from
Chandra were matched to the rest of the catalog with a
position-dependent search radius. There are no new sources that do not
already have an X-ray detection (from ROSAT, XMM, and/or Chandra) in
the literature (R15). Just 13\% of objects with light curves have an
X-ray counterpart. However, half the standard set of members (and
$\sim$40\% of the augmented set of members) have $L_x$. 

It is well-known that YSOs are often bright in X-rays, and thus we can
use X-ray data to identify YSOs that do not have IR excesses. As
discussed in R14, we improved our inventory of YSOs in NGC 1333 by
identifying objects with X-ray detections, IRAC counterparts, and SEDs
that are consistent with those of stars. We have adopted this YSO
selection mechanism as the other main component of the primary YSOVAR
sample definition; see R14, R15, and Sec.~\ref{sec:members} below.

\subsection{Subaru {\it Suprime-Cam} and CFHT {\it MegaCam}}

NGC 1333 was observed in several optical bands from Subaru and the
Canada-France-Hawaii Telescope (CFHT), and we obtained these data from
the public archives. Optical data from both telescopes cover the full
extent of the region monitored for YSOVAR. As noted in R15, broadband
optical imaging over this region has not been reported for many
sources, so its addition can contribute substantially to
identification of the object nature. 

NGC~1333 was observed in the $i^{\prime}$ and $z^{\prime}$ bands with
the {\it Suprime-Cam} wide-field camera (Miyazaki \etal\ 2002) on
Subaru on 2006-11-18. A total of 1 short exposure (10~s) and 60
dithered long exposures (60~s) were obtained in each filter as part of
the SONYC project (Scholz \etal\ 2009, 2012ab). The raw data and
corresponding calibrations were retrieved from the public archive
(Baba \etal\ 2002). 

The individual raw Subaru images were processed using an updated version of
\emph{Alambic} (Vandame 2002), a software suite developed and optimized for
the processing of large multi-CCD imagers, and adapted for {\it
Suprime-Cam}.  \emph{Alambic} includes standard processing procedures such
as overscan and bias subtraction for each individual  readout ports of each
CCD, flat-field correction, bad pixel masking, CCD-to-CCD gain
harmonization, and fringing correction, registration of the individual
images (using 2MASS as a reference) and stacking. Point-spread-function (PSF)
photometry was extracted from the final mosaics using {\it SExtractor}
(Bertin \& Arnouts 1996) and {\it PSFEx} (Bertin 2011). The photometric
zero-points (Vega-based zero points) were derived using standard fields
obtained the same night.  

NGC~1333 was observed in the $ugri$ filters at several epochs (see
Table~\ref{tab:megacam}) with the {\it MegaCam}  wide-field camera
(Boulade \etal\ 2003) mounted on the Canada France Hawai'i Telescope
(CFHT) as part of the DANCE survey (Bouy \etal\ 2013). The individual
CFHT images were retrieved from the public archives maintained at the
Canadian Astronomy Data Centre. 

The raw CFHT images were processed and calibrated with the recommended
{\it Elixir} system (Magnier \& Cuillandre 2004), which includes
detrending (darks, biases, flats and fringe frames). Nightly magnitude
zero-points (Vega-based zero points) were derived by the CFHT team
using Sloan Digital Sky Survey (SDSS) observations of standard star
fields (Smith \etal\ 2002, Landolt 1992). A precise astrometric and
photometric registration of all the individual images was obtained by
first extracting PSF photometry from the individual images using {\it
SExtractor} and {\it PSFEx}, and then registering and aligning the
individual catalogs on the same photometric scale using {\it Scamp}
(Bertin 2006). The final deep mosaics were produced using {\it SWarp}
(Bertin \etal\ 2002), and the sources astrometry and PSF photometry
were measured using {\it SExtractor}. 

The resultant photometry from the CFHT and Subaru telescopes is
well-matched to each other (where there is overlap), and on average
well-matched to the rest of the assembled SEDs. The data appear in
Table~\ref{tab:singlevaluedata}. Note that these are AB magnitudes,
consistent with SDSS convention, whereas the other magnitudes reported
here are Vega magnitudes. The reported errors on these data are purely
statistical, not systematic, and thus the errors are likely
underestimated. 

Histograms of the CFHT $r$ and the Subaru $z$ measurements for those
objects with light curves appear in Figure~\ref{fig:magdist}. The
distributions for all the optical bands peak at $\sim$20th mag and
reach as faint as $\sim$25th mag. Unlike the cryo-era Spitzer or
$JHK_s$ catalogs, the optical measurements were only obtained for 
those sources we thought might be `interesting', so they are strongly
biased towards members and/or variable objects.  This can be seen in
the histograms in Figs.~\ref{fig:magdist}, where the histograms for
all the CFHT $r$ or Subaru $z$ measurements are very similar to just
the members (selected in Sec.~\ref{sec:members}). Similar results are
obtained for the other optical bands.

Just 5-20\% of all the objects with light curves have an optical
measurement from Subaru or CFHT. About half of the standard set of
members (and $\sim$45\% of the augmented set of members) have a Subaru
or CFHT counterpart at any one band. About half of the variables have
a Subaru or CFHT counterpart at any one band.

\begin{deluxetable}{ccc}
\tablecaption{Overview of CFHT {\it MegaCam} observations\label{tab:megacam}}
\tablewidth{0pt}
\tablehead{\colhead{Date} & \colhead{Filters} & \colhead{Exp.~time} \\
\colhead{YYYY-MM-DD} & & \colhead{sec} }
\startdata
2005-09-27  & $g$   & 4$\times$40 \\
2008-12-21  & $i$   & 9$\times$560 \\
2009-03-24  & $i$   & 1$\times$56 \\
2009-03-26  & $i$   & 1$\times$56 \\
2009-09-21  & $i$   & 2$\times$10 \& 2$\times$160 \\
2005-09-27  & $r$   & 4$\times$40 \\
2005-09-27  & $u$   & 3$\times$150 \\
2005-12-23  & $u$   & 2$\times$150 \\
2006-02-28  & $u$   & 3$\times$150 \\
\enddata
\end{deluxetable}

\subsection{Literature Data}
\label{sec:litdata}

R15 assembled literature data from more than 25 published sources,
from 1994 to 2014, including cross-identifications and wavelengths
ranging from X-rays to 3 and 6 cm VLA data, but focusing on $J$ (1.25
\mum) to 24 \mum. For many of the older studies, the cross-IDs
(nomenclature) is the most important thing to retain, since the data
have since been reprocessed or superceded by later observations; see
R15. The catalog presented in R15 provides the literature basis on
which we build now. The coordinates in this catalog are all linked to
2MASS, so they are well-matched to the YSOVAR coordinates. 

We note here that most of the spectral types in the literature come
from searches for brown dwarfs, so the distribution is highly skewed
to mid-M and later.  There are about 100 objects with light curves
that have an estimate of spectral type in the literature, including
coarse ones (e.g., ``$<$M0").  Foster \etal\ (2015), in an analysis of
near-IR spectra obtained with the Apache Point Observatory Galactic
Evolution Experiment (APOGEE; Zasowski \etal\ 2013), obtain \teff\
values for many of their targets based on spectral model fitting.
Target selection for that project included considerations based on the
variability of objects in the YSOVAR data, so many of the objects in
which we are interested in the context of YSOVAR also have
measurements in Foster \etal\ from APOGEE.  The \teff\ values from
Foster \etal\ (2015) can be used to constrain the spectral type, even
though the \teff\ are much more uncertain. Spectral types and \teff\
values appear in R15 and are repeated in
Table~\ref{tab:singlevaluedata} for reference.

\section{Subsets of the Data and Sample Properties}
\label{sec:importantdatasubsets}

In this section, we define the scope of the dataset -- numbers of
sources and categories (members/non-members, variables, etc.).
Table~\ref{tab:countingthings} collects many of the relevant numbers
and fractions.

\subsection{IRAC Light Curves: Spatial and Temporal Distribution}

\begin{figure}[h]
\epsscale{0.8}
\plotone{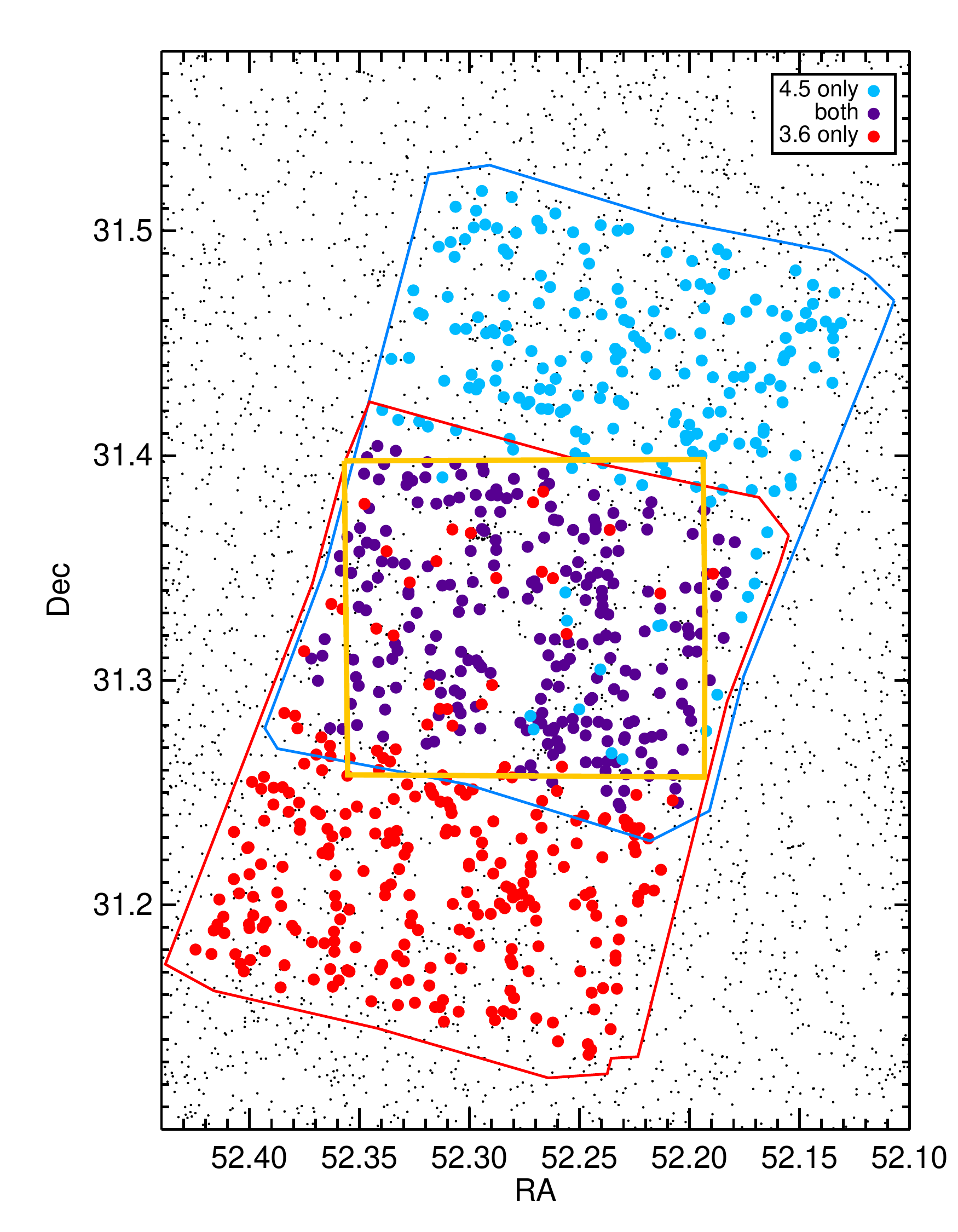}
\caption{Sky locations of objects with YSOVAR light curves. Small
black dots throughout region are objects on the sky in our literature
catalog (from R15). Large blue dots are objects with light curve only
in [3.6]. Large red dots are objects with light curve only in [4.5].
Large purple-black dots are objects with light curves in both [3.6]
and [4.5]. Objects with a light curve in only one band in the region
where there is 2-band coverage are for the most part faint, where one
band is below the detection threshhold and the other is not, or one
band is compromised by instrumental effects whereas the other is not.
Blue polygon is the [3.6] footprint; red polygon is the [4.5]
footprint; yellow square is the Chandra coverage. }
\label{fig:wherelcs}
\end{figure}

There are 701 objects in our NGC 1333 fields for which we have an IRAC
light curve with at least five viable\footnote{`Viable', meaning
obtained with sufficient redundancy on the sky as described in R14,
and not obviously compromised in the images due to artifacts or cosmic
rays; see discussion in R14.} points in 3.6 and/or 4.5 \mum,
194 just in 3.6 \mum, 265 just in 4.5 \mum, and 242 with light curves in both IRAC
channels.  Their locations on the sky
are shown in Figure~\ref{fig:wherelcs}. Objects with a light curve in
only one band in the region where there is 2-band coverage are for the
most part faint, where one band is below the detection threshhold and
the other is not, and the remaining cases are where one band is
compromised by instrumental effects whereas the other is not. 

\begin{figure}[h]
\epsscale{0.8}
\plotone{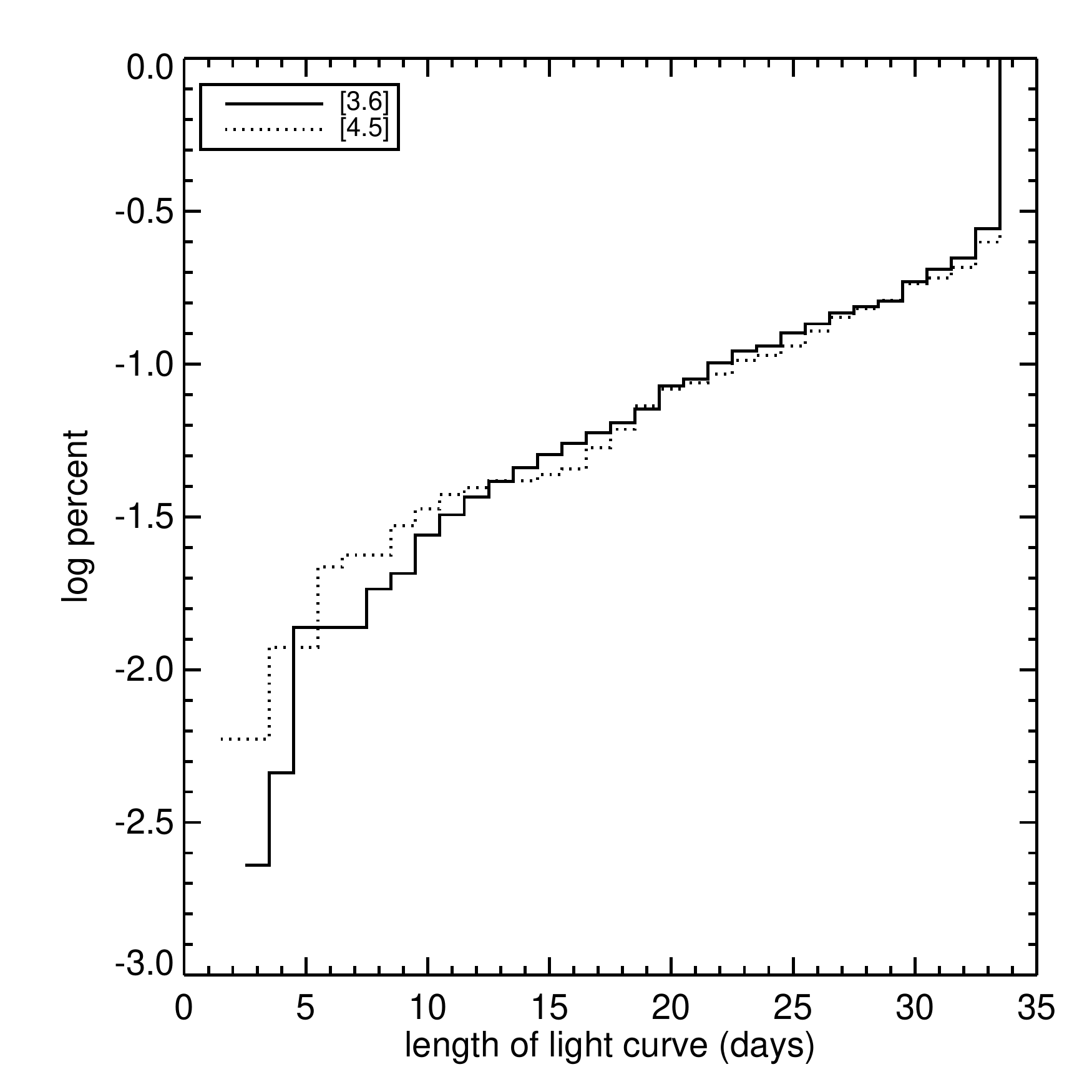}
\caption{Cumulative distribution functions of the length of the light
curves, in days, for the [3.6] (solid) and [4.5] (dotted). The
$y$-axis is the log of the sample fraction. The peaks (at 35 days) of
the histograms reach a total number of 315 for [3.6] and at 379 for
[4.5].  Most ($>$80\%) of the light curves overall, and most ($>$90\%)
of the light curves tagged variable, are longer than 30 days, with
strong peaks at 35 days.}
\label{fig:lclength4}
\end{figure}

As described in R14, all YSOVAR clusters have at least one sequence of
fast cadence observations, used to compute the `standard set for
statistics' to enable fair comparisons between clusters. In NGC 1333,
all of the YSOVAR observations are fast cadence.  We select variables
following R14; the approach is summarized below in
Sec.~\ref{sec:variables}. Figure~\ref{fig:lclength4} shows cumulative
distribution functions of the lengths of the light curves for both
channels. Most ($>$80\%) of the light curves overall, and most
($>$90\%) of the light curves tagged variable, are longer than 30
days, with a strong peak at $\sim$35 days. There are only five light
curves tagged variable with length $<$30 days in either channel; they
are often objects with light curves in both bands, though tagged
variable based on the characteristics of the longer light curve.

\clearpage

\subsection{Identifying Variables in the YSOVAR Campaign}
\label{sec:variables}

As described in detail in R14, we use three different methods of
finding variables in our light curves: Stetson index, $\chi^2$ relative
to a flat nonvariable lightcurve, and periodicity tests. 
We now briefly summarize this approach. 

For those objects where there are both [3.6] and [4.5] data, we can
calculate the Stetson index, given by:\begin{equation}
S=\frac{\sum\limits_{i=1}^p g_i
sgn(P_i)\sqrt{|P_i|}}{\sum\limits_{i=1}^N g_i} \end{equation} where
$p$ is the number of pairs of observations for a star taken at the
same time, $P_i$ is the product of the normalized residuals of two
observations, $g_i$ is the weight assigned to each normalized
residual, and $sgn$ means `the sign of'. In our case the weights are
all equal to one.  
As established in R14, if the Stetson index is $>$0.9, the object 
is variable. The values of the Stetson index for our
objects are given in Table~\ref{tab:singlevaluedata}. Fifty-six
objects in our field of view are identified as variable using the
Stetson index.

For those objects where only one band of data exists at all (or where
one band may be corrupted by instrumental effects), we can identify
variables from the chi-squared test ($\chi^2$), which, for a given
band, is given by \begin{equation}
\chi^2=\frac{1}{N-1}\sum\limits_{i=1}^{N}
\frac{(mag_i-\overline{mag})^2}{\sigma_i^2} \end{equation} where
$\sigma_i$  is the estimated photometric uncertainty. Again, as
established in R14, we take objects with $\chi^2 >5$ as variable. The
$\chi^2$ values for our objects are given in
Table~\ref{tab:singlevaluedata}, and the total counts of objects are
in Table~\ref{tab:countingthings}. For [3.6], 54 objects have
$\chi^2>5$, and for [4.5], 57 objects have $\chi^2>5$. Of these
$\chi^2$ variable objects, however, 45 have light curves in both IRAC
channels, and were selected as variable via their Stetson indices. As
a result, a total of 16 objects were newly identified as variable
by one-band $\chi^2$ calculations. 

Finally, as described in R14, we searched independently for periodic
behavior in the  [3.6], [4.5], and [3.6]$-$[4.5] light curves, between
0.05 and 15 d (from the sampling to $\sim$half the length of the
campaign). We adopt the period obtained from [3.6] as the most
reliable, then [4.5], and then [3.6]$-$[4.5] as the least reliable.
Periods measured for our objects are given in
Table~\ref{tab:singlevaluedata}. Any periodic objects not identified
as variable using the Stetson index or the $\chi^2$ test are added to
the set of variable objects. There are 23 objects for which we can
derive a period in NGC 1333; 6 of them are identified as variable
solely on the basis of the periodic signal.

For each of the variable objects, we manually inspected the light
curves and the images to make sure that the photometry was not
significantly compromised due to residual instrumental effects not
corrected by the processing described above (and more extensively in
R14).  As mentioned above and in R14, IRAC light curves for objects
fainter than $\sim$16 mag in either band are particularly noisy, and
were given special scrutiny. None of the objects in NGC 1333 that we
identify as variable have [3.6]$\sim$[4.5]$\gtrsim$16. One (SSTYSV
032912.05+311305.8) has, from the cryo era, [3.6]$>$16 and [4.5]$<$16,
but a YSOVAR light curve only in [4.5]. 

In the end, there are 78 objects in the NGC 1333 YSOVAR campaign data
that we have identified as mid-IR variable from the YSOVAR
fast-cadence monitoring. These mid-IR variable objects are shown on
the sky in the left panel of Figure~\ref{fig:wherevar}. Although the
variable objects were selected without regard to membership status,
the variable objects are clustered on the sky in a manner similar to
the members, e.g., there is a visible clumping of the variables in the
central region, and far fewer in the regions of single-band coverage. 

Additional information about, e.g., fraction of the standard members
(selected in \S\ref{sec:members} below) that are variable
(67/130=52\%), is provided in Table~\ref{tab:countingthings}. 

We note here explicitly that the remaining 701-78=623 objects are not
detected as variable (NDAV) in the YSOVAR campaign. Monitoring at a
different wavelength or cadence may find variability in these objects.

\subsection{Identifying Cryo-to-YSOVAR (CY) Variables}
\label{sec:cyvar}

Separately from searching for variability {\em within} the YSOVAR
campaign data, R14 looked for objects whose brightnesses at [3.6]
and/or [4.5] had changed significantly ($>$3$\sigma$ based on the
ensemble distribution) between the cryo-era observation and the mean
of the YSOVAR light curves. Out of the entire set of 701 IRAC light
curves in the NGC 1333 field, 92 can be identified as long-term mid-IR
variable in this approach.  In R14, those were referred to as
``long-term variables" or ``variable on the longest timescales". To
limit nomenclature confusion within the YSOVAR suite of papers, we
hereafter refer to these objects as ``Cryo-to-YSOVAR'' variables, or
CY variables. The fraction of CY variables (13\% over everything in
the field; see Table~\ref{tab:countingthings}) represents a higher
fraction of variables than was found in the YSOVAR campaign alone,
which is closer to 11\% (again, over everything in the field).  Just
32 of the 92 CY variables are also shorter-term variables, which
emphasizes that the mechanism for selecting variables (and likely the
underlying nature of the variability) is entirely different between
the two approaches. There is some expectation based on numerous
reports in the literature that larger amplitude variations are found
when analyzing data spanning longer temporal baselines, which might
explain these different fractions.  However, it is important to note
that the fraction of variables out of all objects in the field is
physically very different than the fraction of members (defined in
Sec.~\ref{sec:members} below). Out of the standard set of members,
35/130 ($\sim$30\%; see Table~\ref{tab:countingthings}) are CY
variables, 28 of which are also found as variable over the YSOVAR
campaign. The fact that there is a much higher fraction of member CY
variables, combined with the expectation that more YSOs will be
variable over a longer time baseline, gives us some hope that the CY
variables are not extensively contaminated, despite the fact that the
identification of variability relys upon a single epoch from the
cryo era.

\begin{figure}[h]
\epsscale{1.0}
\plottwo{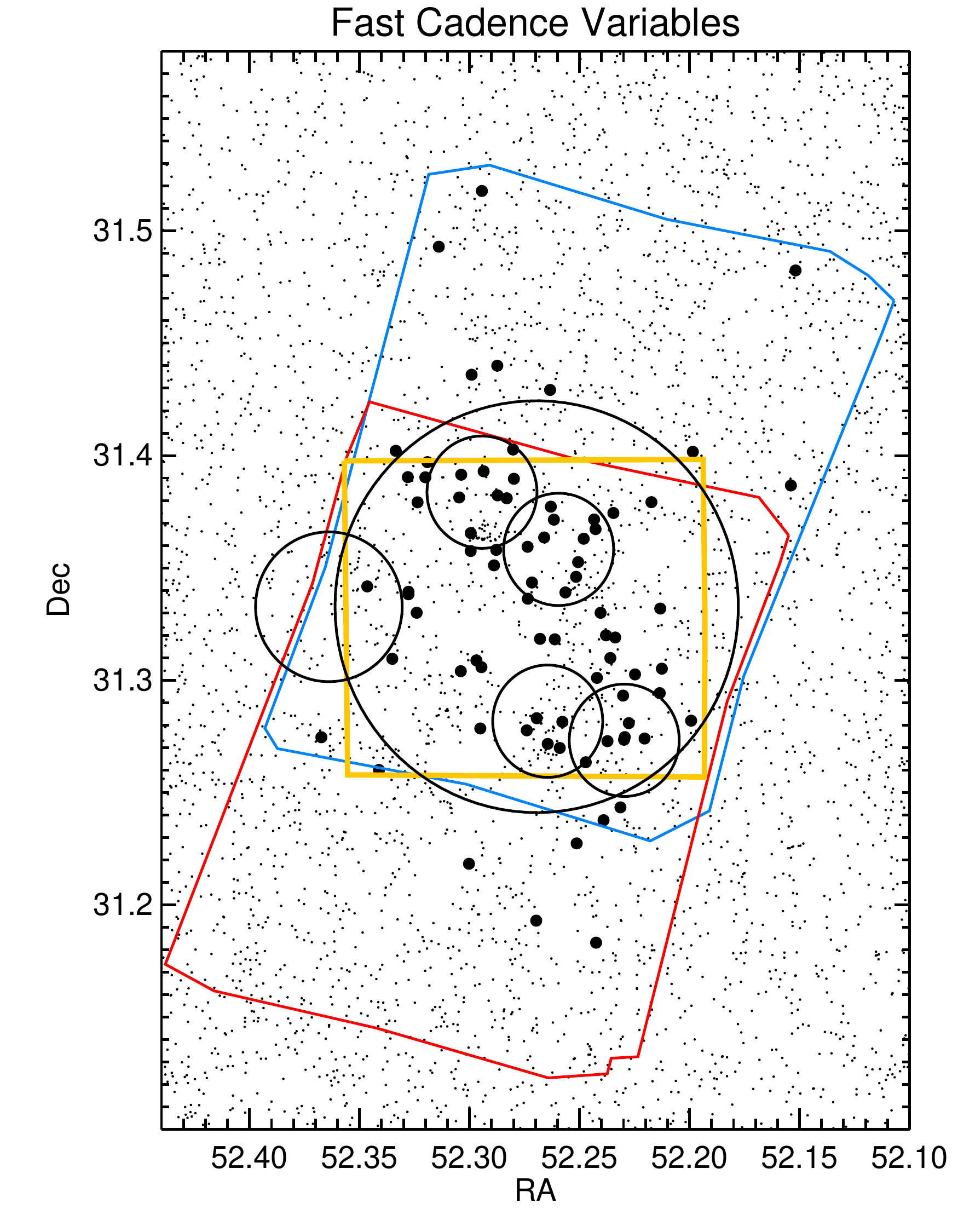}{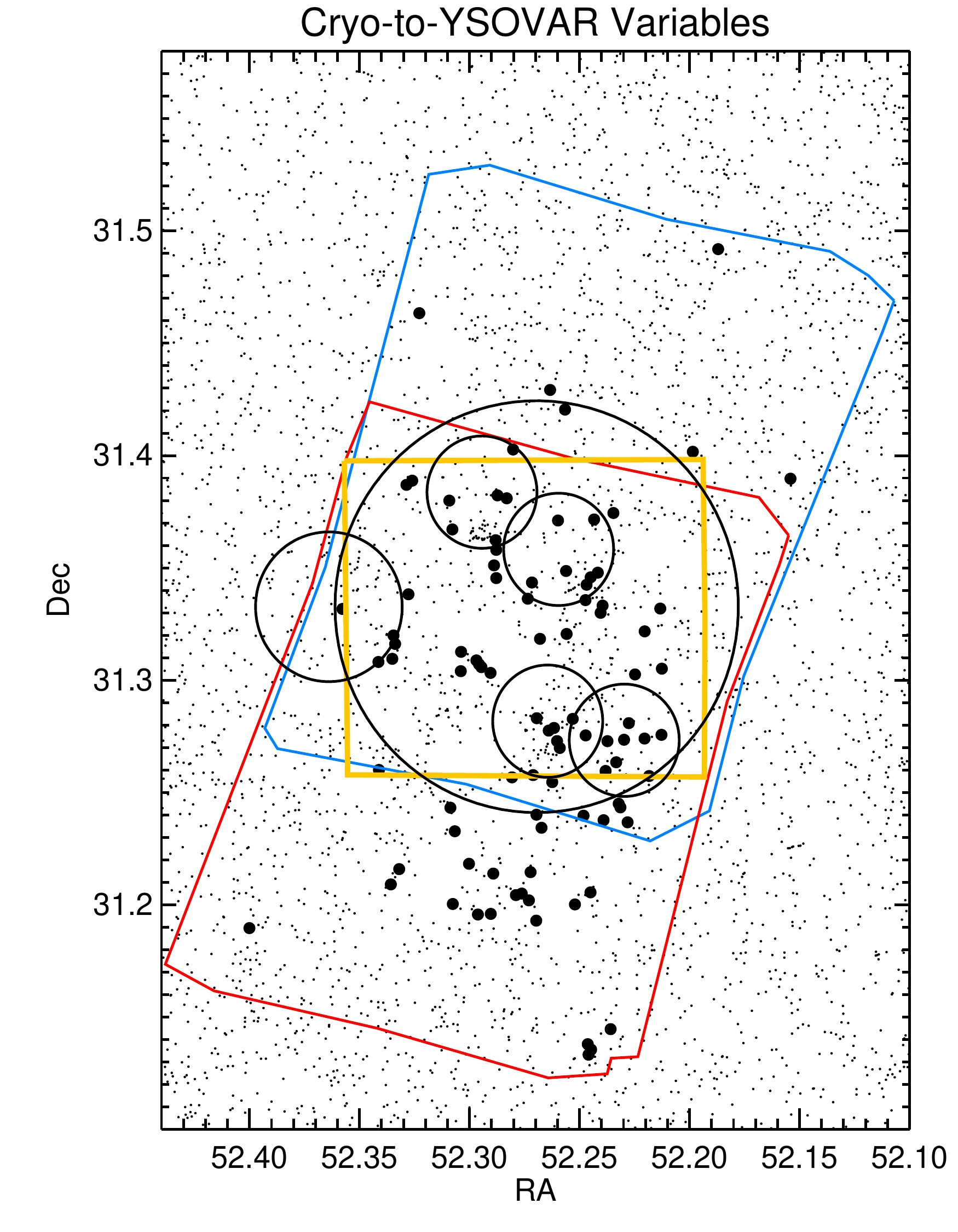}
\caption{Locations of objects identified as
variable in the mid-IR. Notation is similar to
Figure~\ref{fig:wheremem}, but large black dots indicate objects
identified as variable (see text) -- left panel has objects identified
as variable from the fast cadence (YSOVAR) data (timescales of day to
month); the right panel has objects identified as cryo-to-YSOVAR
(CY) variable via a comparison between the cryo-era and YSOVAR data
(timescales of years). Despite having been selected without reference
to membership status, variable objects are clustered in a manner very
similar to NGC 1333 members. The CY variable objects are not quite as
tightly clustered, particularly in the south, suggesting that there
may be a higher rate of non-cluster-member contamination.}
\label{fig:wherevar}
\end{figure}

The distribution of these CY variables on the sky is shown in the
right panel of Figure~\ref{fig:wherevar}. It can be seen to be
clustered in basically the same fashion as the members, which suggests
limited contamination. However, it can be seen by eye that it is also
somewhat less clustered than the standard set of members, particularly
in the south, suggesting that there may be a higher contamination rate
of non-cluster sources in this sample. On the other hand, the
distribution of literature-identified members not selected as standard
members (see \S\ref{sec:members}) also extends towards the south. 

\subsection{SED Classification}
\label{sec:sed}

After assembling all of the YSOVAR and literature photometry,
including the mean [3.6] and [4.5] measurement from our YSOVAR
campaign, we constructed each object's SED using all available data as
collected above and in R15. Some objects have well-defined SEDs, with
data from $u$ to 8, 24, or even 70 \mum; others have far
less-well-defined SEDs, for example, only one point that is the mean
from our YSOVAR campaign.

As discussed in R14 and R15, we defined an internally consistent SED
classification for the YSOVAR sources. In the spirit of Wilking \etal\
(2001), we define the near- to mid-IR (2 to 24 \mum) slope of the SED,
$\alpha = d \log \lambda F_{\lambda}/d \log  \lambda$,  classifying
sources with  $\alpha > 0.3$ as Class I objects, 0.3 to $-$0.3 as
flat-spectrum sources, $-$0.3 to $-$1.6 as Class II sources, and
$\alpha <-$1.6 for Class III sources.  For each of the YSOs and
candidate YSOs in our sample, we performed a simple ordinary least
squares linear fit to  all available photometry (just detections,
not including upper or lower limits)  as observed between 2 and 24
$\mu$m, inclusive.  Note that formal errors on the infrared points
are so small as to not affect the fitted SED slope. Note also that the
fit is performed on the observed SED, e.g., no reddening corrections
are applied to the observed photometry before fitting. Classification
via this method is provided specifically to enable comparison within
this paper (and to other YSOVAR papers) via internally consistent
means; see discussion in R14. We can perform this calculation only for
those objects with points at more than one wavelength in their SED
between 2 and 24 \mum. 

We adopt the SED slopes and classes determined by R15 between 2 and 24
\mum. R15 explicitly compares these 2-24 \mum\ SED classifications to
those determined between a restricted wavelength range (2-8 \mum) and
those SED classes reported in the literature.  The 2 to 24 \mum\
slopes and classes are repeated in Table~\ref{tab:singlevaluedata} for
reference.

\subsection{Standard Members}
\label{sec:members}

As discussed in R14, we took a two-pronged approach to identify a
`standard set of members' selected in the same fashion across all
YSOVAR clusters. We identify members for the standard set based on
their IR colors using the Gutermuth \etal\ (2008, 2010) and G09
selection algorithm, and with X-ray observations -- if they are
detected in X-rays and have a star-like SED (Class III), we add them
to the standard set of members.  Very few of these IR- and
X-ray-selected members have spectroscopic follow-up (or pre-existing
spectra in the literature), and thus, most should be considered YSO
candidates, though we include them all provisionally in the standard
set of members. 

\begin{figure}[h]
\epsscale{0.8}
\plotone{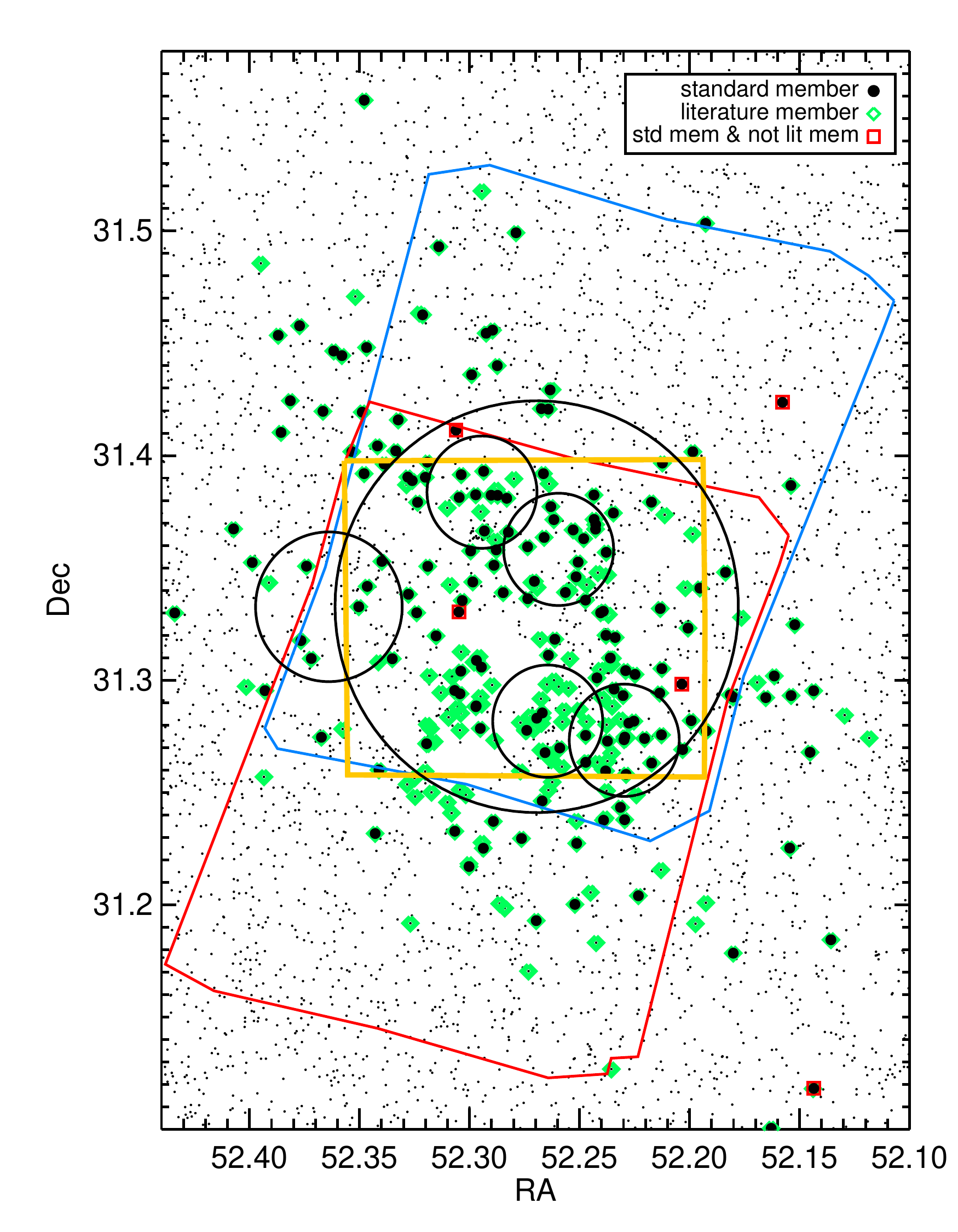}
\caption{Locations of objects identified as
members. Notation is similar to Figure~\ref{fig:wherelcs},
except now large black dots are objects identified as part of our
`standard set of members'; see text. Green diamonds are objects
identified in the literature as confirmed or candidate NGC 1333
members. The five objects shown here as part of the standard set of
members but not already identified in the literature as members have
an additional red square (only four of them are in the region with
light curves). The large black circles are the clusters
identified in Gutermuth \etal\ (2009). Nearly all of the standard
members are identified in the literature; there is a significant
number of literature member objects not selected for the standard set
of members. The literature-selected members are included in the
augmented set of members. Member objects (of either type) are
clustered towards the center of the observed region. }
\label{fig:wheremem}
\end{figure}

Out of 701 objects in NGC 1333 with light curves, 130 are identified
as part of the standard set of members. Most, 102, are identified
using the G09 IR selection method; the remainder are identified via
the X-ray+Class III SED approach (individual objects so identified are
indicated in Table~\ref{tab:singlevaluedata}). There are just four
objects with light curves that are part of the standard set of members
that are not already identified in the literature as YSO candidates;
SSTYSV J032913.16+311949.4 and J032848.83+311753.7 are identified from
IR colors, and SSTYSV J032913.47+312440.7 and J032837.85+312525.3 are
identified via this X-ray+Class III identification. None of these four
are variable, and they are faint.  Figure~\ref{fig:wheremem} shows
that the identified members are clustered towards the central region
of our observations.  Since the X-ray data do not cover the entire
cluster, Class III objects can only be identified in the central
region. Even without these members selected via X-rays, however, the
IR-selected members are still clearly clumped towards the central
region. Gutermuth \etal\ (2008) identified a main cluster with
subclusters within it; these are indicated in
Figure~\ref{fig:wheremem}.  Note that our formal definition of
standard members does not require a light curve, and as such there are
members indicated in Fig.~\ref{fig:wheremem} outside of the region
with light curves. However, because the purpose of our present
analysis is the time series data, we  consider only those members with
light curves for the rest of this paper.

There are 72 objects with YSOVAR light curves that are identified in
the literature as confirmed or candidate YSOs (see R15), but that are
not selected as candidate YSOs by our approach (see
Fig.~\ref{fig:wheremem}).  Some of these objects are not selected by
our approach because the object is undetected in X-rays and has no IR
excess. It is more often the case, however, that the object is not
identified from IR excess because it is missing at least one band; the
G09 YSO selection mechanism requires a minimum signal-to-noise and a
minimum number of bands for classification, and objects missing bands
thus show up as `unclassifiable' in the G09 approach.  These
literature members are not predominantly bright nor faint; they are
scattered through the distribution of brightnesses in [3.6], though
because of the overall shape of the distribution of sources (see
Fig.~\ref{fig:magdist}), they represent a higher fraction of the
brighter sources than the fainter sources. They are, however, often
near bright sources such that measurements in some bands are difficult
(or even impossible) to obtain, which is almost certainly why the
bands are missing in the first place.  These literature-only members
are also clustered, but there are more of them (than the standard
members) slightly towards the south.  These candidate members that are
not part of the standard set of members (but that have light curves)
are included in the `augmented set of members'
(Sec.~\ref{sec:augmembers}).

\subsection{Augmented Members}
\label{sec:augmembers}

As discussed in R14, we have made provisions for establishing an
``augmented set of members'' to encompass objects not selected via the
standard membership approach, but identified as variable and/or as
members in the literature. For NGC 1333, we believe that all the
objects that we identify as variable over the YSOVAR campaign are
likely members.  

The 72 additional literature members are thus included in our
augmented members, as well as the 12 objects identified as variable,
but not part of the standard set of members. Seven of those are
literature-identified members not in the standard set, and five are
newly identified candidate members. Our augmented set of members is,
at 207 objects, significantly larger than the standard set of
members (130 objects). 

The five newly identified candidate members do not stand out in
properties; they have comparable ranges of brightness, light curve
shapes, etc., to the set of more established members. However, as with
a significant fraction of the sample, spectroscopy is required to
establish membership.

\subsection{Contamination and Completeness}
\label{sec:contam}

Ideally, we would like to be able to quantify the degree of
contamination and completeness in our member sample.  To do this
authoritatively requires a substantial number of spectra of the
objects identified as standard members to understand contamination,
and spectra of other objects with similar colors to estimate
completeness. This is beyond the scope of this work. However, we
try to estimate the level of contamination (and completeness) in the
membership.

Our standard set of members uses IR excess and X-rays to identify
members, and nearly all the standard set of members are already
identified in the literature as YSOs. However, the  standard members,
even those that have been previously identified as YSOs, are not
necessarily spectroscopically confirmed members or even confirmed to
be stellar. Since NGC 1333 is relatively close, the chances of
foreground contamination are low, and NGC 1333's own cloud helps limit
background contamination. However, likely sources of contamination
include (but are not limited to) RSCVn or other variable binaries, and
background asymptotic giant branch (AGB) stars that appear faint and
dusty (IR-bright). Active galactic nuclei (AGN) are another possible
contaminant, but most of these should have been removed by our member
selection approach.  Additionally, the X-ray data are limited to the
central region, so there is no way of finding X-ray detected Class III
objects in the outer reaches of the cluster (specifically in the
YSOVAR single-band coverage region). This is an obvious incompleteness
that could be remedied by additional X-ray observations over the
larger region.  

Unlike other YSOVAR clusters, there is copious literature identifying
candidate members in NGC 1333, which is why we defined the standard
set of members, so as to enable comparisons across clusters. However,
our standard set of members in NGC 1333 has missed a substantial
number of literature-identified sources. The literature has used
X-rays and IR excess, but also color-color and color-magnitude
diagrams with follow-up spectroscopy (for brown dwarfs), and
clustering to find members; in the present paper, we can use
variability to find still more members. This inhomogeneous selection
can, at its best, achieve a complete inventory of members, but at its
worst, introduce complicated biases into the sample.

The literature-identified members not in our standard set of members 
probably do not represent a large source of contamination per se, 
given the reasons why they are missing (\S\ref{sec:members}).
But the sheer numbers of literature-identified members not in our
standard set of members suggests that our standard set of members is
highly incomplete.

Because NGC 1333 is relatively close to us, we expect that the known
members will be among the brighter sources in the region, and this is
borne out by Fig.~\ref{fig:magdist} -- the standard members indeed
dominate the bright sides of the histograms. The fainter sources are
where we expect more contamination to be introduced, but this is
greatly complicated by the high (and clumpy) extinction towards NGC
1333, making sources appear fainter than they are.  There are no
obvious indications of contamination in Fig.~\ref{fig:magdist}, such
as large numbers of faint member candidates.  This could, however, be
a circular argument -- the known members are bright because the
cluster is close, but it is also the case that previous investigators
identify the bright members first. New members are more likely to be
fainter if the known objects are also the bright objects.

We also expect that the members will be clustered on the sky (e.g.,
Gutermuth \etal\ 2008), and Fig.~\ref{fig:wheremem} reflects that --
both the standard members and the literature members are clustered on
the sky in the same region identified by Gutermuth \etal\ (2008) as
the main portion of the cluster.   The variables follow the same
overall clustering pattern; the variables from the fast cadence YSOVAR
campaign are more tightly clustered than the CY variables. It is hard
to assess the degree to which the variable selection approach affects
the clustering of the selected variables; there are more variables in
the region with 2-band coverage, but there are also more bonafide
cluster members in that region. The YSOVAR campaign variables use many
epochs of observation; in contrast, the CY variables depend on the
reliability of the cryo-epoch point, so they seem more likely to be
contaminated.  There are more CY variables to the south of the main
cluster than the fast cadence selection. However, the standard set of
members and the literature members both also identify objects to the
south of the main cluster region.

\begin{figure}[h]
\epsscale{0.8}
\plotone{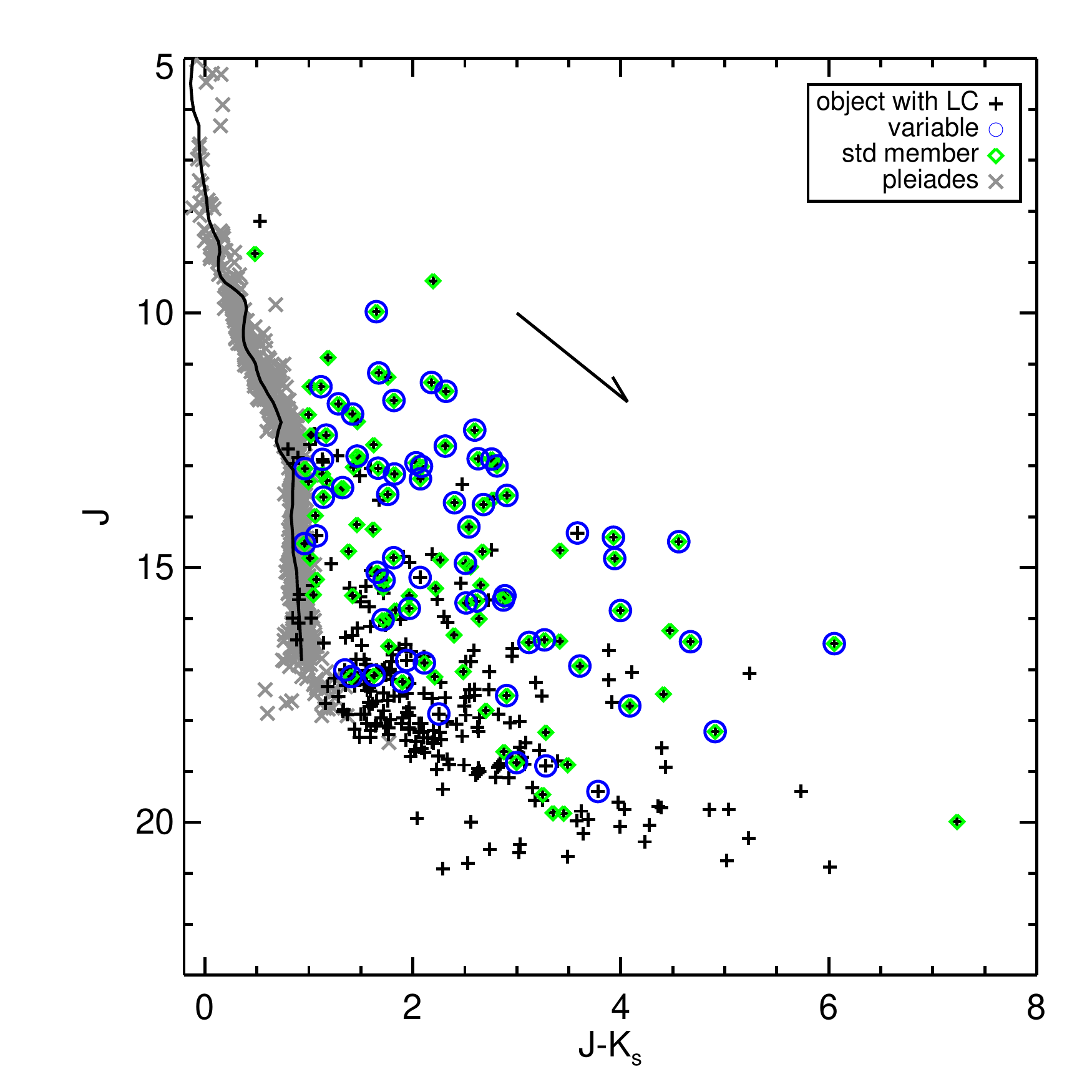}
\caption{$J$ vs.\ $J-K_s$ for the sample. Crosses are objects that
have light curves; an additional blue circle indicates that it was
identified as a variable over the YSOVAR campaign, and an additional
green diamond indicates that it is in the standard set of members.
Grey $\times$ symbols are Pleiades members (moved from 133 pc to 235
pc), and the black line is the expected main sequence (at 235 pc). The
arrow corresponds to \av=6 mag.
This plot does not suggest substantial contamination of the members or
variables by non-members. The variables that are not a priori
selected as members are well- integrated with the distribution
of the rest of the members.}
\label{fig:jhk3}
\end{figure}

We can attempt to assess the degree of contamination by looking at
color-color and color-magnitude diagrams (CMD). The optical catalog is
strongly biased towards `interesting' objects, as well as strongly
biased towards the less embedded objects and affected by spatially
variable \av. Without spectral types to individually deredden the
photometry, an optical CMD is not the best for this purpose.
Figure~\ref{fig:jhk3} shows a NIR CMD for our sample compared to
Pleiades members (Stauffer \etal\ 2007) (moved from 133 pc to 235 pc).
The members and variables are generally bright and red, as expected.
The objects with light curves that are unlikely members are generally
faint, consistent with background stars or galaxies. There are no blue
objects below the main sequence (MS), which would be foreground stars.
This plot does not indicate that contamination is a substantial
problem. There is likely strong reddening affecting many of the
members, which we knew would be an issue. (The \av\ vector as shown
corresponds to \av=6 mags, which is the value for the heart of the
cluster in the large map shown in Ridge \etal\ 2006.) The  variables
not in the standard set of members are all between  $J\sim$12 and 20,
and $J-K_s\sim$1 and 4. They are well-integrated with the standard
member distribution, consistent with those objects being likely
members. 

\begin{figure}[h]
\epsscale{1.0}
\plottwo{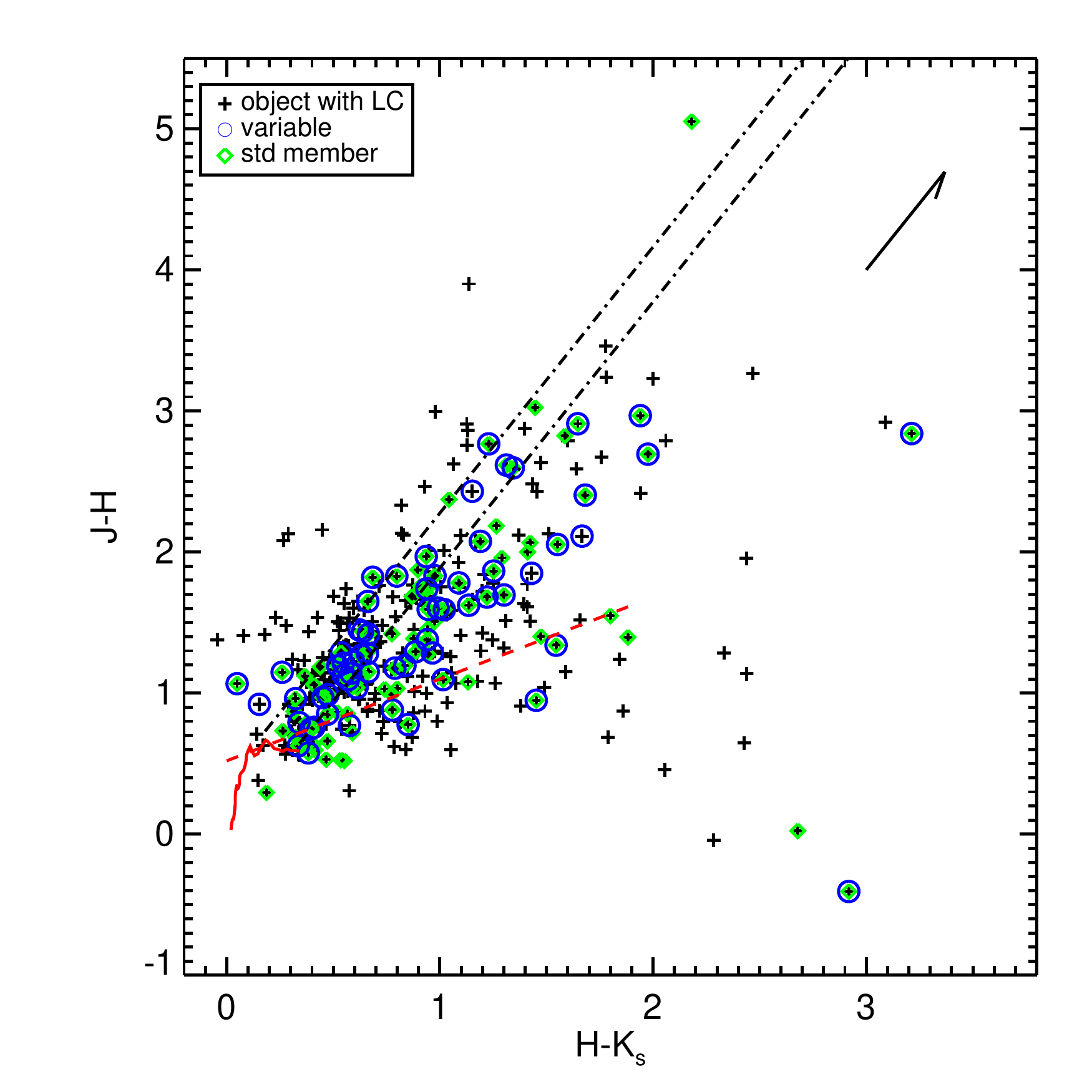}{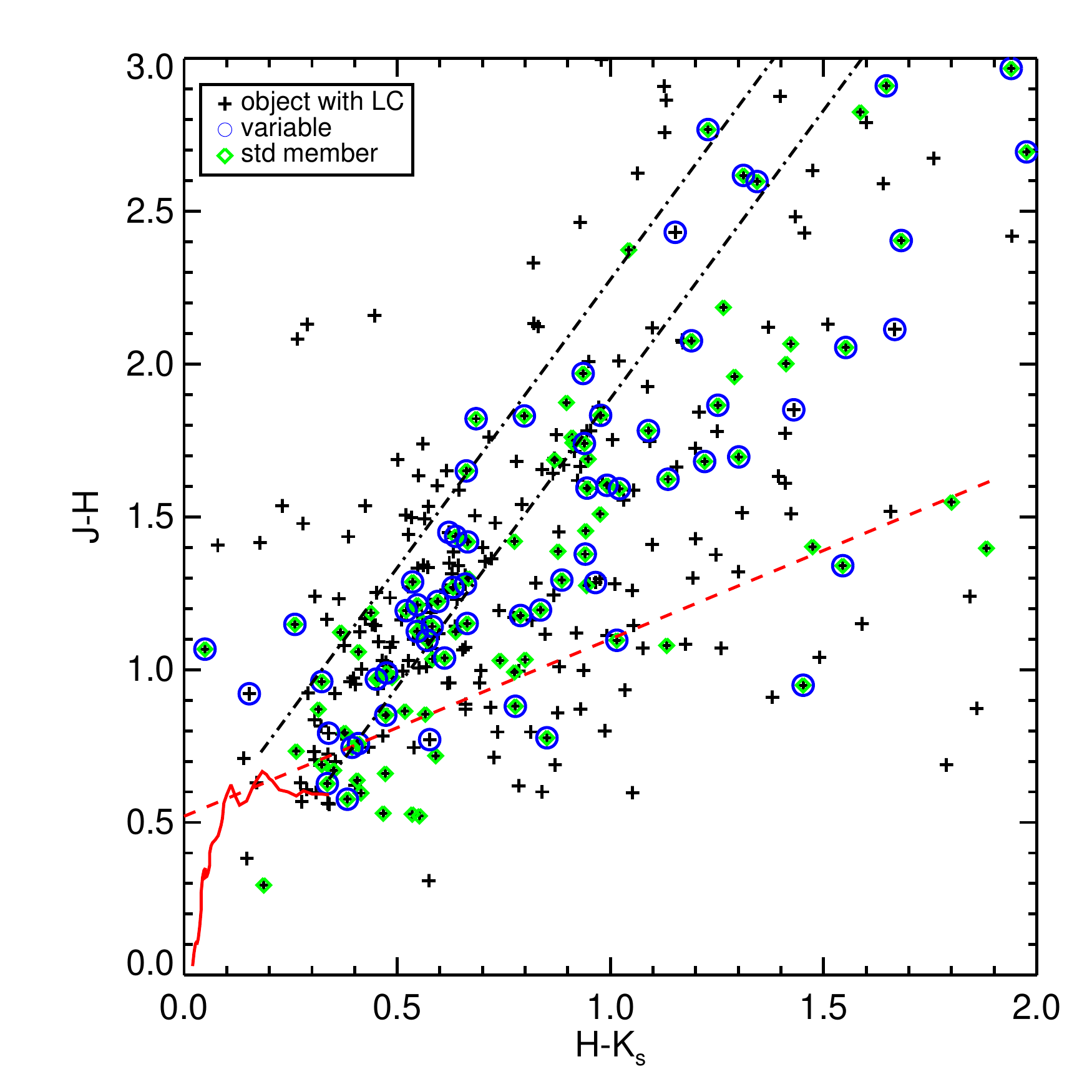}
\caption{$J-H$, $H-K_s$ for objects with light curves in the catalog,
with variables and the standard set of members indicated as shown.
(The right panel is a zoom-in on the densest region from the left.)
The red solid line is the expected ZAMS location; the red dashed line
is the classical T Tauri locus from Meyer \etal\ (1997). A reddening
vector for \av=6 is shown; the black dot-dash line are extensions of
the ZAMS according to this reddening law. This plot suggests that
there is not a large contamination rate from foreground/background
sources; see text.}
\label{fig:jhk}
\end{figure}

Figure~\ref{fig:jhk} is a NIR color-color diagram with the standard
set of members and variables (over the YSOVAR campaign) indicated. 
The standard members (and variables) are located in the expected
location for YSOs. There are few sources below or even on the MS
relation, suggesting at minimum that there is little foreground
contamination.  There are large numbers of objects but, somewhat
surprisingly, relatively few variables or standard members near the
classical T~Tauri locus from Meyer \etal\ (1997);  reddening is
evidently pushing objects off the locus. There are many objects in a
location consistent with substantial reddening of the MS. This could
be suggestive of contamination. The most likely objects in the
standard set of members to be reddened MS non-members would be the
X-ray selected Class III SED objects, but the X-ray selected objects
are reasonably well-distributed through the plot. There are very few
standard members or variables in the upper left corner of the plot,
which would also be suggestive of contamination. We conclude that this
plot, as Fig.~\ref{fig:jhk3}, suggests both that there is not
substantial contamination, but also that there is substantial
reddening affecting some sources.

\begin{figure}[h]
\epsscale{0.8}
\plotone{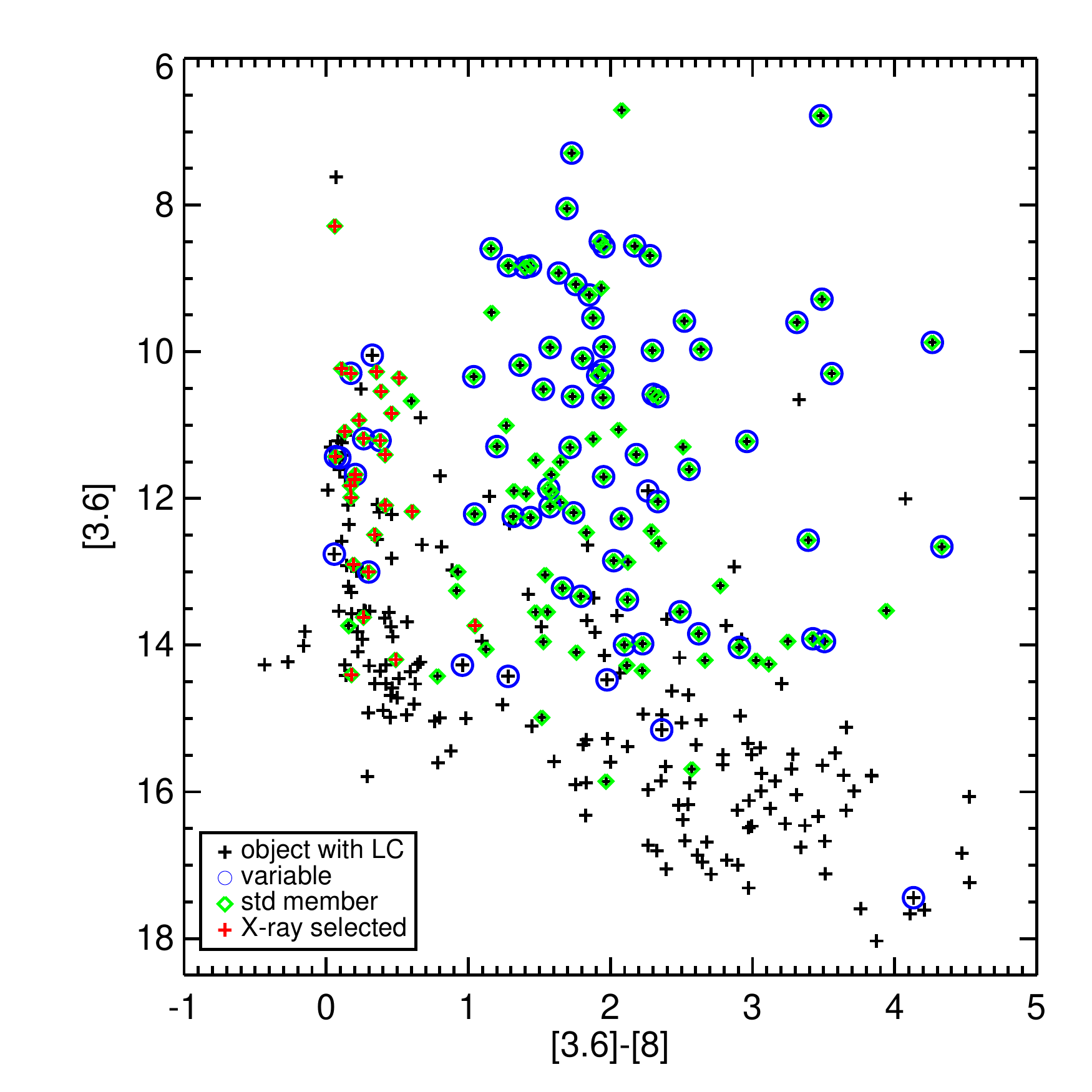}
\caption{[3.6] vs.\ [3.6]$-$[8] for objects with light curves in the
catalog, with variables and the standard set of members indicated as
shown. An overplotted red cross is a standard member selected from
X-ray detection and Class III SED shape. A reddening vector
corresponding to \av$\sim$6 is comparable to the symbol size here, so
is not shown. This plot also suggests that contamination is not a
major issue. }
\label{fig:iraccmd}
\end{figure}

Figure~\ref{fig:iraccmd} is an IRAC color-magnitude diagram for the
sample with [3.6] and [8] detections from the cryo era. Nearly all the
variables and standard members are bright, and most are red,
consistent with the locations expected for YSOs. This plot, as for the
NIR, does not suggest that contamination is a major issue. As a
result of our mechanism for selection of standard members, the members
with [3.6]$-$[8]$\sim$0 mag are nearly all selected on the basis of
X-ray detection. Many -- but not all -- of the
variables that are not also standard members are also faint in [3.6].
This is consistent with expectations that most of the members that are
bright are already identified.  Most of these new objects are brighter
than [3.6]$\sim$15.5 mag; light curves fainter than $\sim$16 mag are
difficult to identify as variable. The one object fainter than
[3.6]$\sim$17 mag is SSTYSV J032912.05+311305.8, and it has a SED such
that [4.5]$\sim$12 mag, and it is identified as variable from the
[4.5] light curve. It is identified in the literature as a YSO only by
c2d. (We note that the object at this location has sometimes been
matched to IRAS4b and SK3, but that identification is uncertain; we
have matched IRAS4b and SK3 more securely to another source; see
R15).  The remaining variables that are not also standard members are
in the photospheric locus near [3.6]$-$[8]$\sim$0 mag, but are not
detected in X-rays.  

The newly identified variable members will require spectroscopic
observation to confirm (or refute) their membership. However, their
properties suggest that they are well-integrated with the rest of the
members (standard or augmented sets). Their light curves do not
generally distinguish them from the other members; they blend in with
the other likely members. (We note that SSTYSV J032911.86+312155.7 may
be an exception -- it is variable, and not a literature or standard
member, but its light curve and for that matter SED are different from
other objects; see discussion below in
Sec.~\ref{sec:032911.86+312155.7}.) 

Summarizing the major issues addressed in this section, to the degree
that we can tell, contamination is not obviously a factor, though
completeness is an issue. There are substantial numbers of YSOs
identified in the literature not selected via our approach, but this
is for the most part because at least 1 band that is needed for
identification is missing. Contamination is most likely to be
background sources, possibly AGN or AGB stars. The standard members
are significantly brighter on average than the rest of the sample. The
objects newly identified as members through variability share
properties with the previously-identified members in every diagnostic
we have used. Spectroscopy and possibly X-ray observations over a
larger region will contribute to our understanding of contamination
and completeness in this region.  Many of the objects are subject to
large and spatially variable \av, which complicates such followup.

\clearpage

\section{Variable Fractions and Other Fractions of Note}
\label{sec:countingthings}

\begin{deluxetable}{cccc}
\tablecaption{Variable fractions and other fractions of note\label{tab:countingthings}}
\tabletypesize{\scriptsize}
\tablewidth{0pt}
\rotate
\tablehead{\colhead{Category} & \colhead{All
objects\tablenotemark{a} } 
& \colhead{Std.~members\tablenotemark{a}} & \colhead{Aug.~members\tablenotemark{a}} }
\startdata
\tableline
\sidehead{Detections and Members}
Light curve exists at [3.6] & 436/701=0.62$\pm$0.04 & 116/130=0.89$\pm$0.11 & 168/207=0.81$\pm$0.08 \\
Light curve exists at [4.5] & 507/701=0.72$\pm$0.04 & 112/130=0.86$\pm$0.11 & 179/207=0.86$\pm$0.09\\
Light curve exists at [3.6] \& [4.5] & 242/701=0.35$\pm$0.03 & 98/130=0.75$\pm$0.10 & 140/207=0.68$\pm$0.07\\
Standard members            & 130/701=0.19$\pm$0.02 & \nodata      & 130/207=0.63$\pm$0.07 \\
\tableline
\sidehead{SED classes}
Object with Class I SED slope   & 100/690=0.14$\pm$0.02 & 21/130=0.16$^{+0.04}_{-0.03}$ & 38/207=0.18$\pm$0.03\\
Object with Flat SED slope      & 82/690=0.12$\pm$0.01  & 22/130=0.17$^{+0.04}_{-0.03}$ & 30/207=0.14$\pm$0.03\\
Object with Class II SED slope  & 197/690=0.29$\pm$0.02 & 56/130=0.43$\pm$0.04          & 82/207=0.40$\pm$0.05\\
Object with Class III SED slope & 311/690=0.45$\pm$0.03 & 31/130=0.24$^{+0.04}_{-0.03}$ & 57/207=0.28$\pm$0.04\\
\tableline
\sidehead{YSOVAR Variability}
Variable via Stetson        & 56/236=0.24$\pm$0.04  & 51/98=0.52$\pm$0.09  & 56/138=0.41$\pm$0.06\\
Variable via $\chi^2$ at [3.6] & 53/436=0.12$\pm$0.02 & 50/116=0.43$^{+0.05}_{-0.04}$ & 54/168=0.32$^{+0.04}_{-0.03}$\\
Variable via $\chi^2$ at [4.5] & 57/507=0.11$\pm$0.02 & 52/112=0.46$\pm$0.05 & 57/179=0.32$\pm$0.05\\
Periodic                       & 23/701=0.033$\pm$0.007 & 19/130=0.15$^{+0.04}_{-0.03}$ & 23/207=0.11$\pm$0.02\\
{\bf Variable over YSOVAR campaign}  & {\bf 78/701=0.11$\pm$0.01} & {\bf 66/130=0.51$\pm$0.08}  & {\bf 78/207=0.38$\pm$0.05} \\
\tableline
\sidehead{YSOVAR Variability and SED classes}
Variable (YSOVAR) object with Class I SED slope   & 17/100=0.17$^{+0.04}_{-0.03}$ & 13/21=0.62$^{+0.09}_{-0.11}$ & 17/38=0.45$\pm$0.08 \\
Variable (YSOVAR) object with Flat SED slope      & 19/82=0.23$^{+0.05}_{-0.04}$ & 16/22=0.73$^{+0.07}_{-0.11}$ & 19/30=0.63$^{+0.08}_{-0.09}$\\
Variable (YSOVAR) object with Class II SED slope  & 34/197=0.17$\pm$0.03 & 32/56=0.57$^{+0.06}_{-0.07}$ & 34/82=0.41$^{+0.06}_{-0.05}$\\
Variable (YSOVAR) object with Class III SED slope & 8/311=0.026$\pm$0.009 & 5/31=0.16$^{+0.09}_{-0.05}$ & 8/57=0.14$^{+0.06}_{-0.04}$\\
{\bf Variable (YSOVAR) Disked objects (I+F+II)} & {\bf 70/379=0.18$\pm$0.02 } & {\bf 61/99=0.62$\pm$0.05} & {\bf 70/150=0.47$\pm$0.04}\\
\tableline
\sidehead{Cryo-to-YSOVAR (CY) Variability}
CY variable             & 92/701=0.13$\pm$0.01 & 35/130=0.27$\pm$0.04 & 48/207=0.23$\pm$0.04\\
Variable over CY and YSOVAR campaign & 32/701=0.045$\pm$0.008 & 28/130=0.22$^{+0.04}_{-0.03}$ & 32/207=0.15$\pm$0.03\\
\tableline
\sidehead{YSOVAR and X-rays}
$L_x$ and IR light curve exists & 89/701=0.13$\pm$0.01 & 70/130=0.54$\pm$0.04 & 79/207=0.38$\pm$0.05 \\
$L_x$ and variable (YSOVAR) & 41/89=0.46$\pm$0.05 & 40/70=0.57$\pm$0.06 & 41/79=0.52$\pm$0.05 \\
GL-vary$>$7 and IR light curve exists\tablenotemark{b}& 17/90=0.19$^{+0.05}_{-0.04}$ & 16/71=0.23$^{+0.06}_{-0.04}$ & 17/80=0.21$^{+0.05}_{-0.04}$ \\
GL-vary$>$7 and variable (YSOVAR) & 13/41=0.32$^{+0.08}_{-0.06}$ & 13/71=0.18$^{+0.05}_{-0.03}$ & 13/80=0.16$^{+0.05}_{-0.03}$ \\
\enddata
\tablenotetext{a}{The last three columns consist of the numbers and
fractions of objects in that category. Ex: 436 objects have a light
curve at [3.6] out of 701 that have a light curve at any band, which
is 62\%; 116 objects out of 130 standard members have a light curve 
at [3.6], or 89\%; 168 objects out of 207 augmented members (81\%)
have a light curve at [3.6].}
\tablenotetext{b}{GL-vary can be calculated for 90 objects, but $L_x$
can only be calculated for 89 objects.}
\end{deluxetable}

\subsection{NGC 1333}
\label{sec:varfrac1333}

Table~\ref{tab:countingthings} collects variability fractions, among
other things, for NGC 1333. Errors in the table are calculated using
the larger of either Poisson or the binomial statistics from the
appendix in Burgasser \etal\ (2003).  

We expect that the bonafide cluster members are more likely to be
variable in the YSOVAR campaign than field stars, and that is borne
out by the data.  The standard set of members is about 20\% of the
entire set of objects for which we have light curves. While $\sim$11\%
of all the light curves are variable, we find that about half,
$\sim$52\%, of the standard set of members are variable over the
YSOVAR campaign. We also expect that a higher fraction of the disked
YSOs with disks will be variable (compared to all YSOs), and we find
this; $\sim$62\% of the disked standard members are variable in the
MIR.

In Sec.~\ref{sec:contam} above, we argued that contamination in the
standard set of members is likely not substantial, but we should also
consider the `missing' members that appear in the literature but not
in our standard set of members. Among the augmented member sample
(including the literature-identified members and new variable
members), the variable fraction is lower than that for the standard
members, just $\sim$38\%. Only $\sim$10\% of these
literature-identified members not included in our standard set of
members are variable in the YSOVAR campaign.  A higher fraction, 47\%,
of the disked augmented members are variable, but it is still lower
than the analogous number for the standard members. Since we have a
baseline expectation that YSOs will be variable, this is surprising,
and is perhaps suggestive of a high contamination rate among these
literature-identified members. 

We considered whether we have properly identified all of the
variables.  R14 discussed the variability detection limits, so we do
not repeat it here.  In R14, we set conservative bounds on the
$\chi^2$ limits for variablity in the hopes that we would only include
high-confidence variables (at the expense of likely missing some
lower-level legitimate variables). In cases where we have light curves
in 2 bands, and the Stetson index identifies a variable, the
likelihood that it is really a variable is higher than if we have only
one band and can only identify the object as variable based on the
$\chi^2$. In NGC 1333, we have some hope that the use of the $\chi^2$
metric for selecting variables is indeed not unduly polluting the
variables. Over the entire set of light curves, the variable fraction 
using the Stetson metric is larger by about a factor of two than the
fraction based on $\chi^2$. Among the standard set of members, the
fractions are all reasonably comparable.  This suggests that we are
not adding many false variables by relying on $\chi^2$ to find some
variables. However, we have no easy way of assessing how many
variables we may be missing.

Another possible influence on our standard set of members is the use
of X-rays to identify members.  We do not have complete X-ray
coverage; only 13\% of the objects with light curves have aX-ray
counterparts. However, in part because of our mechanism for defining
variables, but also in part because young stars are brighter in X-rays
than older stars, 54\% of the standard set of members have an X-ray
counterpart. Of the 70 with a calculated log $L_x$, almost 60\% are
variable over the YSOVAR campaign. As discussed above, we also checked
for variability in X-rays using the GL-vary metric; just 18\% of
standard members with light curves are variable in YSOVAR and also
variable in X-rays. The length of the monitoring campaigns is very
different between the IR and X-ray observations, admittedly.  But the
different variability fractions are consistent with the origin of
IR variability being very different from the origin of X-ray
variability (see, e.g., Flaherty \etal\ 2014).

As mentioned above and in R15, only some of the NGC 1333 members
(standard or augmented) have spectroscopic follow-up, but perhaps we
can gain insight into the variability fraction by considering a
limited number of members. The most recent spectroscopic study is
Foster \etal\ (2015), who observed in the NIR. Out of their sample,
there are 150 members for which we also have light curves, and 67
(44$\pm$7\%) are variable, an even lower fraction than among our
standard members. However, the Foster \etal\ (2015) target selection
was influenced by our variability work; perhaps it is biased. Arnold
\etal\ (2012) reported on MIR spectroscopy from Spitzer, with targets
selected based on MIR excess from Spitzer (Gutermuth \etal\ 2008;
Winston \etal\ 2009, 2010), and is necessarily biased towards targets
bright enough to be observed spectroscopically. There are 61 members
identified in Arnold \etal\ (2012) for which we have light curves, and
45 (74$\pm$14\%) are variable. This significantly higher fraction of
variables could be a result of the bias towards brighter objects,
and/or suggests that there very well could be contamination among our
standard and augmented members. On the other hand, there is only one
non-member identified in Arnold \etal\ (2012); we have a light curve
for it, and it is not variable. None of the analysis above suggests a
clear and obvious population of contaminants. 

We note that there are substantial numbers of objects with light
curves that are neither standard nor augmented members. For example,
there are 98 objects with a Stetson index not suggestive of
variability that are also not identified as members (standard or
augmented). Eighteen of those objects also have an SED slope
consistent with Class I objects. Especially if there is limited
foreground/background contamination, what are these objects?
Spectroscopy is warranted to identify these objects. 

Perhaps a variability fraction between 50 and 75\% is within
expectations for young clusters. We now compare NGC 1333 to the other
clusters that have been examined in detail in the context of YSOVAR.

\subsection{NGC 1333 and Other YSOVAR Clusters}
\label{sec:varfrac1333context}

\begin{deluxetable}{p{5cm}ccccccc}
\tablecaption{Variable fractions across published YSOVAR clusters\label{tab:acrossclusters}}
\tabletypesize{\scriptsize}
\rotate
\tablewidth{0pt}
\tablehead{\colhead{Item} & \colhead{NGC 1333\tablenotemark{a} } 
& \colhead{GGD 12-15\tablenotemark{b}} & \colhead{L1688\tablenotemark{c}} & 
\colhead{IRAS 20050+2720\tablenotemark{d}} & \colhead{Mon R2\tablenotemark{e}} & \colhead{Sum} }
\startdata
Standard Members &	130	&	141	&	54	&	156	&	174	&	655			\\
Variable standard members & 	67	&	106	&	40	&	106	&	106	&	425			\\
Variable fraction &	0.52$\pm$0.08&	0.75$\pm$0.10&	0.74$\pm$0.15&0.68$\pm$0.09&0.61$\pm$0.08 &	0.65$\pm$0.04\\
\tableline
Standard Members with 2-band LCs	&	98	&	122	&	37	&	109	&	89	&	455			\\
Stetson variable standard members with 2-band LCs &	51	&	93	&	26	&	76	&	66	&	312			\\
Variable fraction &	0.52$\pm$0.09&	0.76$\pm$0.10&	0.70$\pm$0.18&	0.70$\pm$0.10&	0.74$\pm$0.12&	0.69$\pm$0.05 \\
\tableline
Standard Members with 2-band LCs, Class I &	13	&	18	&	11	&	34	&	20	&	96			\\
Standard Members with 2-band LCs, Class Flat&	19	&	20	&	13	&	27	&	22	&	101			\\
Standard Members with 2-band LCs, Class II&	46	&	57	&	10	&	38	&	35	&	186			\\
Standard Members with 2-band LCs, Class III&	20	&	27	&	3	&	10	&	12	&	72			\\
\tableline
Standard Members with 2-band LCs, Class III/total standard members with 2-band LCs &	0.20$\pm$0.05	&	0.22$\pm$0.05&	0.08$\pm$0.05&	0.09$\pm$0.03&	0.13$\pm$0.04&	0.16$\pm$0.02\\
\tableline
Stetson Variable standard members with 2-band LCs, Class I&	10	&	17	&	10	&	23	&	18	&	78	&var frac I:	0.81$\pm$0.12\\
Stetson Variable standard members with 2-band LCs, Class Flat &	14	&	19	&	10	&	23	&	16	&	82	&var frac F:	0.81$\pm$0.12\\
Stetson Variable standard members with 2-band LCs, Class II&	27	&	45	&	6	&	29	&	28	&	135	&var frac II:	0.73$\pm$0.08\\
Stetson Variable standard members with 2-band LCs, Class III&	0	&	12	&	0	&	1	&	4	&	17	&var frac III:	0.24$\pm$0.06\\
Standard Members with 2-band LCs, disked &78 & 95 & 34& 99 & 77 & 383\\
\tableline
Stetson Variable Members with 2-band LCs, disked & 51& 81 & 26 & 75 & 62 & 295\\
Variable fraction & 0.65$\pm$0.12 & 0.85$\pm$0.13 & 0.76$\pm$0.20 &
0.76$\pm$0.12 & 0.81$\pm$0.14 & 0.77$\pm$0.06 \\
\enddata
\tablenotetext{a}{This work}
\tablenotetext{b}{Wolk \etal\ (2015)}
\tablenotetext{c}{G\"unther \etal\ (2014)}
\tablenotetext{d}{Poppenhaeger \etal\ (2015)}
\tablenotetext{e}{Hillenbrand \etal\ in prep.}
\end{deluxetable}

\begin{figure}[h]
\epsscale{0.6}
\plotone{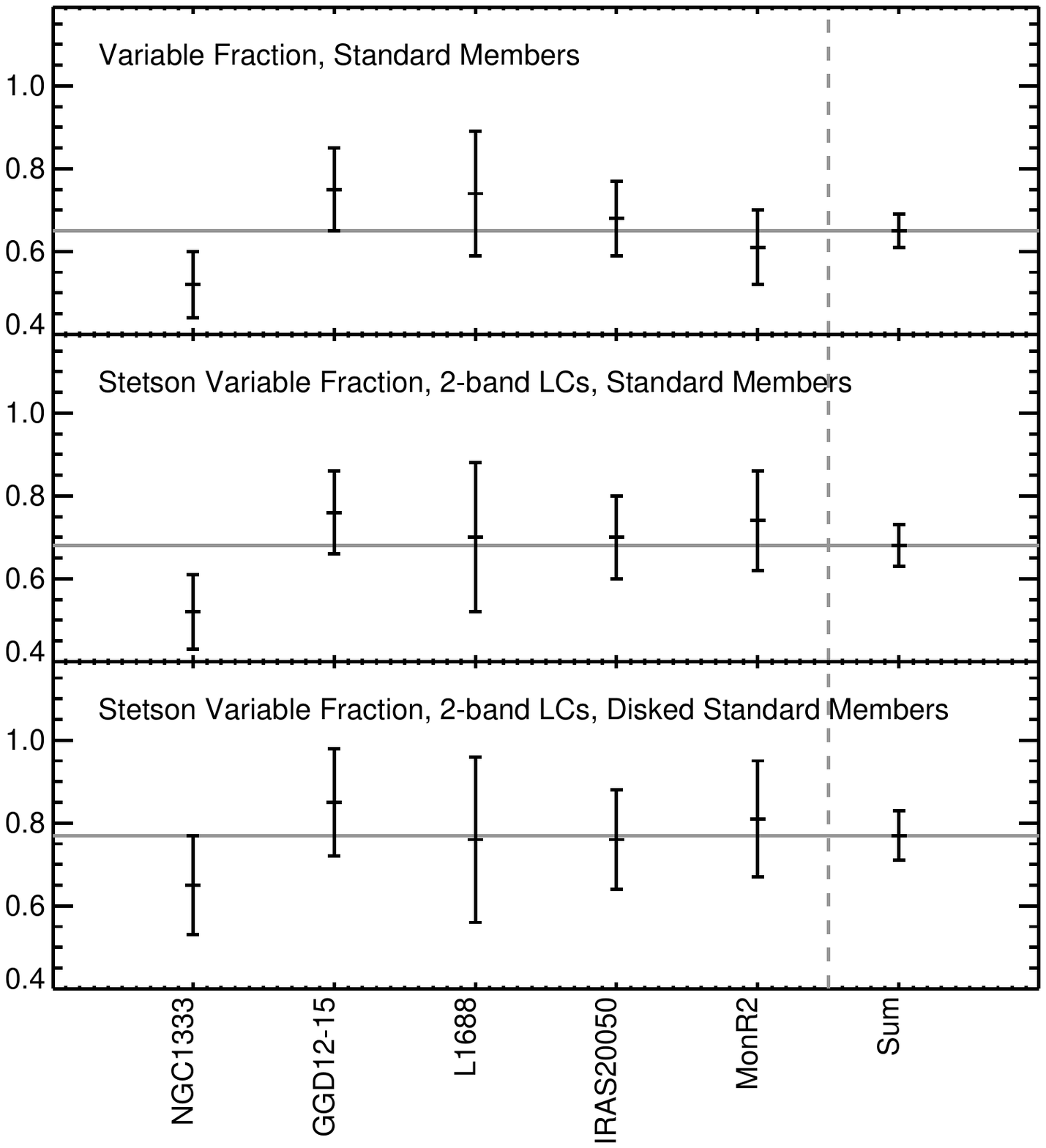}
\caption{Variability fractions for several clusters (listed across the
$x$-axis) for the standard set of members (first row), just those that
have light curves in 2 bands and are identified as variable from the
Stetson index (middle row), those that are Stetson-variable among the
2-band light curves for standard members with disks (Class I, Flat,
and II). The last column sums up all of the corresponding data across
these five clusters. The horizonal grey line corresponds to the value
from this sum across clusters. Error bars are calculated assuming
Poisson statistics. NGC 1333 has a slightly lower variability fraction
than most of the other clusters, but GGD 12-15 has a slightly higher
variability fraction than most of the other clusters.  }
\label{fig:varfrac2}
\end{figure}

About half (52$\pm$8\%) of the NGC 1333 standard set of members is
variable in the MIR, and we want to compare this to other clusters.
Morales-Calder\'on \etal\ (2011) found that  $\sim$70\% of Orion
members are variable in the IR; Cody \etal\ (2014) find that 90\% of
NGC 2264 members (with disks) are variable. In both of these cases,
these clusters are well-studied, with well-established sets of
members, so the set of members (and even variables) is not selected in
the same fashion as it is in NGC 1333.  

For the smaller-field YSOVAR clusters, where we have defined the
`standard set of members' in the same way across clusters, four
clusters besides NGC 1333 have been investigated in detail.
Table~\ref{tab:acrossclusters} compares the same numbers for NGC 1333
and GGD 12-15 (Wolk \etal\ 2015), L1688 (G\"unther \etal\ 2014), IRAS
20050+2720 (Poppenhaeger \etal\ 2015), and Mon R2 (Hillenbrand \etal\ in
prep). Fig.~\ref{fig:varfrac2} presents some of the information from
Table~\ref{tab:acrossclusters} in graphical form. 

NGC 1333 has the lowest fraction of standard member variables among
all 5 of these clusters. The mean across all clusters is within
$\sim$1$\sigma$ of the values for all of the other clusters; NGC 1333
is $>$1$\sigma$ below this line. (We note, however, to the extent that
NGC 1333 may be low, perhaps GGD 12-15 is high.) It is not clear why
NGC 1333 should be different than the other clusters. NGC 1333 is
considerably closer than all of those clusters except L1688, which
affects the mass range to which we are sensitive; using the distances
as reported in R14, NGC~1333 is at 235 pc, to be compared with Mon R2
at 830 pc, GGD 12-15 at 830 pc, IRAS 20050+2720 at 700 pc, and L1688
at 120 pc. In NGC 1333, we should be sensitive to considerably lower
masses than all of those clusters except L1688, where we should be
quite comparable. But, the member variability fraction in L1688 is
much higher than in NGC 1333.  

Both NGC 1333 and L1688 contain known Class 0s, thought to be the
youngest cluster members. Searches for Class 0s have only been
conducted in nearby clusters, but nonetheless, given the available
information, the YSOVAR campaigns in NGC 1333 and L1688 should include
several of these youngest objects. Naively, then, since both of these
clusters have several known Class 0s, one might expect similar
variability fractions between these two clusters, but they are very
different. However, when investigated in more detail, the relative
number of Class 0s in these clusters is different, among other things,
so perhaps differences are expected.  

Some possible sources of contamination (see above and in R14 as well
as G\"unther \etal\ 2014) are that it is easier to `fool' the $\chi^2$
variability test, and that X-ray data does not cover the whole field
for which there are light curves (which affects identification of
Class IIIs in the regions where there are no X-ray data). Therefore,
we looked at subsamples designed to limit these influences. First, we
selected variables solely using the Stetson index out of the standard
set of members where there are light curves in both IRAC bands. The
variability fractions for this sub-sample (see
Tab.~\ref{tab:acrossclusters} and the second row of
Fig.~\ref{fig:varfrac2}) are very similar to that for the overall
standard set of members. With generally fewer stars involved, the
errors increase, and most variability fractions increase slightly. 

Secondly, Class IIIs are generally going to be the most difficult to
identify reliably in these clusters. The most likely contaminant will
be background/foreground stars (since galaxies are unlikely to have
photosphere-like SEDs). YSOs seen through the NGC 1333 cloud though
still belonging to Per OB 2 would appear as Class IIIs, though Per OB
2 members should be $\lesssim$6 Myr (e.g., Bally \etal\ 2008), and
therefore should also have IR excesses. Both NGC 1333 and GGD 12-15
have standard members with 2-band light curves that are composed of
$\sim$20\% Class IIIs, whereas the other clusters in
Tab.~\ref{tab:acrossclusters} are closer to $\sim$10\%. GGD 12-15 has
by far the highest variability fraction among the Class IIIs. We then
considered the variability fraction among just those standard members
with 2-band light curves (and Stetson index to identify variables),
but just for those that are Class I, flat, or II (see
Tab.~\ref{tab:acrossclusters} and the third row of
Fig.~\ref{fig:varfrac2}). All of the clusters are still roughly
consistent with each other, but NGC 1333 is still slightly low, and
now GGD 12-15 is slightly high. 

Summing over all the clusters in Tab.~\ref{tab:acrossclusters}, the
(Stetson) variability fraction for Class I members with 2-band light
curves is the same as that for Flat class members, 81$\pm$12\%. The
variability fraction for Class II objects is significantly lower at
73$\pm$8\%, and the variability fraction for Class III, largely driven
by the GGD 12-15 sources, is 24$\pm$6\%. So our expectation that more
embedded objects are more likely to be variable is met across all
clusters, suggesting that most of the MIR variability arises from dust
in the disk or from variable accretion from the disk to the star. (For
the Class III sources, which should have little or no dust disk, the
IR variability must be coming from the photosphere.)

To summarize, NGC 1333 may have legitimately a slightly lower fraction
of variables. NGC 1333 may have some contamination in its standard set
of members, although such contamination is not immediately apparent
from the diagnostics above in Sec.~\ref{sec:contam}.  This could be
random error, or it could be a sign of different environments in
different clusters. More analysis of this is deferred until detailed
analysis of each of the remaining YSOVAR clusters is complete.

\section{Sizes of Brightness and Color Changes}
\label{sec:amplitudes}

In this section, we investigate the amplitude of brightness and color
changes over the YSOVAR campaign, and identify some of the objects
with the largest brightness changes in the CY variables.

\subsection{Light Curve Amplitudes Over the YSOVAR Campaign}

\begin{figure}[h]
\epsscale{0.9}
\plottwo{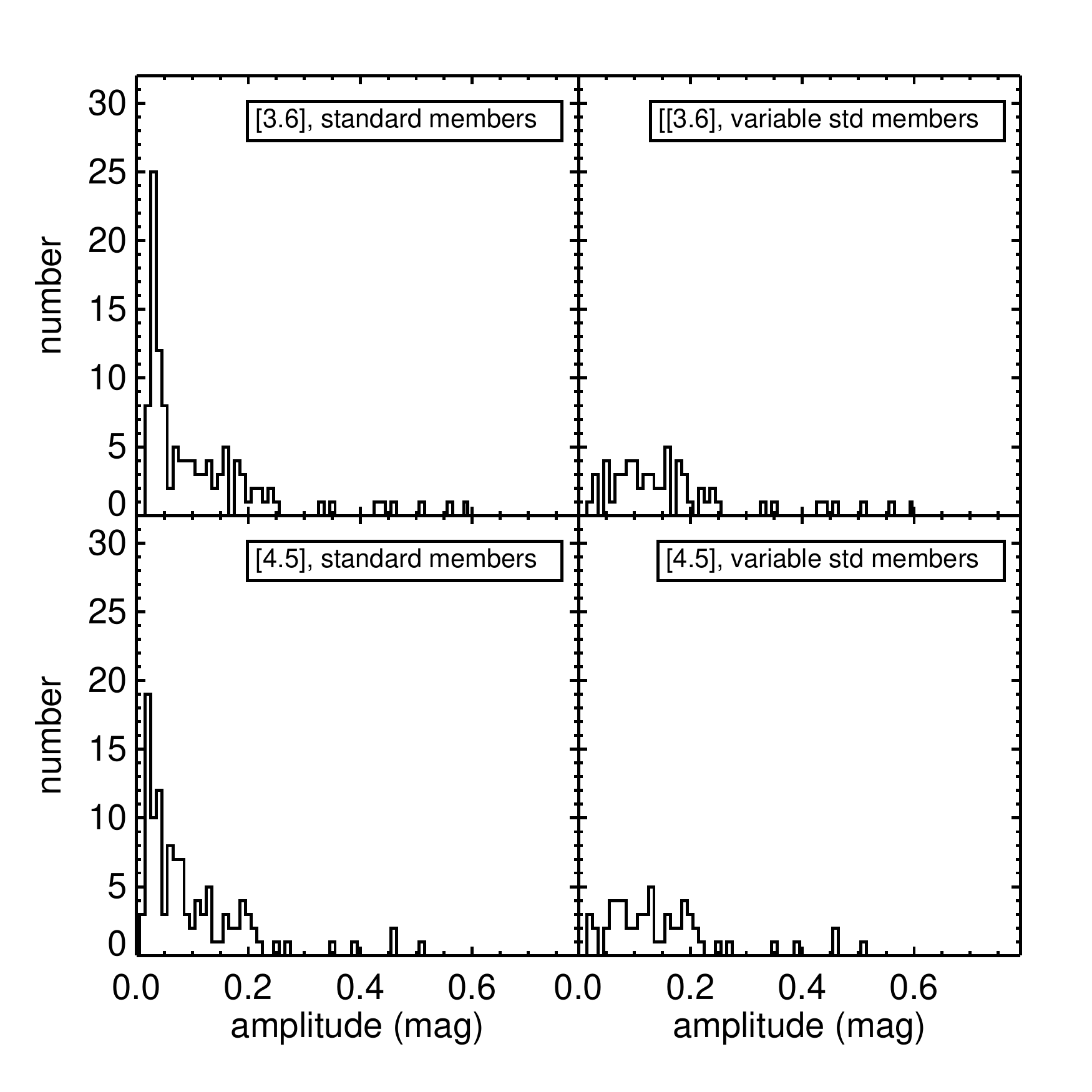}{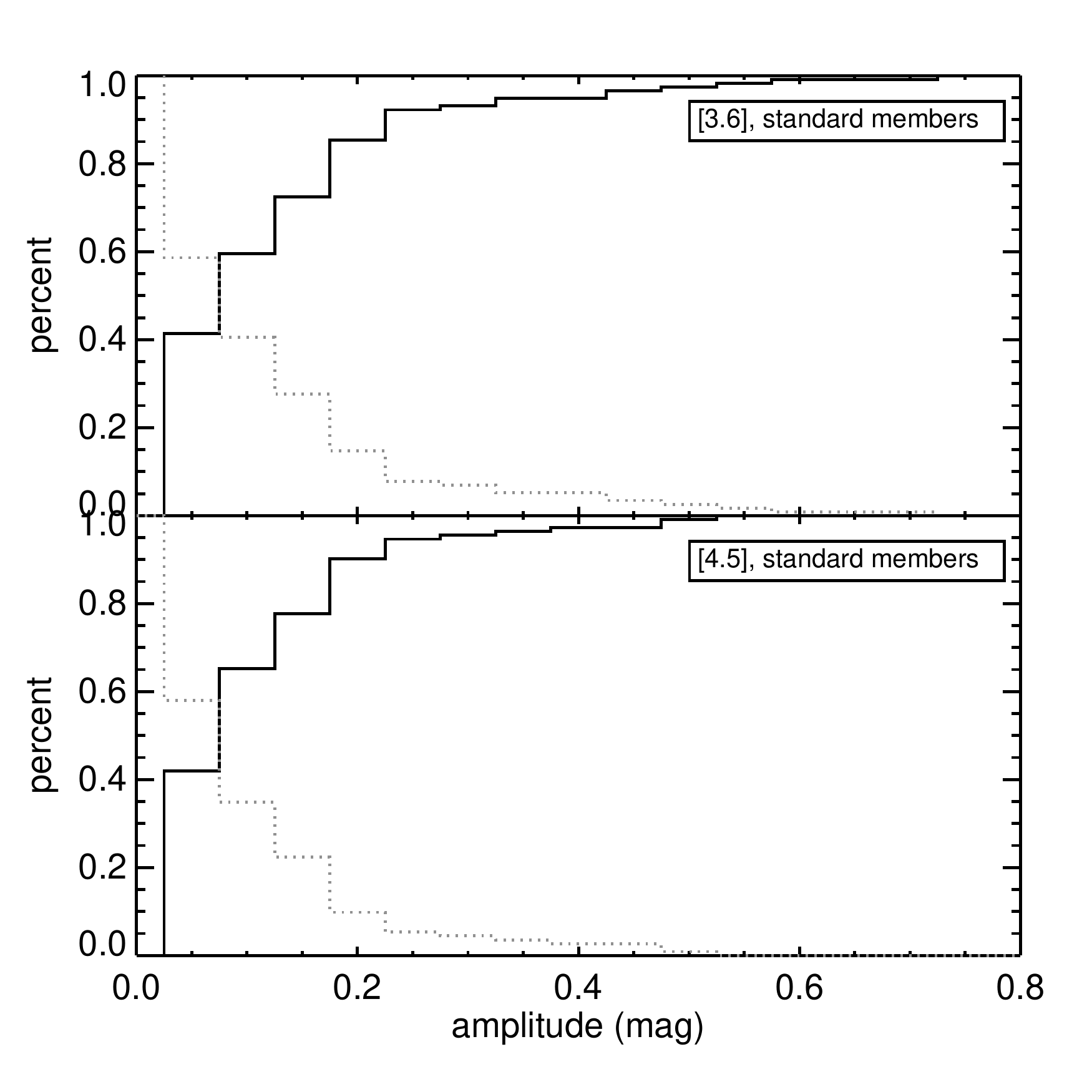}
\caption{LEFT SIDE: Histogram of the amplitudes of the light curves. 
Upper left: [3.6] light
curves, for standard set of members; upper right: [3.6] light curves,
for all variable (standard) members;  lower left: [4.5] light curves,
for standard set of members; lower right, [4.5] light curves, for all
variable (standard) members. RIGHT SIDE: Cumulative distributions (in
both directions of increasing and decreasing magnitudes -- black solid and
grey dotted lines) for [3.6] (top) and [4.5] (bottom), standard set of
members (variable and non-variable). About 20\% of the variable 
members vary at 0.2 mags or more; about 10\% of the standard set of
members vary at 0.2 mags or more.  }
\label{fig:amplitudes}
\end{figure}

\begin{figure}[h]
\epsscale{0.5}
\plotone{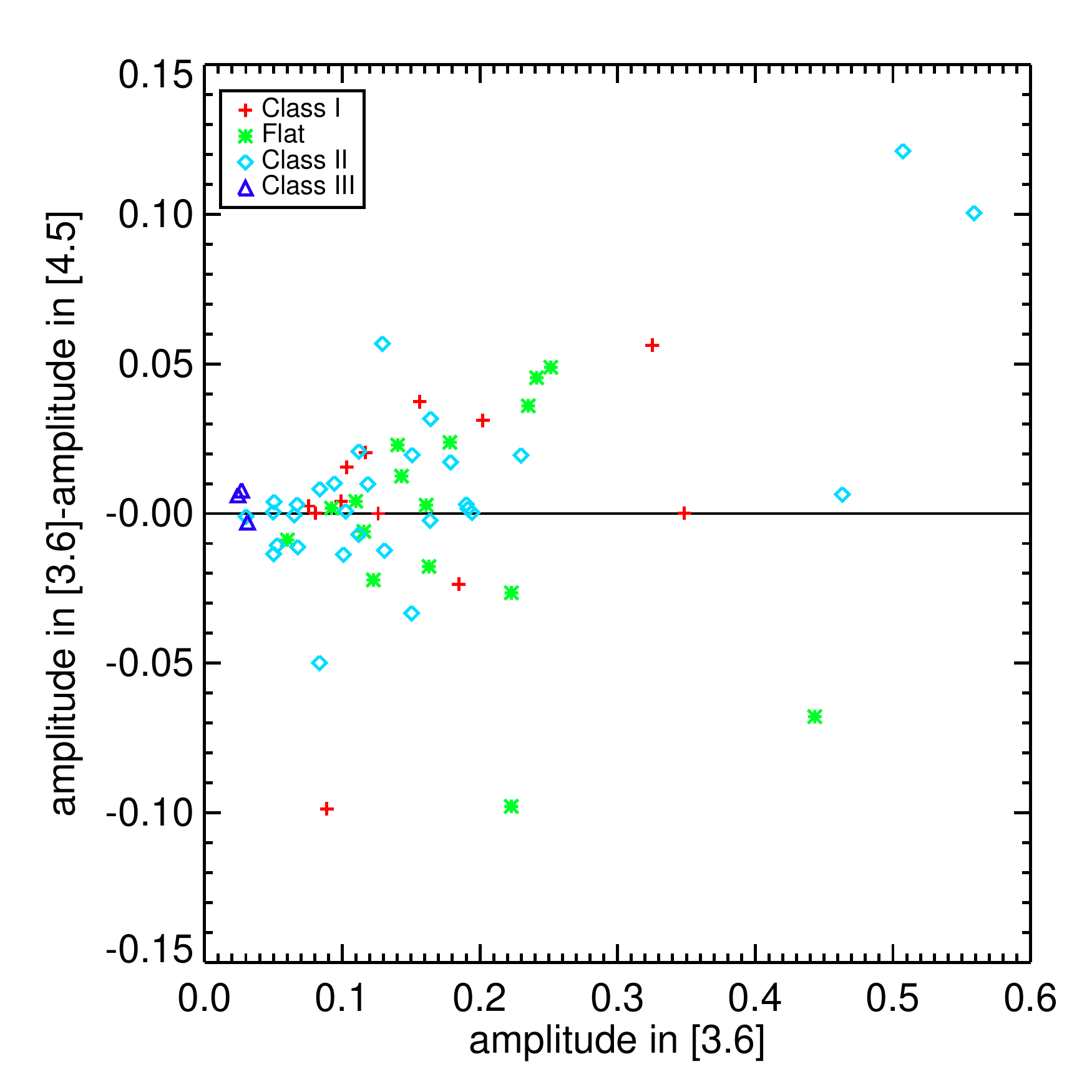}
\caption{Amplitude in [3.6] against the difference of amplitude in
[3.6] $-$ that in [4.5] for the all of the variables that have data in
both channels. Colored symbols
correspond to SED classes: red plus = Class I, green asterisk = Flat,
cyan diamond = Class II, dark blue triangle = Class III.  The
amplitudes found for the Class III objects are smaller than most of
the rest of the amplitudes. The distributions of amplitudes for the
rest of the classes are statistically indistinguishable, though there
may be a slight tendency for the Class I objects to have larger
amplitudes on average. The solid line is the line along which the
points would fall if the amplitudes in both bands were identical. The
points are symmetrically distributed about that line.  }
\label{fig:ampl3}
\end{figure}

We define light curve amplitude over the YSOVAR campaign to be the
difference between the 10th and 90th percentile in the distribution of
points. This effectively omits flares or other few-epoch excursions,
and is the same definition used in several other YSOVAR papers.
Histograms of light curve amplitudes for both IRAC channels are shown
in Figure~\ref{fig:amplitudes}. Distributions are shown for the entire
standard set of members (variable or `not detected as variable,'
NDAV), as well as just the variable standard members. Some of the
variable members have very low amplitudes, comparable in size to the
NDAV objects; the distribution of amplitudes for all members likely
includes some objects that are legitimately variable but not
identified here.  A few objects vary with amplitudes more than 0.2
mag, but a more typical amplitude for the variables is 0.1-0.15 mag.

Fig.~\ref{fig:amplitudes} includes distribution functions for the
entire standard set of members.  Out of the variable members, at
[3.6], 15/62 or 24($\pm 6$)\% vary at 0.2 mags or more, and at [4.5],
11/60 or 18($\pm6$)\% vary at 0.2 mags or more. Out of the entire
standard set of members, at [3.6], 17/116 or 14$\pm3$\% vary at 0.2
mags or more, and at [4.5], 11/112 or 10$\pm3$\% vary at 0.2 mags or
more.  About 65\% of the variable members vary at 0.1 mags or more in
either channel, $\sim$30-40\% of the standard set of members vary at
0.1 mags or more in either channel, and $\sim$65\% of the augmented
set of members vary at 0.1 mags or more in either channel.   Out of
the entire set of variables (not just the standard set of members),
$\sim$7\% at either channel vary at amplitudes $>$0.4 mag. 

Looking at the distributions of amplitudes for each SED class, the
distribution of the amplitudes of Class I, Flat, and Class II objects
are statistically indistinguishable.  Figure~\ref{fig:ampl3} plots the
amplitudes of the variables in [3.6] against [4.5], with color coding
by class.  There may be a very slight tendency for the Class I objects
to have larger amplitudes on average. However, the mean amplitude in
[3.6] or [4.5] (for all variables) for Class I, flat, and II are all
$\sim$0.16 mag. So any tendency towards larger amplitudes in the more
embedded SEDs must be subtle, or lost in low-number statistics.  
Class III objects are significantly different (in either IRAC
channel); the mean for Class III variables is $\sim$0.03 mag,
distinctly different from the rest of the distribution. 

There are no discernible trends of amplitude of variation with
effective temperature or spectral type, for those objects where we can
look for such a correlation.  Our sample is not large or long enough
to determine the fraction of sources in a burst or elevated state
(see, e.g., Hillenbrand \& Findeisen 2015).

\begin{deluxetable}{ccp{8cm}}
\tablecaption{Large-amplitude variables\label{tab:largeampl}}
\tabletypesize{\scriptsize}
\tablewidth{0pt}
\tablehead{\colhead{SSTYSV} & \colhead{sample} & \colhead{Notes} } 
\startdata
032847.63+312406.1 & YSOVAR 1-band & Identified in literature as YSO, std member. SED Class II. [3.6] only, but light curve much like other YSOs. Deep dip.\\
032851.01+311818.5 & YSOVAR 2-band & $>$0.4 mag in both channels. Identified in literature as YSO, std member. SED Class II. Light curve shaped as almost a step up.\\
032859.54+312146.7 & YSOVAR 2-band & $>$0.4 mag in both channels. Identified in literature as YSO, std member. SED Class II. Large-amplitude color variability (nearly 0.1 mags) entirely consistent with reddening.\\
032903.13+312238.1 & YSOVAR 2-band & $>$0.35 mag in both channels. Identified in literature as YSO, std member. SED Class II. Periodic with an overall trend down.  \\
032908.97+312624.0 & YSOVAR 1-band & Identified in literature as YSO, std member. SED class Flat. [3.6] only, but light curve much like other YSOs. Deep dip.\\
032909.32+312104.1 & YSOVAR 2-band & $>$0.4 mag in both channels. Identified in literature as YSO, std member. SED Class Flat. Broad, sharply peaked `burst'.  \\
032911.86+312155.7 & YSOVAR 1-band & Not lit YSO or std member. [4.5] only. See text. \\
032904.31+311906.3 & YSOVAR color  & $>$0.2 mag color change as part of large dip in light curve. Identified in literature as YSO, not std member. SED Class Flat. \\
032910.70+311820.9 & CY var & $\sim$2 mag change between cryo and YSOVAR. SED class I (but identified in literature as an SED class 0). Reddened SED. YSOVAR campaign itself relatively unremarkable.
\enddata
\end{deluxetable}

\label{sec:032911.86+312155.7}

\begin{figure}[h]
\epsscale{0.8}
\plotone{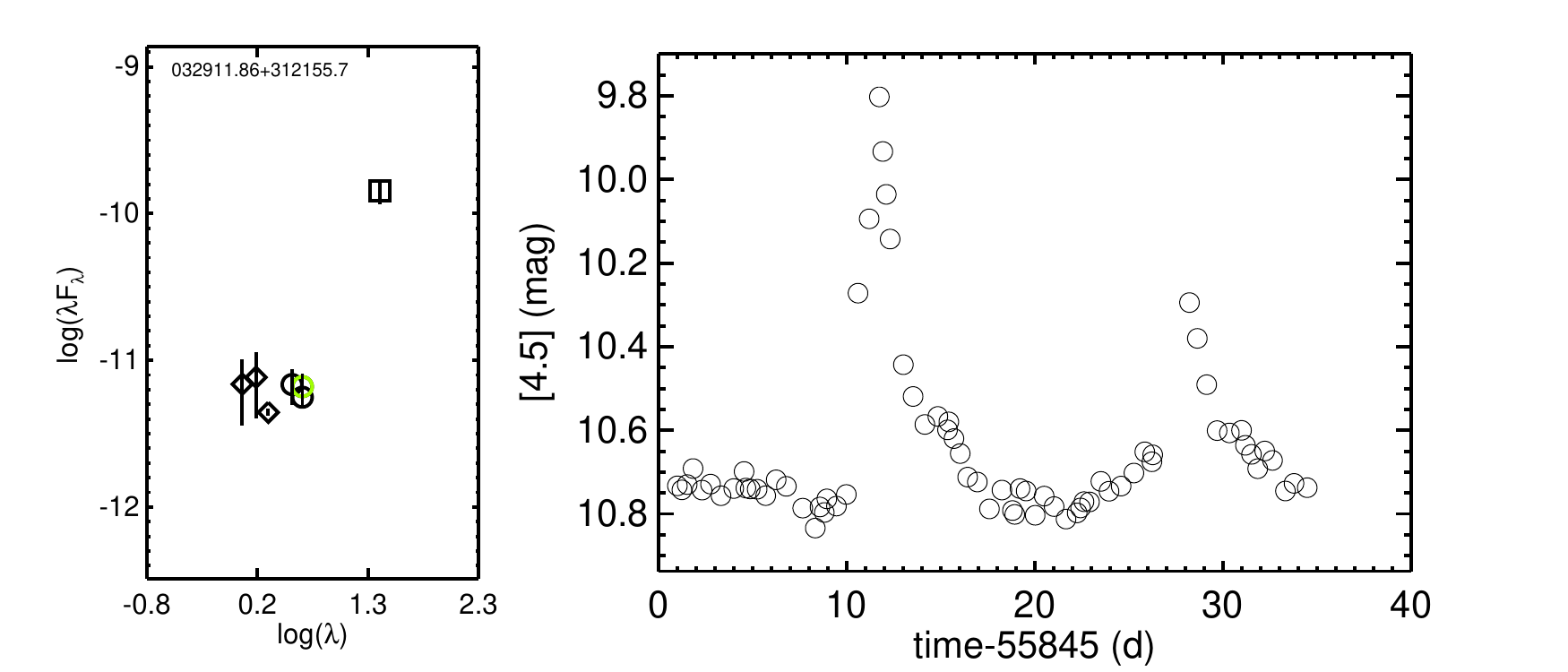}
\caption{SED and light curve for source SSTYSV J032911.86+312155.7
(=LAL244=MBO81=Foster 140). In
the SED, log $\lambda F_{\lambda}$ is plotted with $\lambda
F_{\lambda}$ in cgs units (erg s$^{-1}$ cm$^{-2}$) and log $\lambda$
is plotted with $\lambda$ in microns. Diamonds are 2MASS, circles are
IRAC (with the green being from the YSOVAR campaign), and square is
MIPS. Vertical black lines are error bars (or amplitude of YSOVAR
campaign in the case of the green IRAC point, where amplitude is
defined as in the text).  This source may not be
a YSO; see discussion in text. }
\label{fig:032911.86+312155.7}
\end{figure}

There are seven variables with amplitude $>$0.4 mag in either channel.
They are listed in Table~\ref{tab:largeampl}. They are nearly all
literature-identified YSOs and part of our standard set of members.
One object, SSTYSV J032911.86+312155.7, merits more discussion than
can easily be fit in Table~\ref{tab:largeampl}. It is not a literature
YSO or a standard member set YSO. It has an unusual SED that is quite
flat from $J$ to 4.5 \mum, admittedly with large errors, and the SED
rises abruptly at 24 \mum\ to more than an order of magnitude more
energy density; see Figure~\ref{fig:032911.86+312155.7}. Under our
scheme, it is an SED Class I. In the images, it is in a very messy
region, with high surface brightness, which precludes a WISE
detection. There is definitely a source at 24 \mum, but whether all
the flux should be assigned to this object is an open question. The
SED does not particularly look like a YSO SED, but its location
amongst other sources in a high surface brightness region is
circumstantial evidence that it is a YSO. Foster \etal\ (2015) discard
it as a non-member based on NIR spectroscopy, but looking at the
spectrum, the NM assessment was based on low signal-to-noise, as
opposed to high-quality radial velocity (RV) measurements that were
inconsistent with membership. The formal RV is 25 $\pm$75 km s$^{-1}$,
and is within 0.3 sigma of the cluster's velocity of $\sim$8 km
s$^{-1}$, and the spectrum is dominated by sky lines (telluric and
nebular). Thus, we conclude that the APOGEE spectrum is not of
sufficient quality to rule out this object as an NGC 1333 member. 
There is no X-ray detection from any instrument. Its light curve
exists only at [4.5] and consists of a flat continuum with two large
($\sim$1 mag) flares (see Figure~\ref{fig:032911.86+312155.7}), where
the flares are defined by several points (not just one per flare,
which might suggest, e.g., cosmic rays). This is different from the
other light curves in NGC 1333; this object is either an unusual young
star, or a contaminant object.

\clearpage

\subsection{Color Amplitudes Over the YSOVAR Campaign}

\begin{figure}[h]
\epsscale{1.0}
\plottwo{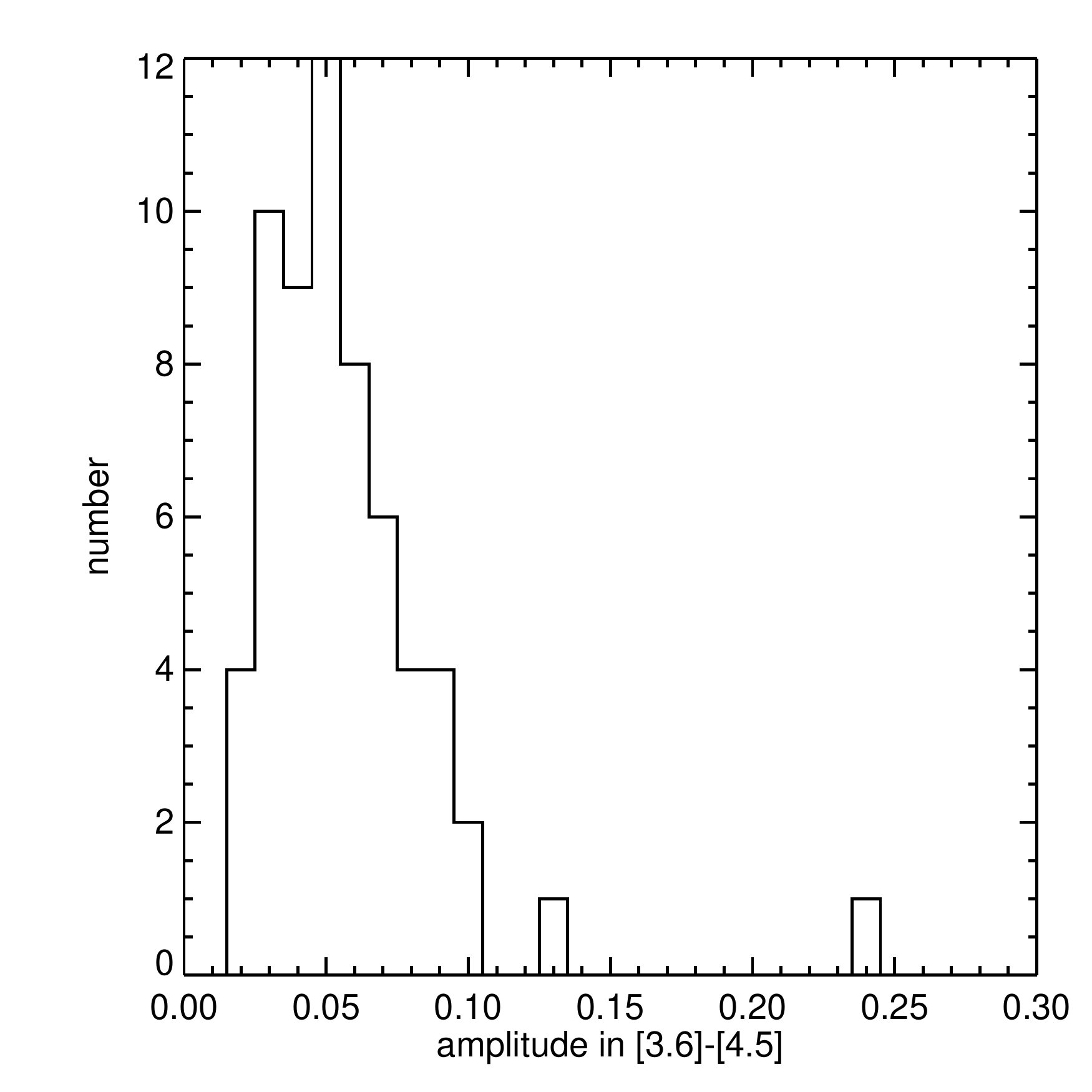}{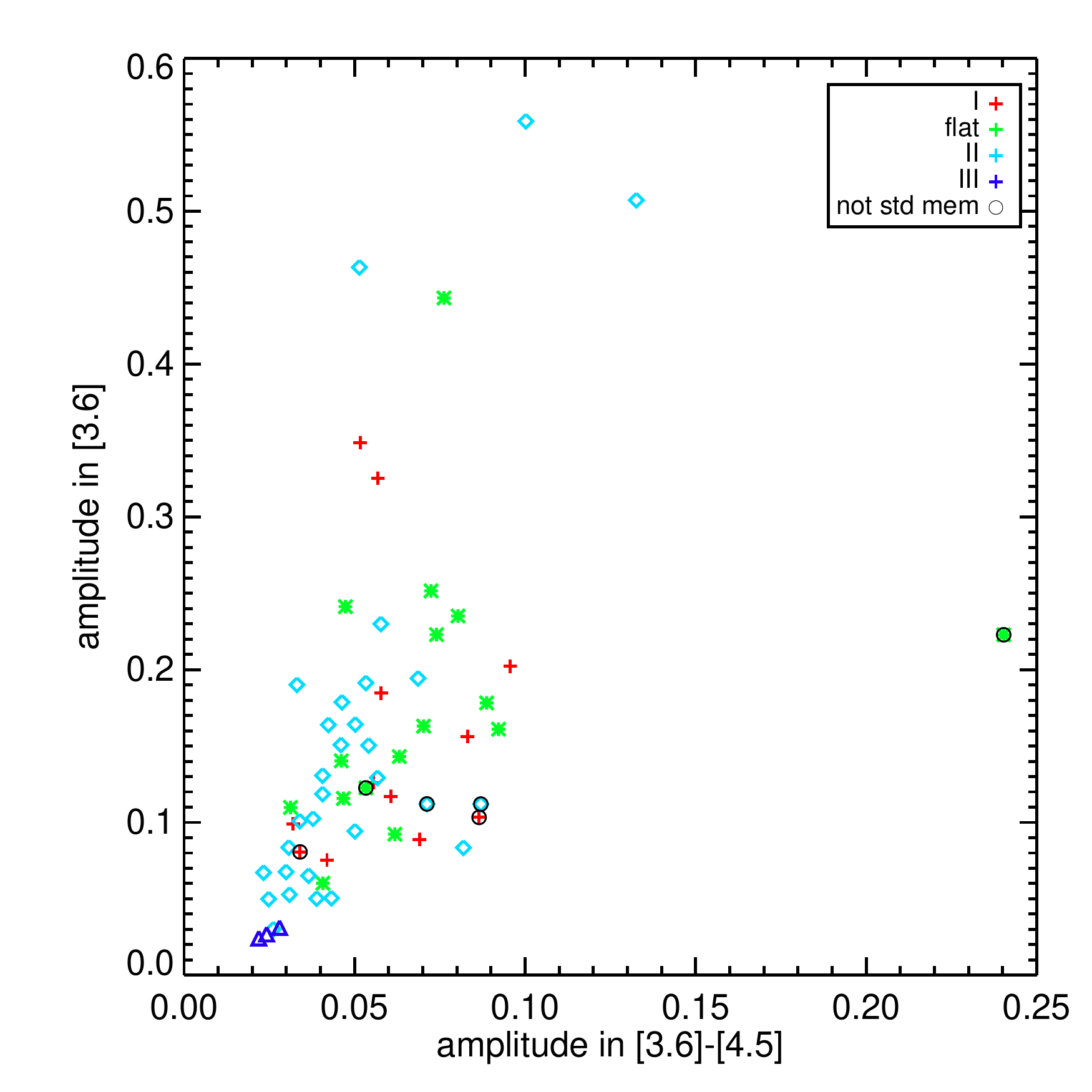}
\caption{These plots include all variables with light curves in both
channels. LEFT: distribution of the amplitude of [3.6]$-$[4.5], which
peaks at $\sim$0.05 mag. There are some extreme color variables, but
most have small changes in color. RIGHT: amplitude of [3.6]$-$[4.5] against
the amplitude in [3.6], both in magnitudes.  Colored symbols
correspond to SED classes: red plus is Class I, green asterisk is Flat,
cyan diamond is Class II, dark blue triangle is Class III.  An
additional circle means that the variable is NOT in the standard set
of members -- most of the variables are in the standard set of
members. Typically, the largest single-band amplitudes are accompanied
by the largest color amplitudes, but there is a lot of scatter.}
\label{fig:amplcolor}
\end{figure}

Figure~\ref{fig:amplcolor} shows the distribution of the amplitude
(10th to the 90th percentile of the distribution) for [3.6]$-$[4.5] for
the variables with light curves in both channels. The distribution of
color changes peaks at $\sim$0.05 mag.  Figure~\ref{fig:amplcolor}
also shows the amplitude of [3.6]$-$[4.5] against the amplitude in
[3.6]. Larger color changes generally (although not always) also
translate to larger single-band changes.  There are no discernible
trends of amplitude of color change with effective temperature.

\begin{figure}[h]
\epsscale{0.8}
\plotone{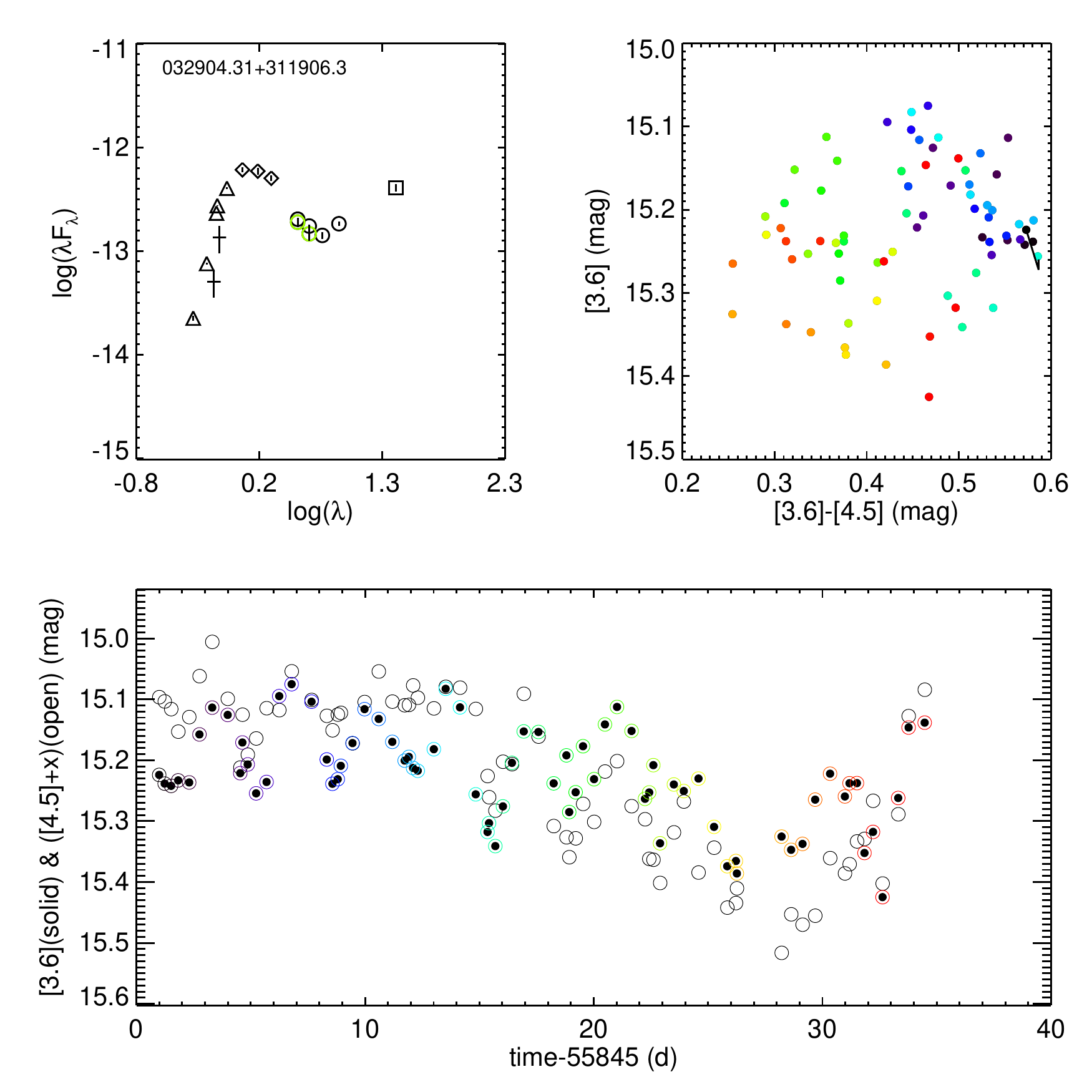}
\caption{SED, CMD, and light curve for source SSTYSV
032904.31+311906.3 (=ASR61=MBO156=Foster 100, an M0-3 spectral type).
In the SED, log $\lambda F_{\lambda}$ is plotted with $\lambda
F_{\lambda}$ in cgs units (erg s$^{-1}$ cm$^{-2}$) and log $\lambda$
is plotted with $\lambda$ in microns. Triangles and crosses are
optical data, diamonds are 2MASS, circles are IRAC (with the green
being from the YSOVAR campaign), and square is MIPS. Vertical black
lines are error bars (or amplitude of YSOVAR campaign in the case of
the green IRAC points, where amplitude is defined as in the text). The
CMD is in the upper right, with colored points corresponding to the
circled epochs in the bottom light curve (that is, black/purple points
are from early in the campaign, and orange/red points are from late in
the campaign). There is an \av=1 vector shown from the first point in
the campaign. In the light curve on the bottom, solid points are
[3.6], and open black circles are [4.5], shifted (represented by `+x')
such that the mean [4.5] matches the mean [3.6]. The additional
colored circles correspond to the colors used in the CMD. This source
exhibits the largest color change in the entire YSOVAR dataset; see
discussion in text. }
\label{fig:032904.31+311906.3}
\end{figure}

The most extreme color change is $>$0.2 mag change; see
Figure~\ref{fig:032904.31+311906.3}. It is SSTYSV J032904.31+311906.3
and is not part of the standard set of members. Few papers in the
literature identify it as a member, but Aspin \etal\ (1994) does (it
is ASR 61). The G09 identification process classifies it as a
broad-line AGN. Its SED looks stellar (see
Fig.~\ref{fig:032904.31+311906.3}); spectroscopy is needed. Early in
the YSOVAR campaign, the trend in the CMD is along an overall slope
consistent with that expected from reddening; this can be seen in the
CMD as the black through blue points.  Then, as the lightcurve takes
an overall dip (which is at its lowest at $\sim$28 d), the object
becomes, on the whole, bluer and fainter. No obvious image artifacts
are responsible for this dip. The smaller amplitude and timescale
variations, on their own, move in a direction in the CMD very roughly
consistent with a reddening slope. But the larger amplitude and larger
timescale motions are different. The shorter timescale variations
would require \av\ variations of $\sim$10-15 mag (in the optical),
whereas the longer timescale variation would require $\sim$30 mag (in
the optical). This is not physically impossible, just somewhat
unexpected. The disk around this object does not dominate the SED,
though it is clearly a substantial disk; large amounts of obscuring
dust might be expected if the dust dominated the SED. If variations in
the inner disk, close to the star, account for the structure changes
on shorter timescales (moving in the general direction consistent with
reddening), and if the star is viewed close to edge on, such that
structures in the outer disk occult the line of sight by different
amounts on timescales of weeks to months, then this object may be
seen mostly in scattered light as it becomes fainter, which could
cause the overall bluer-when-fainter behavior.

\clearpage

\subsection{The Largest CY Variables}

\begin{figure}[h]
\epsscale{0.8}
\plotone{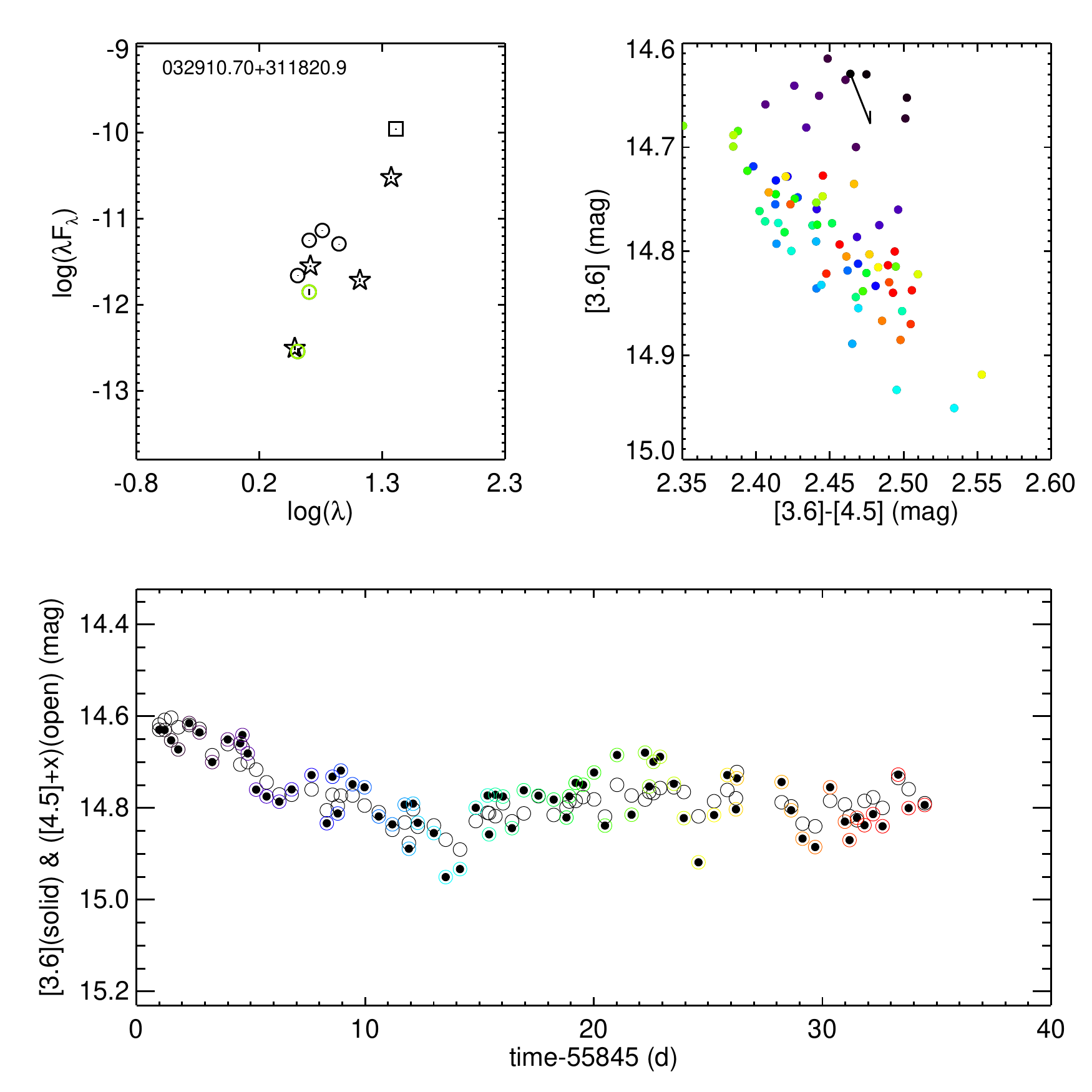}
\caption{SED, CMD, and light curve for source SSTYSV
032910.70+311820.9 (=J07-23=Gutermuth 35). Notation is as in prior
figure, except stars in the SED are WISE data. This source exhibits
the largest CY change in the entire YSOVAR dataset; see discussion in
text. It is also thought to be a jet driver; see
Sec.~\ref{sec:class0jets}. }
\label{fig:032910.70+311820.9}
\end{figure}

R14 looked for objects whose color appeared to change significantly,
specifically looking for objects that appeared to have transient IR
excesses (e.g., color appearing or vanishing between the cryo epoch and the
YSOVAR epochs; see, e.g., Meng \etal\ 2012,  Melis \etal\ 2012, Rice, Wolk,
\& Aspin 2012) One object from NGC 1333 was so identified: SSTYSV
032910.70+311820.9, shown in Figure~\ref{fig:032910.70+311820.9}. 

SSTYSV J032910.70+311820.9 has an ususual SED in comparison to other
objects with light curves here - it appears to be heavily reddened,
with the peak of the SED shorter than 15 \mum\ appearing at 5.8 \mum\
(most other sources peak in the near-IR). The 24 \mum\ point rises
very abruptly to have an order of magnitude more energy than the 5.8
\mum\ point. The WISE data are further confusing; by inspection of the
images, the AllSky (rather than AllWISE) catalog seems to do a better
job of characterizing the flux from this object (R15). The flux
density at 22 \mum\ is lower than that measured at 24 \mum; the 12
\mum\ point appears to be falling roughly on a Rayleigh-Jeans line
from the 8 \mum\ point, resulting in a very abrupt rise from 12 to
$\sim$20 \mum. It may be that 100\% of the flux density at 22 and 24
\mum\ should not be associated with this point source, though the
source is bright ([24]=2.27, [22]=3.9) and point-like at these bands,
and sufficiently separated from another comparably bright source
$\gtrsim$10$\arcsec$ away (SSTYSV J032911.24+311831.8, which appears
in Fig.~\ref{fig:032911.24+311831.8} below). The IRAC images suggest
that there may be some nearby nebulosity.  It is a SED Class I.  The
mean 3.6 and 4.5 \mum\ points from our campaign are very different
than the cryo-era measurements -- in the cryo era,
$[3.6]-[4.5]=12.57-10.82=1.75$, while the YSOVAR means are 
$<[3.6]>-<[4.5]>=14.77-12.32=2.45$. The two WISE measurements at 3.4
and 4.6 \mum\ were obtained at a mean epoch in time between the cryo
and YSOVAR epochs, and they fall between these points in the SED
([3.4]=14.84, [4.6]=11.49), lending credence to this variation despite
its unusual amplitude of $\sim$2 mag in each IRAC channel between the
cryo and YSOVAR epochs ($\Delta[3.6]=2.2$, $\Delta[4.5]=1.5$). 
However, the light curve over the YSOVAR campaign itself is relatively
unremarkable; see Fig.~\ref{fig:032910.70+311820.9}. It is, at least,
identified as variable from the Stetson index and both $\chi^2$
values. The total peak-to-peak variation is $\sim$0.3 mag, and the
amplitude as defined above is $\sim$0.2 mag; it has a standard
deviation $\sigma$ of 0.075 mag and a median absolute deviation (MAD)
of 0.060 mag in [3.6] and, for [4.5], $\sigma$=0.063 mag and MAD=0.044
mag. Over the YSOVAR campaign, the MAD [3.6]$-$[4.5] color is 0.03
mag, so it is not a large color change, as can be seen in
Fig.~\ref{fig:032910.70+311820.9}. Given the observed variability in
the YSOVAR campaign, one would not necessarily have selected this
object as a likely large amplitude variable on timescales of years,
but it nonetheless appears to have a very significant variation on
those timescales. This source is identified in the literature as a jet
driver (see Sec.~\ref{sec:class0jets}), which is consistent with it
being very embedded. 

Three other objects were identified as potential large-amplitude
CY variables in NGC 1333, but none of them were also identified
as variable over the YSOVAR campaign.  For all three of them, we have
reason to question the size of the change between the cryo era and the
YSOVAR campaign. 

SSTYSV J032855.72+311442.1 is clearly detected in the YSOVAR epochs,
but is close to (possibly technically part of) IRAS 2A, which is a
very bright source. It has been matched, perhaps incorrectly, to
sources with a measured flux density at 8 and 24 \mum\ that result in
a very steep SED (nearly monotonically increasing, slope of $\sim$3).
The YSOVAR light curve is not detected as variable, though viable data
really only exist at 4.5 \mum\ (there are a few epochs at 3.6 \mum\
near the end of the campaign). There are indeed large differences
between the cryo-era measurements ([3.6]=16.09, [4.5]=14.51) and the
mean YSOVAR value ($<[3.6]>$=15.69, $<[4.5]>$=13.4). However, the
measured brightness is likely affected by the nearby nebulosity and
point sources, and moreover is near the faint limit at [3.6], beyond
which light curves become very noisy.  We suspect that the apparent
changes between the cryo and YSOVAR eras are spurious.

SSTYSV J032858.78+312044.9 has a reasonably well-defined SED over the
IRAC bands, despite having image morphology suggestive of a nebulous
clump. However, while the 3.6 \mum\ point between cryo and the YSOVAR
campaign is a fairly good match, the 4.5 \mum\ cryo-era point is
inconsistent (by about half a dex) with the rest of the SED; the
measurement in our database was obtained from the c2d catalog and has
a large error bar. This object is also not identified as variable over
the YSOVAR campaign. Since the cryo-era measurement at 4.5 \mum\ is
evidently unphysical, we remove this object from the list of objects
with large changes on timescales of years.

SSTYSV J032918.88+312313.0 has a strange SED, which is not
monotonically increasing but includes a 24 \mum\ point at $\sim$0.5
dex more energy in the SED than the peak at 4.5 \mum. The mean [4.5]
from the YSOVAR campaign is well-matched to the cryo-era measurement;
however, the mean [3.6] from the YSOVAR campaign (15.02)  is
significantly different than the cryo-era [3.6] (15.85). It is in a
region of very high surface brightness, so as a result of the high
background, 15th magnitude is probably close to the detection limit,
and may be a nebulous clump. We strongly suspect that either the
cryo-era [3.6] or the mean from the YSOVAR campaign is incorrect, and
thus we also remove this object from the list of objects with large
changes on timescales of years.

Out of all 242 objects (not just members) with light curves in both
bands, then, we have at most 1 object that has legitimate very large
changes at 3.6 and 4.5 \mum\ over timescales of mulitple years, or at
most a $\sim$0.4\% rate of occurrence. (By collecting objects over all
YSOVAR clusters, R14 estimated a rate of at most $\sim$0.02\%.)


\section{Timescales of Variability}
\label{sec:timescales}

\subsection{Overview}

Characterizing the timescales of the YSOs in our sample is important
for tying the variations we see to physical properties in the
star-disk system. For example, we expect very short timescale
variability to be related to events on or near the surface of the
star, and longer timescale variability to be related to changes
further out in the disk. However, characterizing the diversity of
YSOVAR lightcurves with a single timescale has proven challenging.

There are several different ways in which we could define timescales
of the variability seen in our light curves. Each of the methods we
considered has advantages and disadvantages, and in all cases we
considered, we could find light curves (often many of them) for which
the timescale by eye did not match the numbers resulting from a given
approach. Determining the `best' timescale to use is beyond the scope
of this paper (see, e.g., Findeisen \etal\ 2015). 

In other YSOVAR papers, timescales related to the autocorrelation
function (ACF) have been used. Using the ACF requires evenly spaced
times. We linearly interpolated the light curve onto evenly spaced
times, and then calculated the ACF using the following expression
where $L$ is a lag in days, and $x$ is the light curve (with elements
$x_k$): \begin{equation}
ACF_{x}(L)=ACF_{x}(-L)=\frac{\sum\limits_{k=0}^{N-L-1}
(x_k-\overline{x})(x_{k+L}-\overline{x})}{\sum\limits_{k=0}^{N-1}(x_k-\overline{x})^2}
\end{equation}  McQuillan \etal\ (2013) and Cody \etal\ (2014) used an
approach where they selected the larger of the first two local maxima
in the ACF to determine the period of the light curve.  For those
objects with signficant periods, this timescale is well-matched to the
period.  For objects without significant periods, this value seems to
give an indication of the timescale found in the light curve; it is
often well-matched to the variations we see by eye.  This calculation
can fail, however, if the ACF is not well-behaved. Where the ACF is
well-behaved, we use this timescale here. 
Sec.~\ref{sec:overalltimescales} characterizes the timescale
distribution, and Sec.~\ref{sec:tased} discusses relationships between
timescales, amplitudes, and SED classes.

For periodic objects, the timescale is perhaps the best defined in
theory. In practice, though, for the objects that we determine to be
periodic, sometimes there are variations in addition to the period (on
shorter and/or longer timescales) that call into question whether the
period is actually the dominant timescale.  The period, however, is
well-matched in most cases to the timescale derived from the ACF
above. The periodic objects in NGC 1333 are discussed in
Sec.~\ref{sec:periods}.

\subsection{Characterizing Overall Timescales}
\label{sec:overalltimescales}

\begin{figure}[h]
\epsscale{0.8}
\plotone{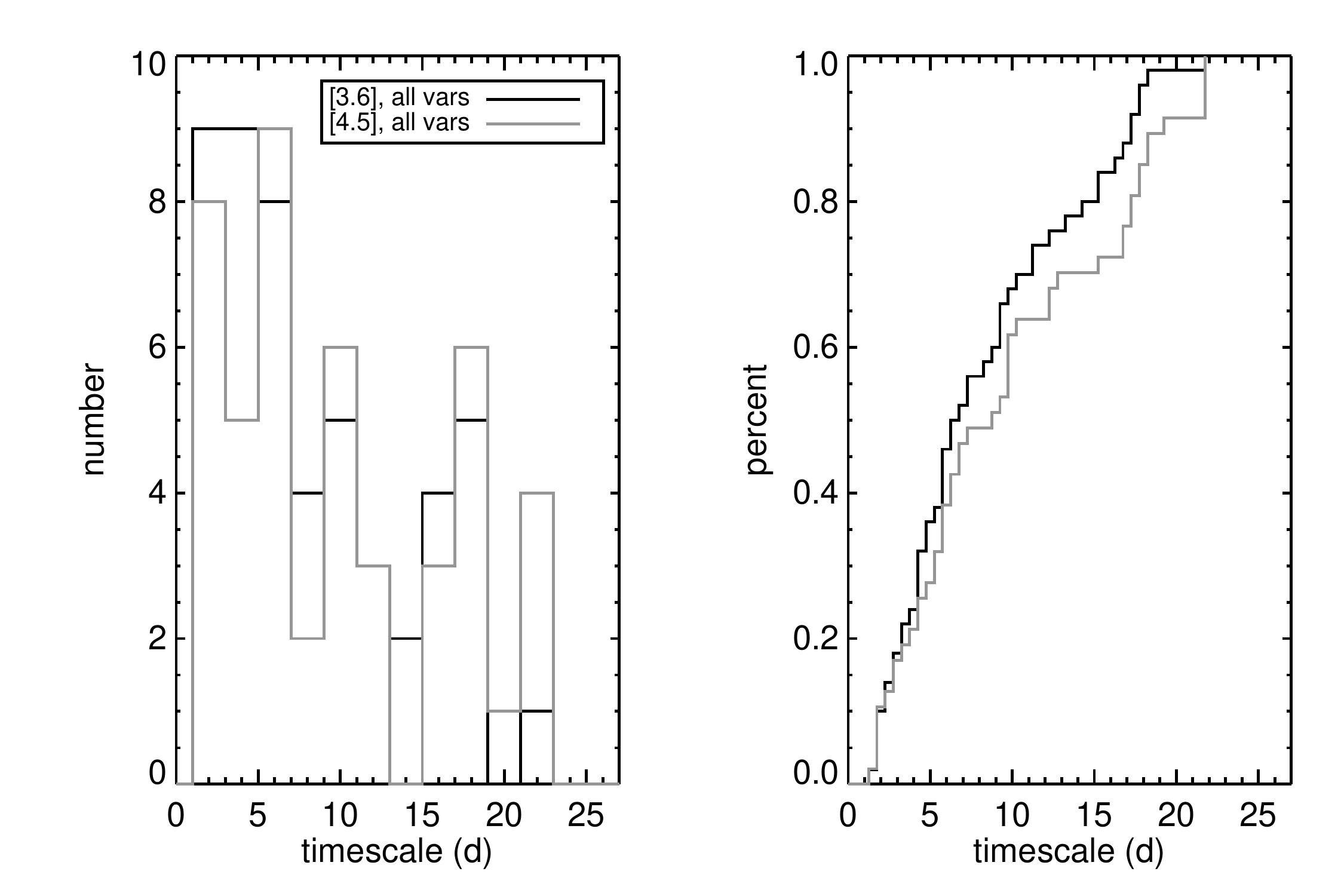}
\caption{Timescales for variable standard member objects. 
Black line is [3.6], and grey line is [4.5]. Left is histograms of
values, right is the distribution functions. The timescales are
typically short, on the order of a few days.   }
\label{fig:timescales}
\end{figure}

Figure~\ref{fig:timescales} has histograms and distributions of the
timescales for the variable standard members in NGC 1333. There are 55
variable members with measured timescales in either (or both) of the
bands.  The timescales are typically reasonably short, with medians
near 6.5 days, but a mode (most frequent value) near 4.5 days. Note
that this is just the standard members -- there are 7 more mid-IR
variable objects that are not part of the standard set of members, so
addition of those objects does not appreciably change
Figure~\ref{fig:timescales}.  We cannot characterize the distributions
of timescales for objects not considered to be members (as distinct
from members) because there are just too few of them. 

\begin{figure}[h]
\epsscale{0.6}
\plotone{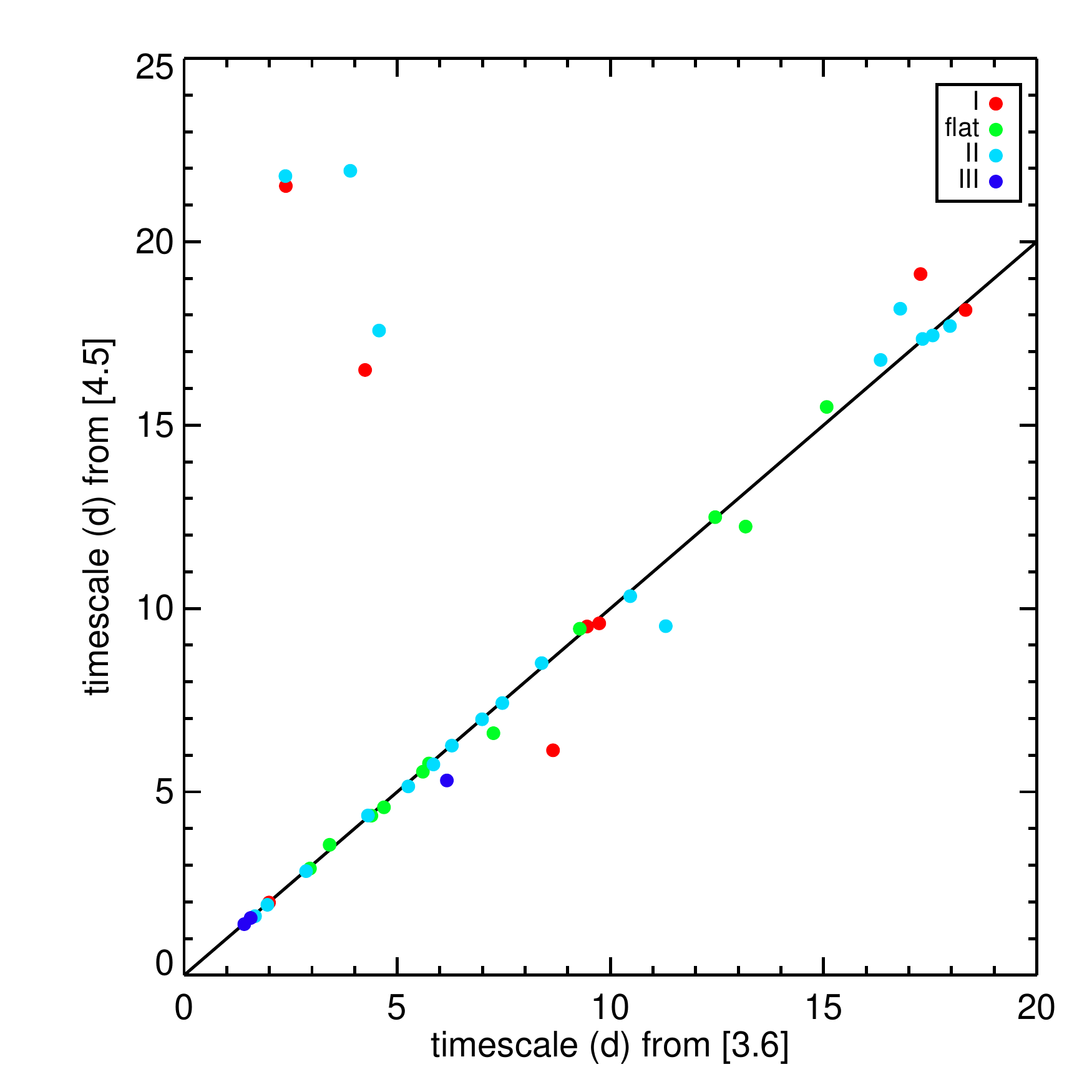}
\caption{Comparison of the timescales derived from the two channels
for all variable standard member objects. The colors of the points
correspond to the SED class as shown. The timescales derived from the
two channels are generally well-matched to each other.}
\label{fig:timescales2}
\end{figure}

For those cases where light curves exist in both channels, we can
compare the timescales obtained from both channels. We expect that the
two IRAC channels should be well-correlated, so we expect that the
timescales so derived will be similar. Figure~\ref{fig:timescales2}
compares the timescales derived for the two channels. They are,
indeed, well-matched for most objects. The lightcurves where they do
not match are typically very complex.

\subsection{Timescales, Amplitudes, and SED Class}
\label{sec:tased}

G\"unther \etal\ (2014), Poppenhaeger \etal\ (2015), and Wolk \etal\
(2015) all found longer timescales for steeper (positive) SED slopes,
that is, that Class I objects had on average longer timescales than
Flat class objects, which was longer than Class II, which was longer
than Class III. For NGC 1333, among the standard set of members, just
the variables for which we can calculate a timescale, and then 
dividing them by class, there are very few objects left per class to
analyze. Since we have shown that the timescales for [3.6] and [4.5]
are generally well-matched, we can combine timescales derived via the
channels, but even then, there are still very few objects per class. 
The medians of these combined timescale distributions show a weak
trend; the median of the 10 class I objects is 9.5 d, the 14 flat
class members have a median of 7.3 d, the 26 Class II members have a
median of 7.5 f, and the median of the 5 class III members is 2.4 d.  
Since there are literature-identified Class 0 objects in this cluster,
we had hoped we could pull out these objects separately, but only one
has a timescale in either IRAC channel, and it is 2.4 d in [3.6] and
21 d in [4.5].  Due to small number statistics, an analysis of this
sort will need to be repeated once all the YSOVAR clusters are
analyzed, so that we have larger numbers of objects in each class.

The low-number statistics similarly make it difficult to look for
correlations between timescale and amplitude, especially as a function
of class. There is very little correlation between timescale and
amplitude for the distribution as a whole. On average, the Class III
objects not only have smaller timescales but also smaller
amplitudes.   Similarly, the Class Is tend to have larger timescales
and larger amplitudes, but there is rather a lot of scatter. In terms
of color amplitude, the Class IIIs tend to have a smaller color change
on average, but the rest of the classes cover similar, large color
changes; the largest color changes are not the Class Is. Some of the
largest color changes are in objects that have small timescales.

\subsection{Periods}
\label{sec:periods}

\begin{figure}[h]
\epsscale{1.0}
\plottwo{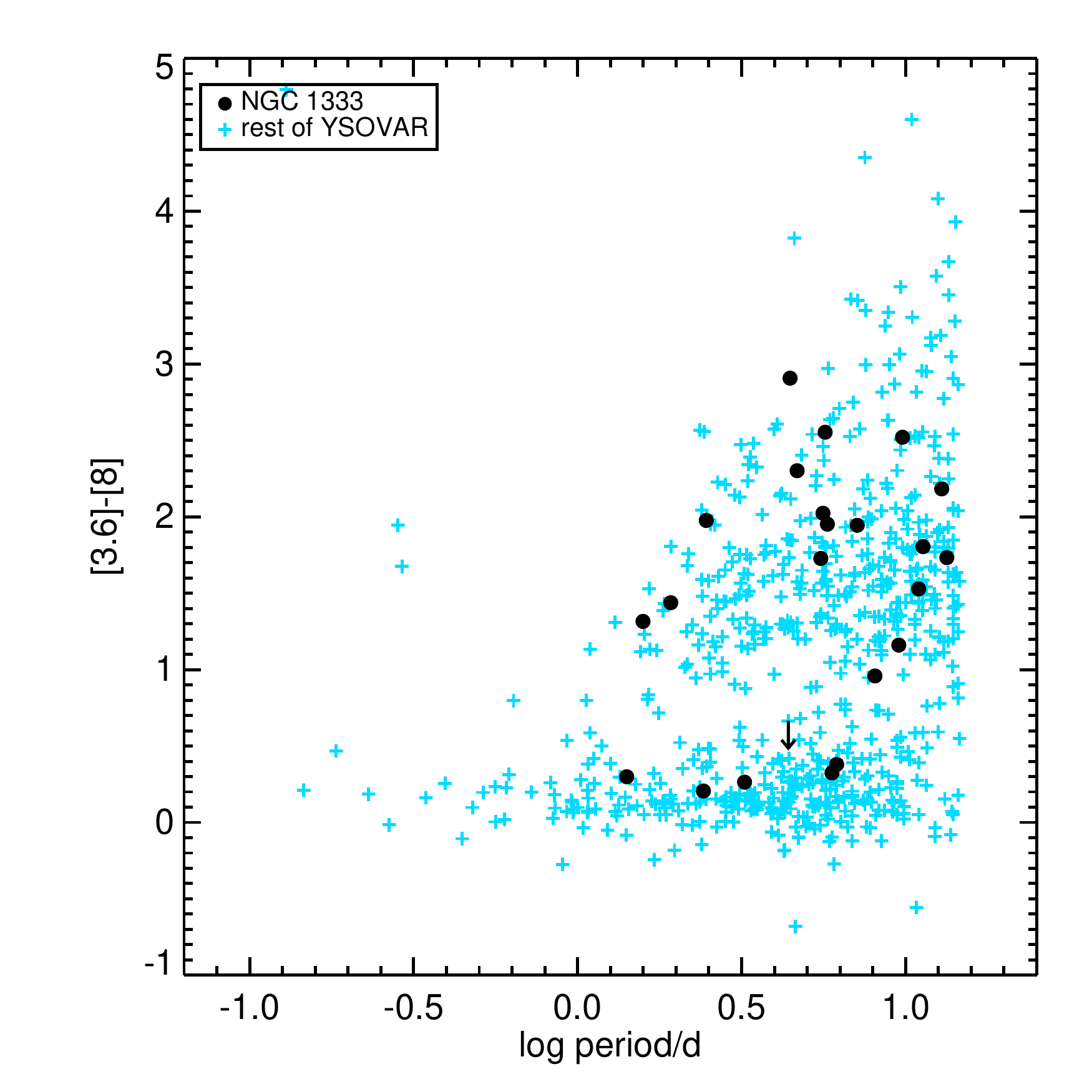}{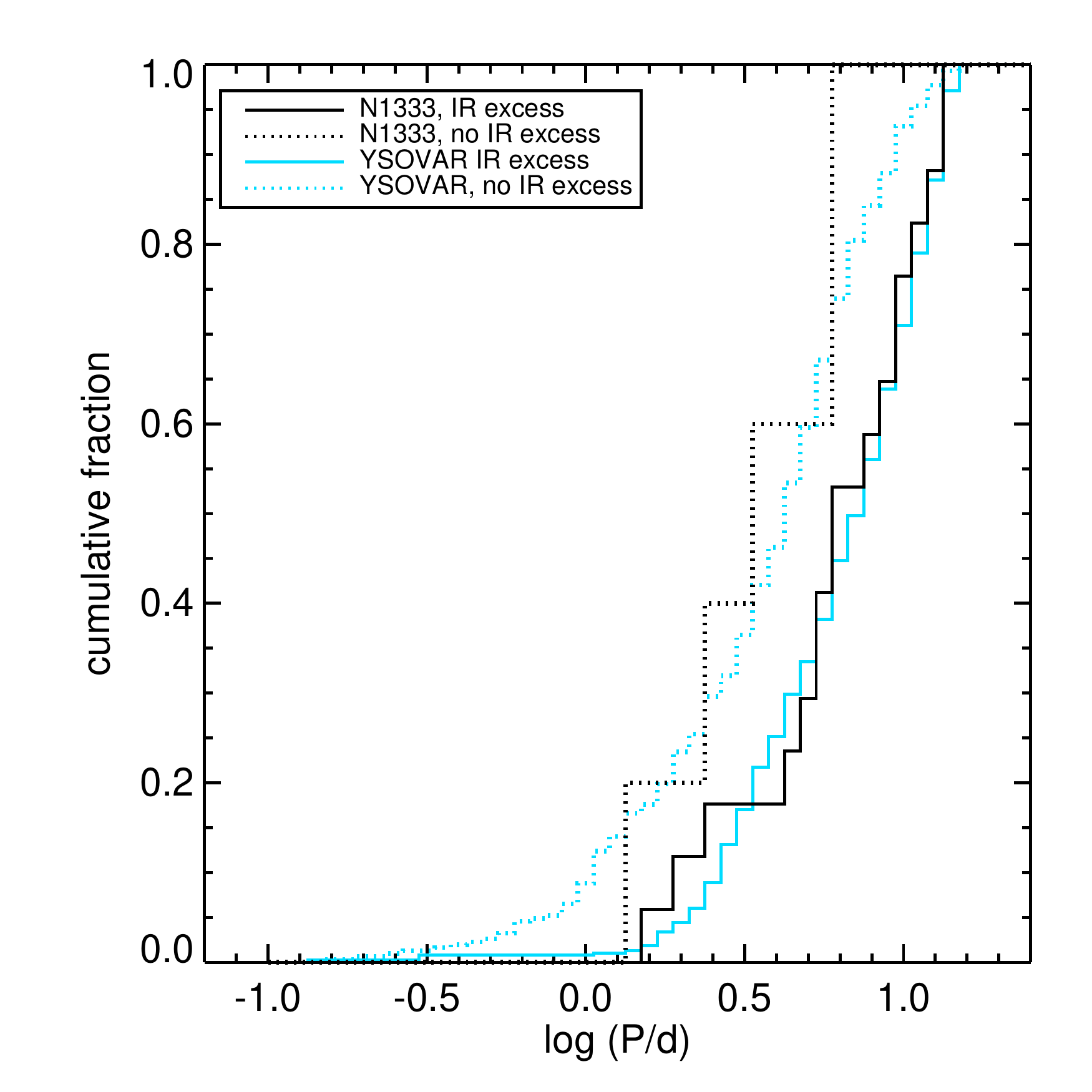}
\caption{LEFT: Plot of log period (in days) against [3.6]$-$[8] as an
indicator of IR excess. Solid circles (and one arrow) are points (and
upper limits) for stars in NGC 1333; blue $+$ are the rest of the
points from the rest of the YSOVAR clusters (R14).  There are no very
fast rotators here in NGC 1333, and there is a range of IR excesses
for the slower rotators. There is one [8] limit seen here, which does
not appear in R14 (because it was limited to detections). RIGHT:
cumulative distribution of period (in days) for NGC 1333, for
[3.6]$-$[8]$>$0.8 (solid) and $<$0.8 (dotted), black lines. The blue
lines are the equivalent for the entire YSOVAR sample (solid line is
[3.6]$-$[8]$>$0.8 and dotted line is [3.6]$-$[8]$<$0.8). }
\label{fig:periods2}
\end{figure}

In NGC 1333, we find 23 objects with periodic light curves, 19 of
which are in the standard set of members. R14 discussed the
relationship between period and IR excess for all the periodic objects
over all 12 YSOVAR clusters; the NGC 1333 points are highlighted in
Figure~\ref{fig:periods2}.  There are no very short periods in this
cluster; all of them are $>$1 d. These objects may not have had enough
time yet to spin up. Figure~\ref{fig:periods2} also has the cumulative
distribution of periods. Half the periods are between 1.4 and 5.6
days; the other half are between 5.6 and 15 d. 

\begin{figure}[h]
\epsscale{0.8}
\plotone{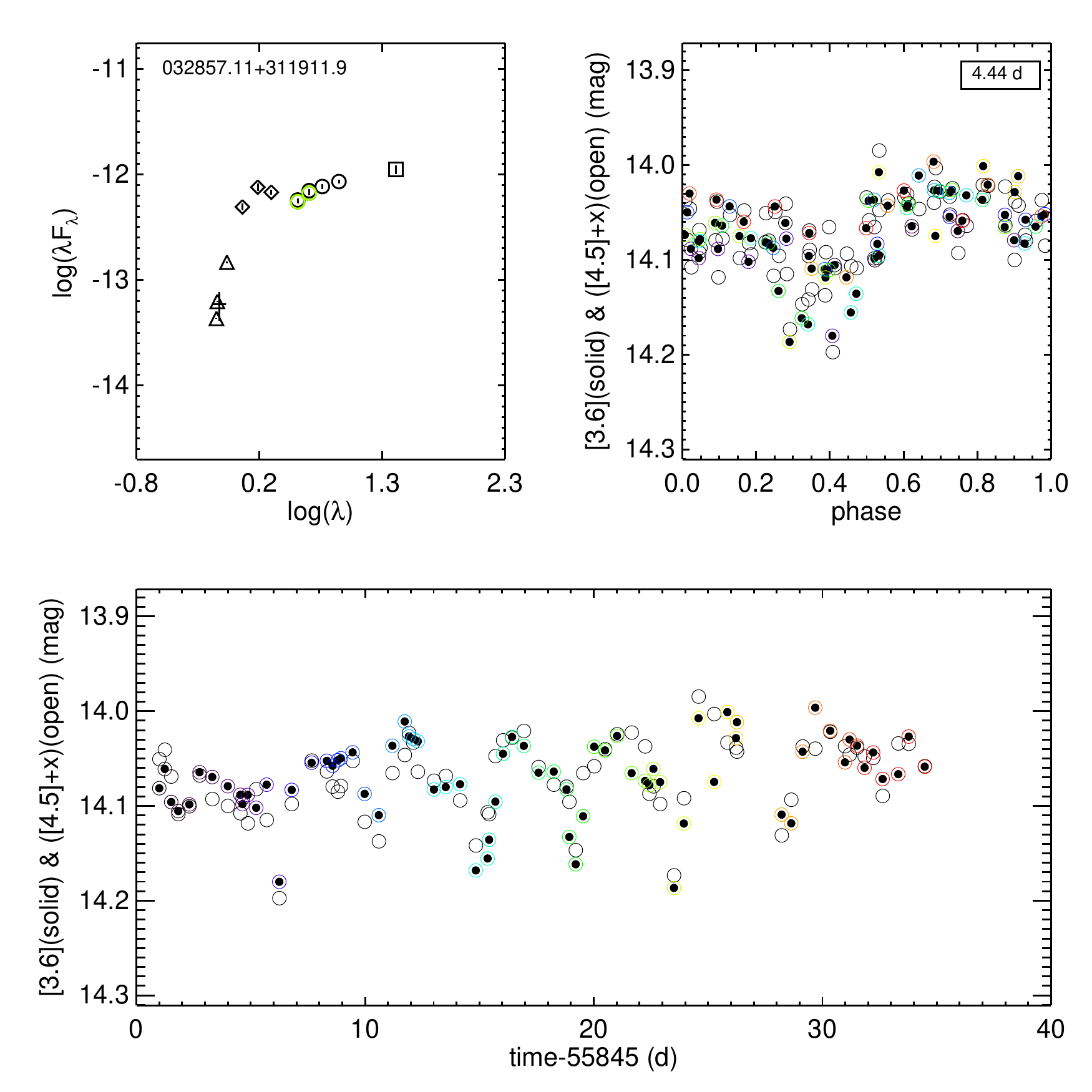}
\caption{Plots for the periodic source SSTYSV J032857.11+311911.9
(=ASR64=MBO148=S-6). The SED in the upper left has the same notation
as prior SEDs. The upper right is a phased light curve, where the
period is 4.44 d, and the dot colors correspond to the circle colors
shown on the full light curve on the bottom. In both the phased light
curve and the full light curve on the bottom, solid circles are [3.6],
open circles are [4.5], and the [4.5] light curve has been shifted
such that the mean [4.5] matches the mean [3.6].  This periodic object
is also a dipper -- see Sec.~\ref{sec:dippers} -- and a brown dwarf at
M7.3 -- see Sec.~\ref{sec:bds}.}
\label{fig:032857.11+311911.9}
\end{figure}

Out of the objects with periods, there is a surprising fraction of
objects that have relatively embedded SEDs.  One object (SSTYSV
032901.87+311653.3) has a slope between 2 and 24 \mum\ consistent with
a Class I, but it has that slope (and therefore class) only because of
a very bright but not obviously spurious 24 \mum\ point. However, the
2-8 \mum\ SED classification is Class III.  This is clearly a case
where a simple fit to all points between 2 and 24 \mum\ does not
adequately capture the nature of this source.  This source would
therefore be considered a transition object. Of the remaining periodic
objects, 9 are flat class, 8 are Class II, and 5 are Class III. This
is unusual compared to the other clusters as yet examined in detail,
where the periodic sources are predominantly Class IIIs. This could
mean that the periodic signal is coming from the inner disk (e.g.,
Artemenko \etal\ 2013), not the photosphere. As a representative from
these embedded periodic sources, Figure~\ref{fig:032857.11+311911.9}
shows the SED, phased light curve, and original light curve for SSTYSV
J032857.11+311911.9. It has a flat SED class, but is still clearly
periodic. The light curve morphology best matches expectations for an
AA Tau analog, or `dipper' -- variable extinction from structure in
the inner disk (e.g., Bouvier \etal\ 1999, McGinnis \etal\ 2015).
(Dippers will be discussed more in the next subsection.)

\begin{figure}[h]
\epsscale{0.8}
\plotone{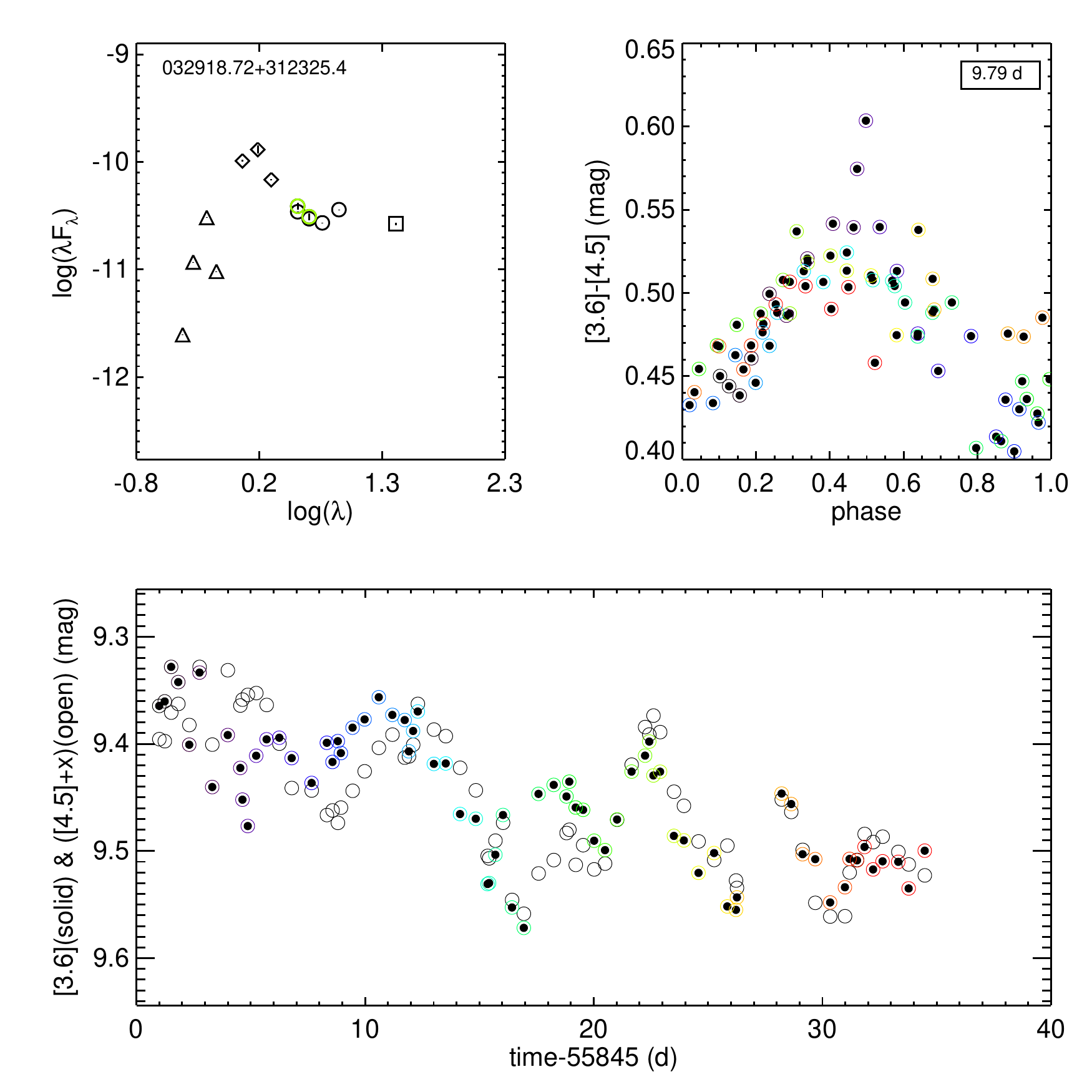}
\caption{Plots for the periodic source SSTYSV J032918.72+312325.4
(=J070-29=LAL293=Getman 81=Preibisch 18=MBO16=Winston 106=Foster 167,
an M0 spectral type). The SED in the upper left has the same notation
as prior SEDs. The upper right is a phased {\em color} light curve,
where the period is 9.79d. (Notation for this and the light curve is
otherwise the  same as Fig.~\ref{fig:032857.11+311911.9}.)  This
source was determined to be periodic from the change of [3.6]$-$[4.5]
vs.\ time; there was no significant period determined from [3.6] or
[4.5] alone.}
\label{fig:032918.72+312325.4}
\end{figure}

One of the periodic objects (SSTYSV J032916.81+312325.2) has a strong
5.6 d period found independently in [3.6], [4.5], and [3.6]$-$[4.5].
Two of the other periodic objects are found from the [3.6]$-$[4.5]
color, not the [3.6] or [4.5] light curves alone; they are SSTYSV
J032910.82+311642.7 and SSTYSV 032918.72+312325.4, the latter of which
is shown in Figure~\ref{fig:032918.72+312325.4}. In both cases, there
is a long term trend with texture on top, which may be why we see the
signature in color but not the individual channels -- subtracting the
channels removes the long-term trend, leaving the color signature of a
repeated pattern. SSTYSV J032918.72+312325.4 has a larger amplitude
fluctuation than SSTYSV J032910.82+311642.7. For the former, while the
Lomb-Scargle approach finds a significant period only in the color,
the ACF analysis finds this timescale in [4.5] and [3.6]$-$[4.5].  The
net amplitude is $\Delta$[3.6]$-$[4.5]$\sim$0.25 mag, and the mean
[3.6]$-$[4.5]=0.48 mag.  In the other case of SSTYSV
J032910.82+311642.7, the phased light curve is messy, but multiple
approaches retrieve the same period. There is a net amplitude
$\Delta$[3.6]$-$[4.5] of $\sim$0.1 mag, and a mean [3.6]$-$[4.5] of
0.85 mag.  

None of the periodic objects are particularly suggestive of eclipsing
binaries. Several are very clear, repeatable signals suggestive of
cold spots on the photosphere. Some have large-scale downward or
upward trends, and/or the periodic signal is changing slightly with
every repetition; in these cases, it may be that we are seeing
signatures of accretion streams or hot spots on the photosphere. In a
few cases, we see `dippers' and `bursters'; see next section.

\begin{figure}[h]
\epsscale{0.5}
\plotone{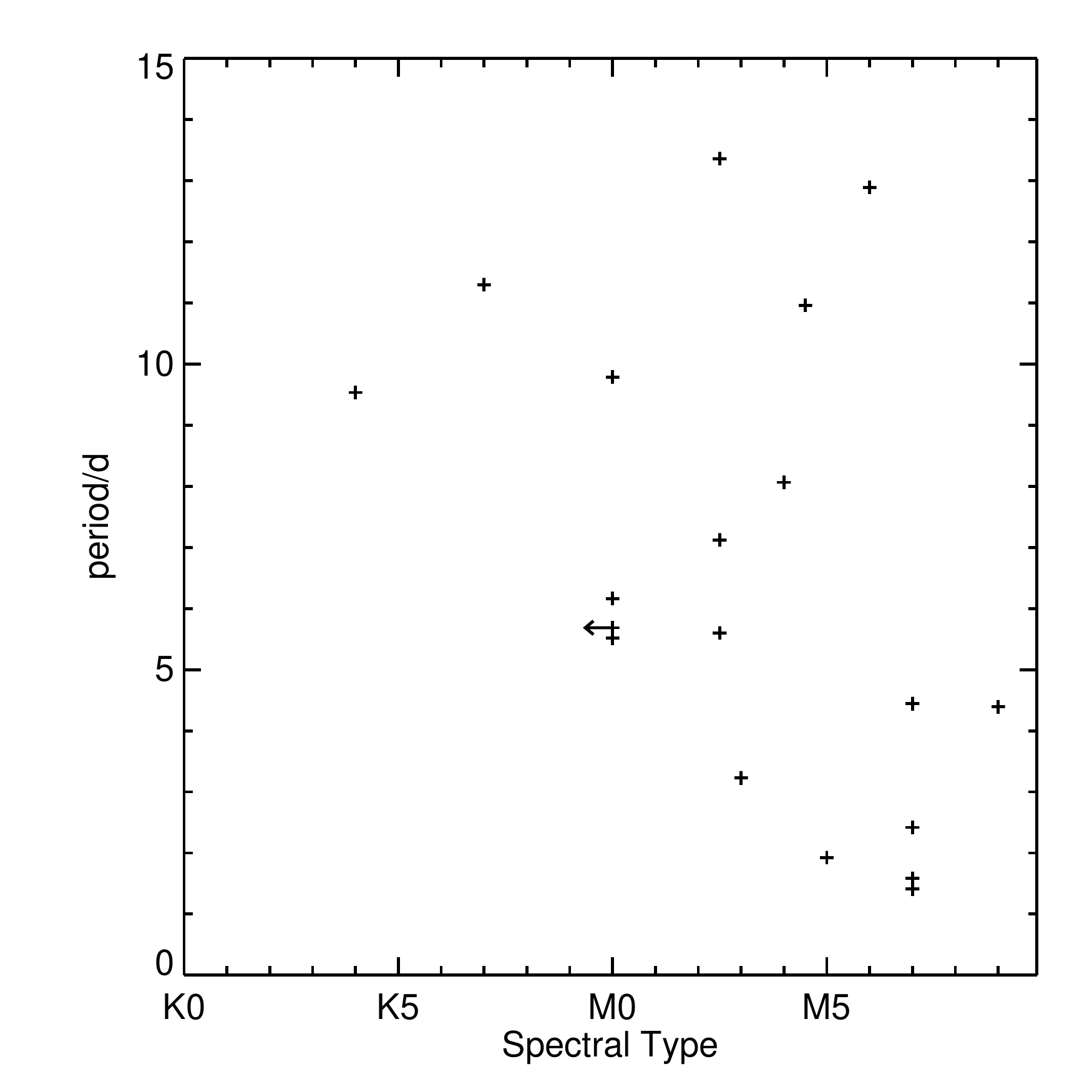}
\caption{Period (in days) against the spectral type. There is a weak
tendency for the later types to be rotating (or otherwise creating
repeated texture in their light curves) more quickly.  }
\label{fig:pspty}
\end{figure}

We can investigate correlations between period and spectral type. Just
four of the objects with periods are missing spectral types. If the
object is unobscured enough that we can derive a period, then it is
also unobscured enough that it has a measured spectral type.
Figure~\ref{fig:pspty} shows the period against the spectral type
(K0=5.0, K2=5.2, M0=6.0, etc). There is a weak tendency for the later
types to have a shorter period, suggesting that they are rotating (or
have inner disks rotating) more quickly. There is no evidence for a
similar trend in Orion or NGC~2264, though the sample sizes in those
clusters are 15-20 times larger than that for NGC 1333.

\clearpage

\section{Dippers and Bursters}
\label{sec:dippers}

Morales-Calder\'on \etal\ (2011), Cody \etal\ (2014), Stauffer \etal\
(2014, 2015), and McGinnis \etal\ (2015) identified and discussed objects
from Orion and NGC 2264 that they categorized as `dippers' (or AA Tau
analogs) and `bursters.' Dippers have light curves where there is a
`continuum' from which one can measure intermittent decreases in the
observed flux, of typically a few hours to days. Bursters have light
curves that are almost inverted dippers in that there is a `continuum'
from which one can measure intermittent increases in the observed
flux. In both Orion and NGC 2264, there was copious additional
contemporaneous data taken at a variety of wavelengths, which allowed
an interpretation of dippers being largely due to stellar occultations
by ``texture" in the circumstellar dust disk (e.g., disk warps or
overdensities), and that of bursters to be accretion instabilities
(see, e.g., Kulkarni \& Romanova 2009 or Romanova \etal\ 2011).  We
have far less ancillary data available here, but we can use the same
metrics presented in Cody \etal\ (2014) to help identify the bursters
and dippers, namely $Q$ and $M$. The $Q$ parameter is a measure of 
pattern repetition, and $M$ is a measure of up/down symmetry.  (Values
for $Q$ and $M$ appear in Table~\ref{tab:singlevaluedata}.)

To calculate $M$, Cody \etal\ (2014) compare the mean and median of
the light curves (having removed the highest and lowest 10\% of
points), and divide by the standard deviation of the light curve.
Following the limits set in Cody \etal\ (2014), values of $M<-$0.25
suggests bursters, and $M>$0.25 suggests dippers. The `continuum' is
thus defined as the mean of the light curve, and asymmetries are
measured with respect to that mean.

Dippers and bursters as identified in earlier work are often but not
always periodic. The NGC 1333 periodic sources listed in
Sec.~\ref{sec:periods} above include 3 periodic bursters, and 1
periodic dipper, the latter of which is shown in
Fig.~\ref{fig:032857.11+311911.9}.  There are 8 aperiodic dippers. Two
more bursters may be periodic, but on longer timescales than we can
measure. The remaining 6 bursters (11 bursters - 3 periodic - 2
candidate periodic) are aperiodic.  

\begin{figure}[h]
\epsscale{0.8}
\plotone{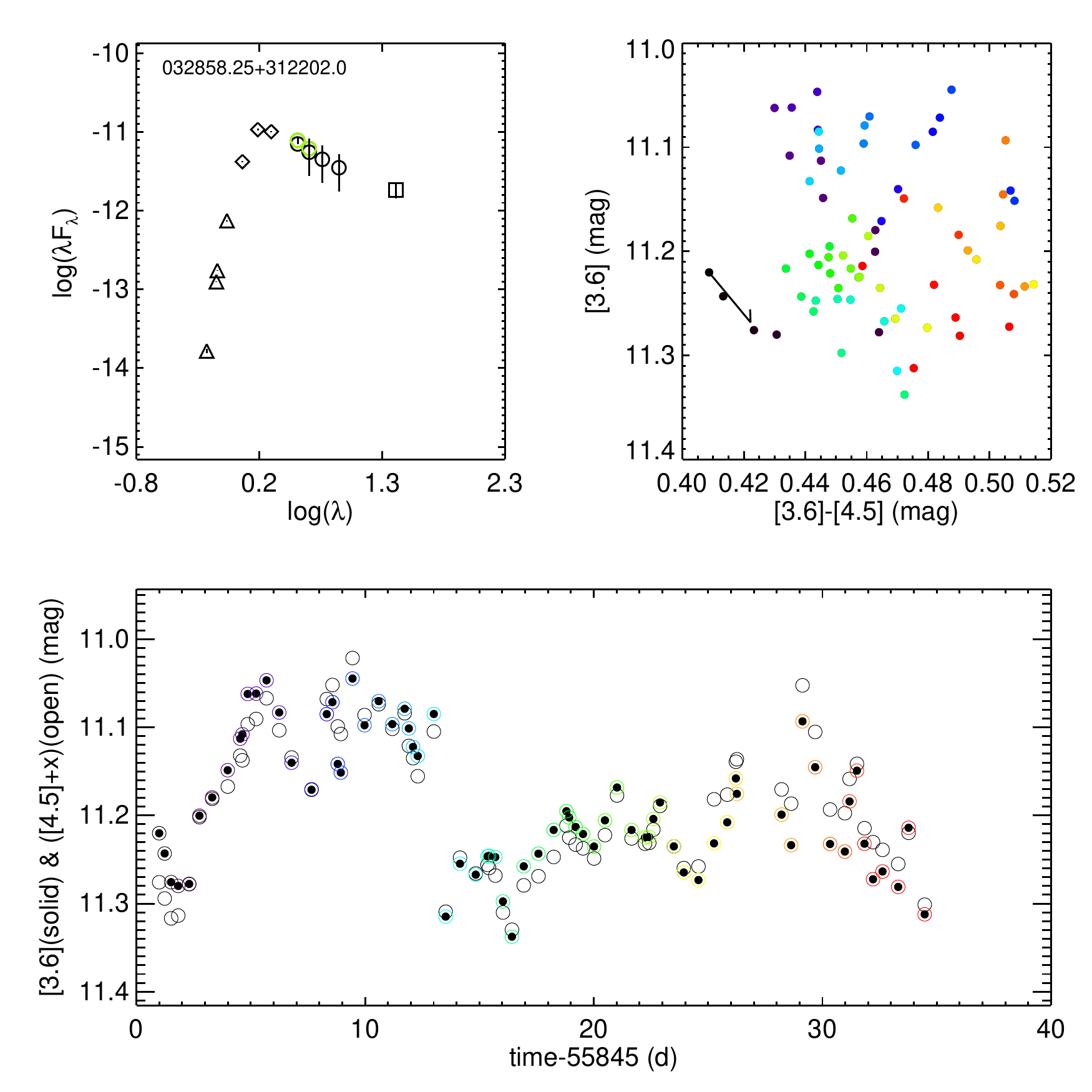}
\caption{Plots for the aperiodic burster source SSTYSV
032858.25+312202.0 (=LAL163=MBO47=Foster 65).  Notation is as in prior
similar figures. This object is also a brown dwarf at M6 -- see
Sec.~\ref{sec:bds}.}
\label{fig:032858.25+312202.0}
\end{figure}

Figure~\ref{fig:032858.25+312202.0} (SSTYSV J032858.25+312202.0) is an
example of an aperiodic burster. The light curve shows the
characteristic behavior of having an apparent lower `continuum' level
with brighter flux excursions (bursts) of varying length superposed
on top of the continuum level. However, it does not have a significant
trend in the CMD; see next section.

\section{Color Trends}
\label{sec:colortrends}

For those objects with well-populated light curves at both bands, we
can investigate the color trends over the YSOVAR campaign. G\"unther
\etal\ (2014), Poppenhaeger \etal\ (2015), and Wolk \etal\ (2015) all
find that many objects become bluer when brighter (as would be
expected for variations in extinction), but that a few objects become
redder when brighter.  We look for these color trends in a slightly
different way than the prior three studies. We describe our approach,
followed by the results of this analysis in NGC 1333.

\subsection{Finding Significant Color Trends}

For most objects in the YSOVAR fields that have light curves at both
bands, the [3.6] and the [4.5] observation at a given epoch were
obtained within 12 minutes of each other. While this is not strictly
simultaneous, for the timescales in which we are interested, we regard
them as simultaneous.  For each object that has light curves at both
bands, we can construct both the [3.6] vs.\ [3.6]$-$[4.5]
CMD and the [4.5] vs.\ [3.6]$-$[4.5] CMD. 
We calculate a linear Pearson correlation coefficient for the
distribution of points. However, this calculation does not take into
account the intrinsic errors on each point. We used a `safe
correlation' routine by P.\ Lloyd found in the IDL Astronomy
Library\footnote{http://idlastro.gsfc.nasa.gov/homepage.html and
http://parkeloyd.com/output/code/safe\_correlate/} to calculate 10$^4$
realizations of the data, where each point can vary its location
corresponding to the error on the point (in both dimensions). It then
recalculates the Spearman rank correlation coefficient for each
realization of the data. From the distribution of the correlation
coefficients, the routine determines the probability that the original
distribution could be uncorrelated. We calculated these values for
both CMDs, for all objects that had light curves in both bands. By
inspection of several hundred CMDs, we determined that a significant
correlation could be found in the CMD if the absolute value of the
correlation coefficient was $>$0.45 and at the same time the
probability calculated by the safe correlation routine was $<$15\%. 
Based on empirical tests, between 1-5 outlying points can affect these
statistics, but the influence of these outliers is limited. Usually if
there is a strong correlation, it stays strong even if outliers are
removed. 

Following the reddening law of Cardelli, Clayton, \& Mathis (1989), in
the [3.6] vs.\ [3.6]$-$[4.5] CMD, the slope of the reddening vector
should be $\sim$3.5, and in the second [4.5] vs.\ [3.6]$-$[4.5], it
should be $\sim$2.5. Using the much more recent Indebetouw \etal\
(2008) values for IRAC bands and the interstellar medium, the slope of
the reddening vector should be $\sim$4.5 and $\sim$3.5, respectively.
For either one, the slope expected from reddening is larger in a [3.6]
vs.\ [3.6]$-$[4.5] diagram than in a [4.5] vs.\ [3.6]$-$[4.5]
diagram.  Thus, variations consistent with reddening are slightly more
easy to find (given our approach) in the [3.6] vs.\ [3.6]$-$[4.5]
diagram than in the other one.

\subsection{Color Trends in NGC 1333}

\begin{figure}[h]
\epsscale{0.8}
\plotone{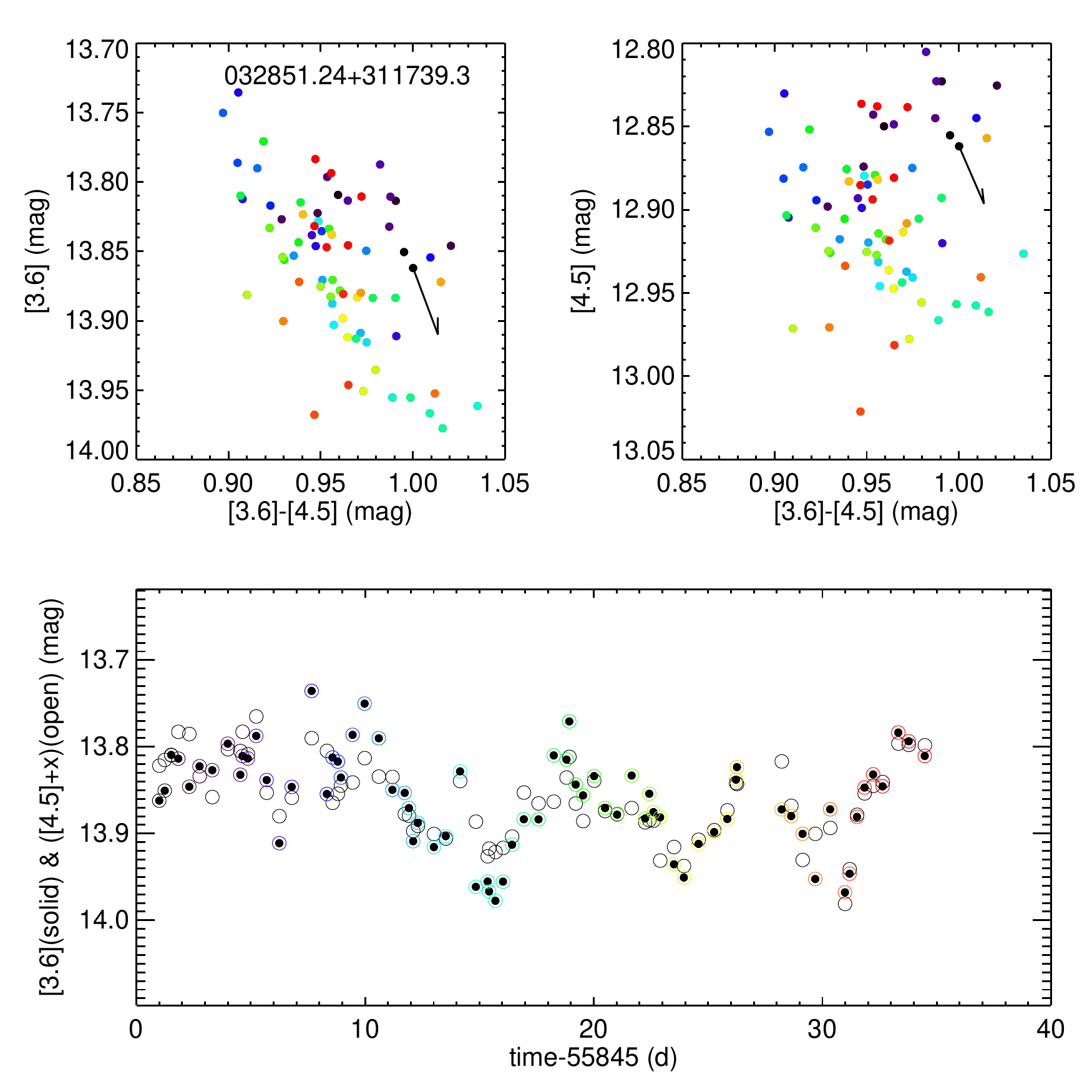}
\caption{Color-magnitude diagrams and light curve for SSTYSV
032851.24+311739.3 (=ASR41=LAL111, and a known edge-on disk -- Hodapp
\etal\ 2004), a source
that has a significant `red' trend (redder when fainter) in [3.6] vs.\
[3.6]$-$[4.5] but not in [4.5] vs.\ [3.6]$-$[4.5]. Notation is as in
prior CMDs and light curves.  A reddening vector with \av=1 is shown
from the first point in each CMD. The correlation coefficients
calculated as per the text is 0.56 for the first CMD and 0 for the
second. The `safe correlate' routine finds that there is a 0.6\%
chance that the points are not correlated in the first CMD, and a 64\%
chance that the points are not correlated in the second CMD..  }
\label{fig:032851.24+311739.3}
\end{figure}

\begin{figure}[h]
\epsscale{0.8}
\plotone{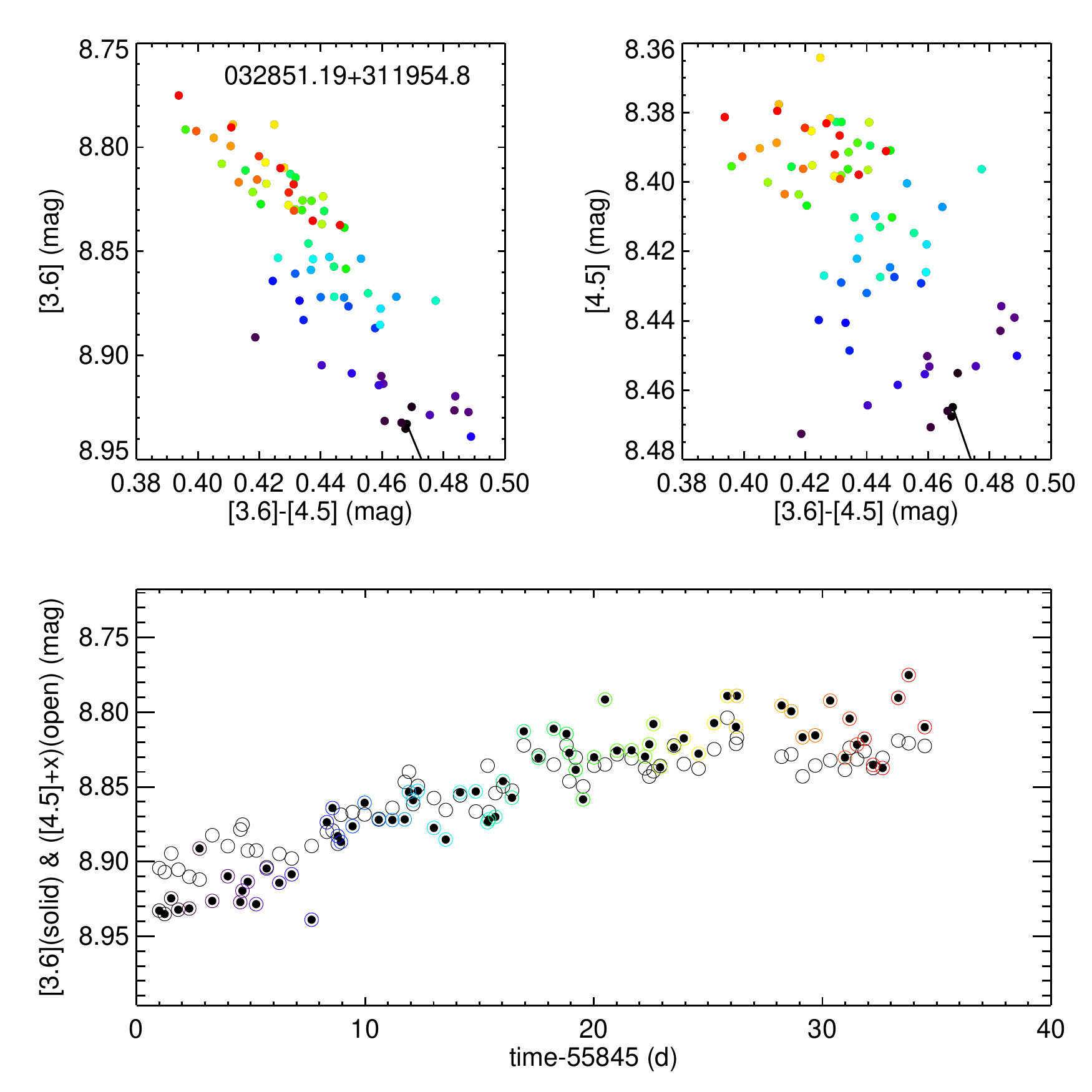}
\caption{Color-magnitude diagrams and light curve for SSTYSV
032851.19+311954.8 (=ASR125=LAL110=Getman 21=Preibisch 7=MBO14=Winston
53=Foster 37, a K7 spectral type), a source that has a significant
`red' trend  (redder when fainter) in both CMDs. Notation is as in
prior CMDs and light curves. The correlation coefficients calculated
as per the text is 0.86 for the first CMD and 0.60 for the second. The
`safe correlate' routine finds in both cases that there is a $<$0.05\%
chance that the points are not correlated. }
\label{fig:032851.19+311954.8}
\end{figure}

\begin{figure}[h]
\epsscale{0.8}
\plotone{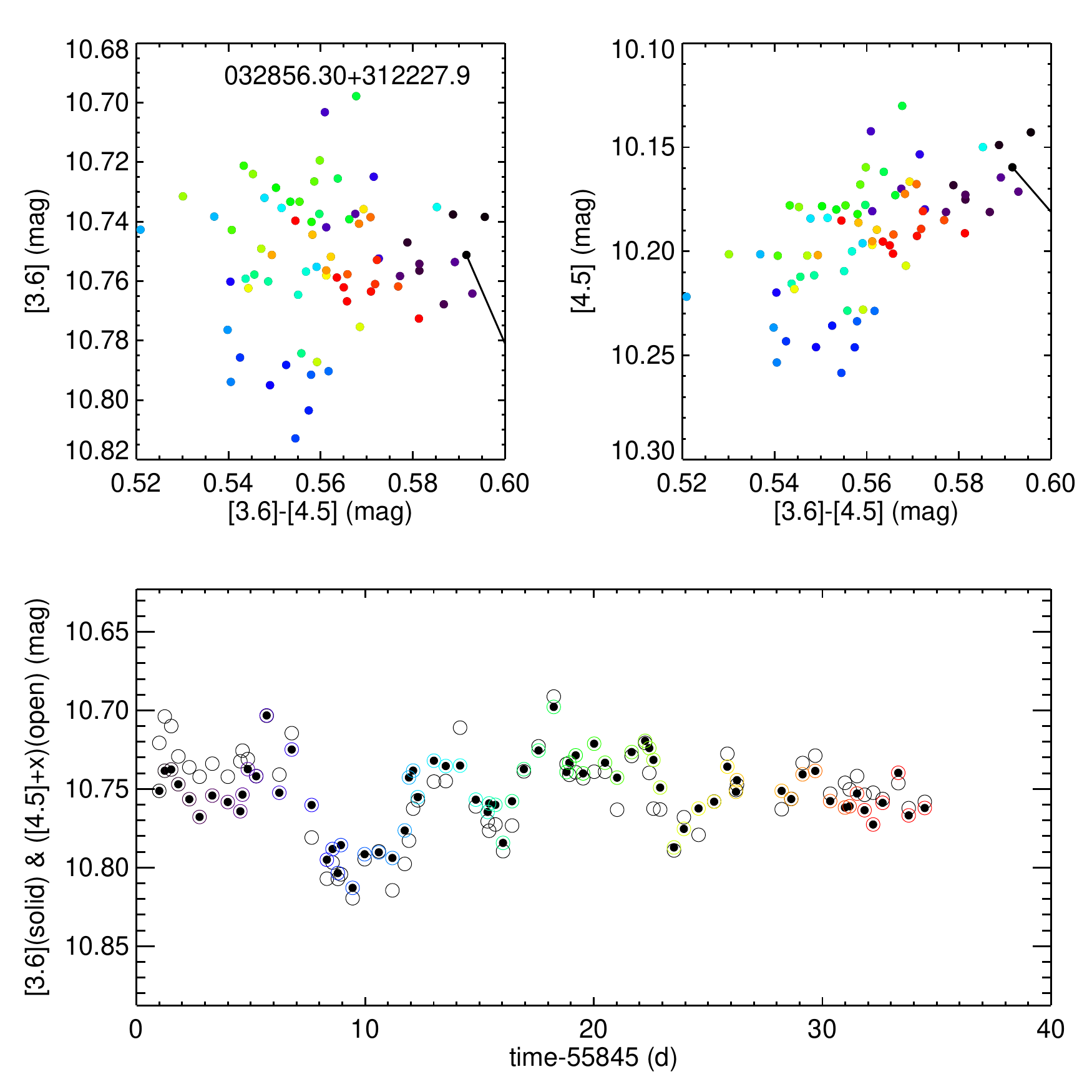}
\caption{Color-magnitude diagrams and light curve for SSTYSV
032856.30+312227.9 (=LAL147=MBO37=Winston92=Foster 51, an M2 spectral
type), a source that has a significant `blue' trend (bluer when
fainter) in [4.5] vs.\ [3.6]$-$[4.5] but not in [3.6] vs.\
[3.6]$-$[4.5].  Notation is as in prior CMDs and light curves. The
correlation coefficients calculated as per the text is 0.$-$0.05 for
the first CMD and $-$0.59 for the second. The `safe correlate' routine
finds for the first CMD that there is a 58\% chance that the points
are not correlated, but a $<$0.05\% chance that the points are not
correlated in the second. }
\label{fig:032856.30+312227.9}
\end{figure}

\begin{figure}[h]
\epsscale{0.8}
\plotone{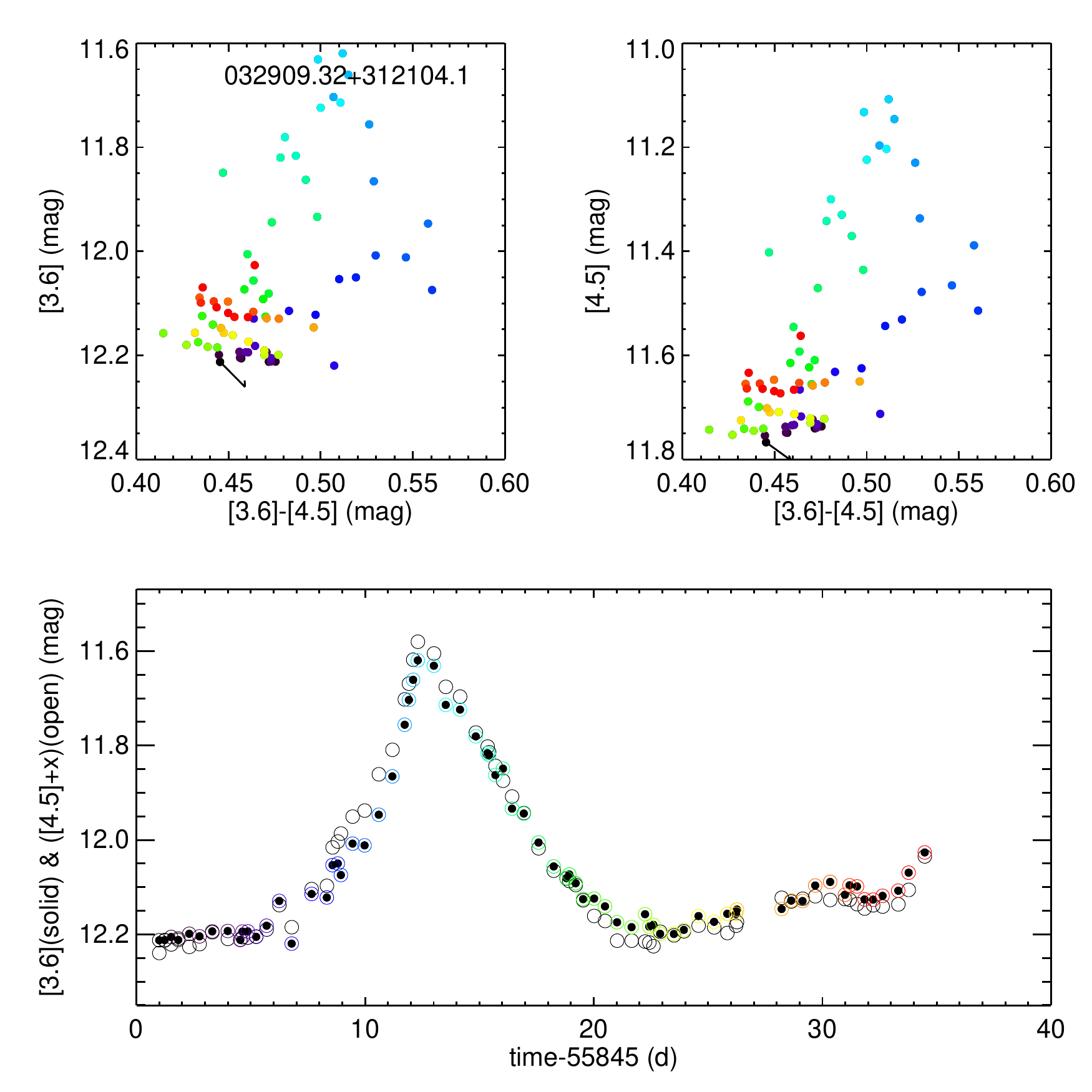}
\caption{Color-magnitude diagrams and light curve for SSTYSV
032909.32+312104.1 (=LAL225=MBO70=Foster 126=S-12), a source that has
a significant `blue' trend  (bluer when fainter) in both CMDs.
Notation is as in prior CMDs and light curves. The correlation
coefficients calculated as per the text is $-$0.52 for the first CMD
and $-$0.64 for the second. The `safe correlate' routine finds in both
cases that there is a $<$0.05\% chance that the points are not
correlated. }
\label{fig:032909.32+312104.1}
\end{figure}

For NGC 1333, there are 17 objects that have a significant correlation
in the [3.6] vs.\ [3.6]$-$[4.5] CMD in the direction of bluer when
brighter and redder when fainter (consistent with extinction
variations), but no significant correlation in [4.5] vs.\
[3.6]$-$[4.5]. There are 6 objects that have a significant correlation
in both CMDs for redder when fainter (bluer when brighter).  There are
15 objects that have a significant correlation in [4.5] vs.\
[3.6]$-$[4.5] in the very roughly orthogonal direction of bluer when
fainter, but no significant correlation in [3.6] vs.\ [3.6]$-$[4.5].
There is only one object that has a significant correlation in both
CMDs (for bluer when fainter). Objects that become redder when
fainter have been referred to as `reddeners', and objects that become
bluer when fainter have been referred to as `bluers' or `blueners'
(G\"unther \etal\ 2014, Poppenhaeger \etal\ 2015, \& Wolk \etal\
2015). Figures~\ref{fig:032851.24+311739.3} --
\ref{fig:032909.32+312104.1} show examples of these four cases --
significant positive correlation in [3.6] vs.\ [3.6]$-$[4.5],
significant positive correlation in both, significant negative
correlation in [4.5] vs.\ [3.6]$-$[4.5], and significant negative
correlation in both.  

SSTYSV J032851.24+311739.3 (Fig.~\ref{fig:032851.24+311739.3}) is an
example of an object that has a significant `red' trend (redder when
fainter) in [3.6] vs.\ [3.6]$-$[4.5] but not in [4.5] vs.\
[3.6]$-$[4.5]. We note here that this is an edge-on disk (Hodapp
\etal\ 2004), which is the right geometry to create a dipper object,
but the light curve for this object is symmetric above and below its
mean, so is not characterized as a dipper.

SSTYSV J032851.19+311954.8 (Fig.~\ref{fig:032851.19+311954.8}) is one
of 6 light curves that have a significant redder-when-fainter trend,
but also one of the $\sim$25\% of variable objects that have a
significant overall trend in the original light curve. This long, slow
trend (sometimes with a concurrent trend in the CMD, sometimes not)
over the entire campaign is very common in this cluster.

SSTYSV J032856.30+312227.9 (Fig.~\ref{fig:032856.30+312227.9}) is an
example of a light curve that has a significant `blue' trend (bluer
when fainter) in [4.5] vs.\ [3.6]$-$[4.5] but not in [3.6] vs.\
[3.6]$-$[4.5]. This object is also identified as a dipper, since there
is a flux asymmetry such that there is a continuum from which downward
excursions can be identified.

SSTYSV J032909.32+312104.1 (Fig.~\ref{fig:032909.32+312104.1}, in
addition to being the only source with signficant trend in the
direction of bluer when fainter, has interesting `tracks' in its CMD.
When the YSOVAR campaign starts (the darkest points in the figure), it
is already moving in the general direction of redder when brighter. As
it climbs the brightness peak towards day $\sim$12 of the campaign
(blue points in the figure), it becomes dramatically redder and
brighter. As it falls back to its original brightness (cyan/green in
the figure), it becomes bluer. Interestingly, the variations near the
end of the fall, and after it (light green, yellow, red), are all back
to moving in the direction of bluer when brighter, more consistent
with the slope of the reddening vector. This is a fascinating source.
While this source is in a region of high surface brightness ($\sim$2
MJy sr$^{-1}$ at 3.6 \mum), it is relatively isolated, so these
variations are unlikely to be the result of an instrumental artifact.
We have 9 epochs of $J$, $H$, or $K_s$ data over this 40 d window, but
the $JHK_s$ variations seem to be independent of the IRAC variations.
It is not at all clear what is happening in/near this source. 

There are 39 sources in NGC 1333 that have one of these four color
trends. Sources of all SED classes are found with either of the `red'
trends (that is, redder when fainter in one or both CMDs). Sources of
all classes except for Class I are found with the `blue' (bluer when
fainter) trends. There are fewer of these `blue' sources (16 vs.\
23), and only one of the sources has a `blue' trend in both CMDs (it
is a Flat class). So, the observation that none of the most embedded
sources have a `blue' trend may be a result of small number
statistics, but is nonetheless interesting. Only one of the `blue'
sources is a class III, and two of the `red' sources are Class III.
More detailed analysis of trends of color with SED class will require
combination of data from this cluster with other YSOVAR clusters.

\clearpage

\section{Special Sources}
\label{sec:specialsources}

\subsection{Class 0 and Jet Drivers}
\label{sec:class0jets}

NGC 1333 is unusual in that it has several very embedded, very young
sources, and sources thought to be driving jets. Unlike most of the
other cluster members with YSOVAR light curves, the flux density in
the IRAC bands for these most embedded sources may very well be
dominated by outflow-driven molecular hydrogen emission, which can be
close to the source (e.g., Arnold \etal\ 2012). Thus, the origin of
the variability in these objects may be very different, and it is
worth looking just at the characteristics of the Class 0s and jet
drivers.

In R15, we identified Class 0 and jet-driving objects from the
literature. Of those sources, 17  have light curves in one or both
bands; they are listed in Table~\ref{tab:class0jets}.  These objects
are relatively faint at [3.6], as seen in Table~\ref{tab:class0jets};
the mean [3.6] from the cryo era is $\sim$14, whereas the mean [4.5]
is $\sim$12.5. Only 6 of the 17 (a third) are variable in either
channel over the YSOVAR campaign; 9 of them (half) are CY variable.
This is perhaps not surprising; maybe the most embedded sources may
only vary on very long timescales because the envelope takes time to
change in response to things going on near the YSO. Very large
variations on long timescales could occur as the YSO accretes at an
unsteady rate, with the reprocessed radiation creating large amplitude
variations on relatively long timescales. Of the 7 that are
specifically identified as jet drivers, 4 ($\sim60$\%) are variable
over the YSOVAR campaign, and 5 ($\sim70$\%) are CY variable. Of the
15 that are specifically identified as Class 0s, 4 ($\sim25$\%) are
variable over the YSOVAR campaign, and 7 ($\sim50$\%) are CY variable.
Since the jet driving mechanism is thought to be accretion-driven, and
accretion is highly unlikely to be a smooth, isotropic process, it
makes sense that there is a higher fraction of variables among the
jet-driving sources. Or, we could be seeing variation in the molecular
hydrogen emission close to the source.

\begin{figure}[h]
\epsscale{0.8}
\plotone{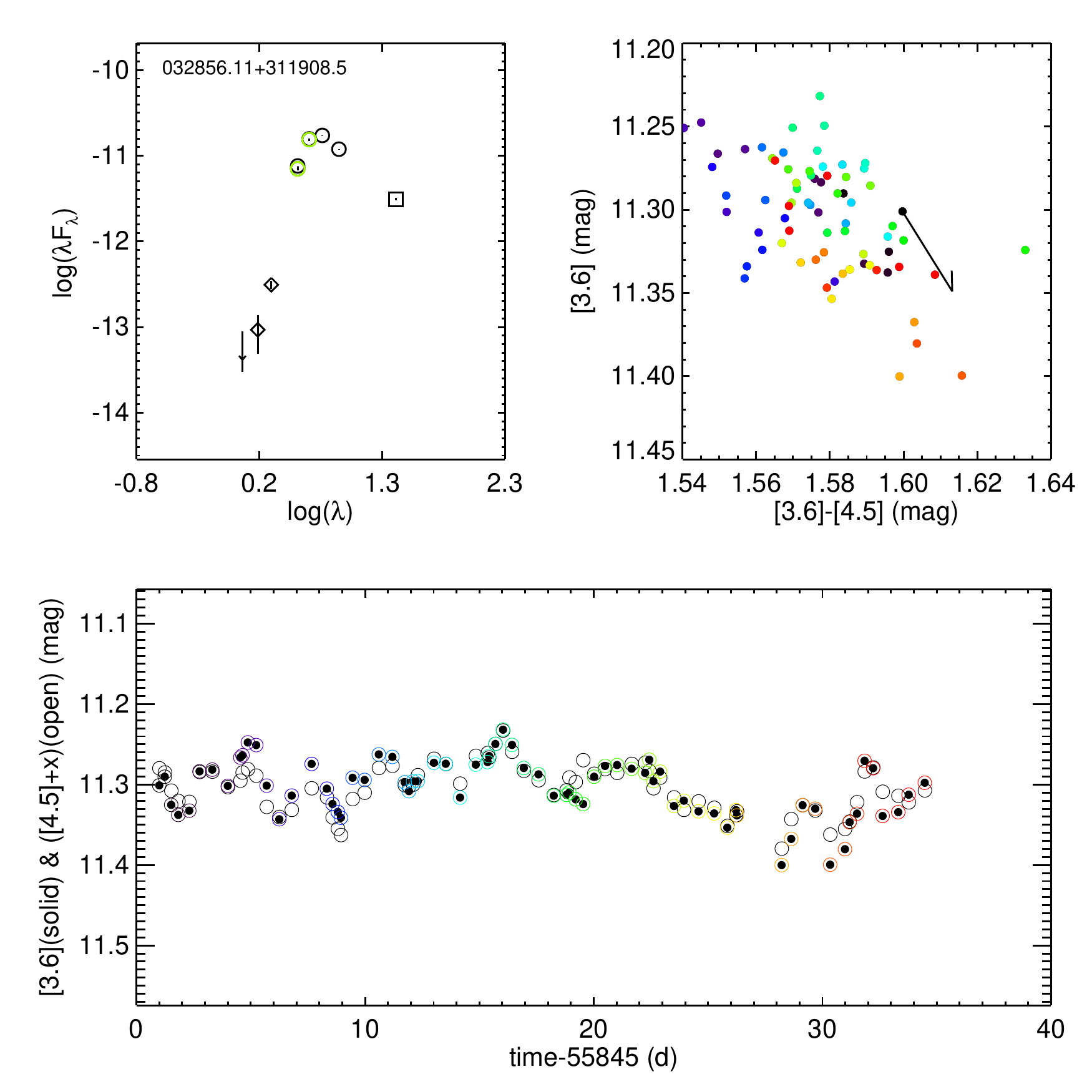}
\caption{Plots for the literature-identified Class 0 source SSTYSV
032856.11+311908.5 (=MBO146=Winston43). The SED in the upper left has
the same notation as prior SEDs. The upper right is a CMD, and the
light curve is at the bottom.}
\label{fig:032856.11+311908.5}
\end{figure}

\begin{figure}[h]
\epsscale{0.8}
\plotone{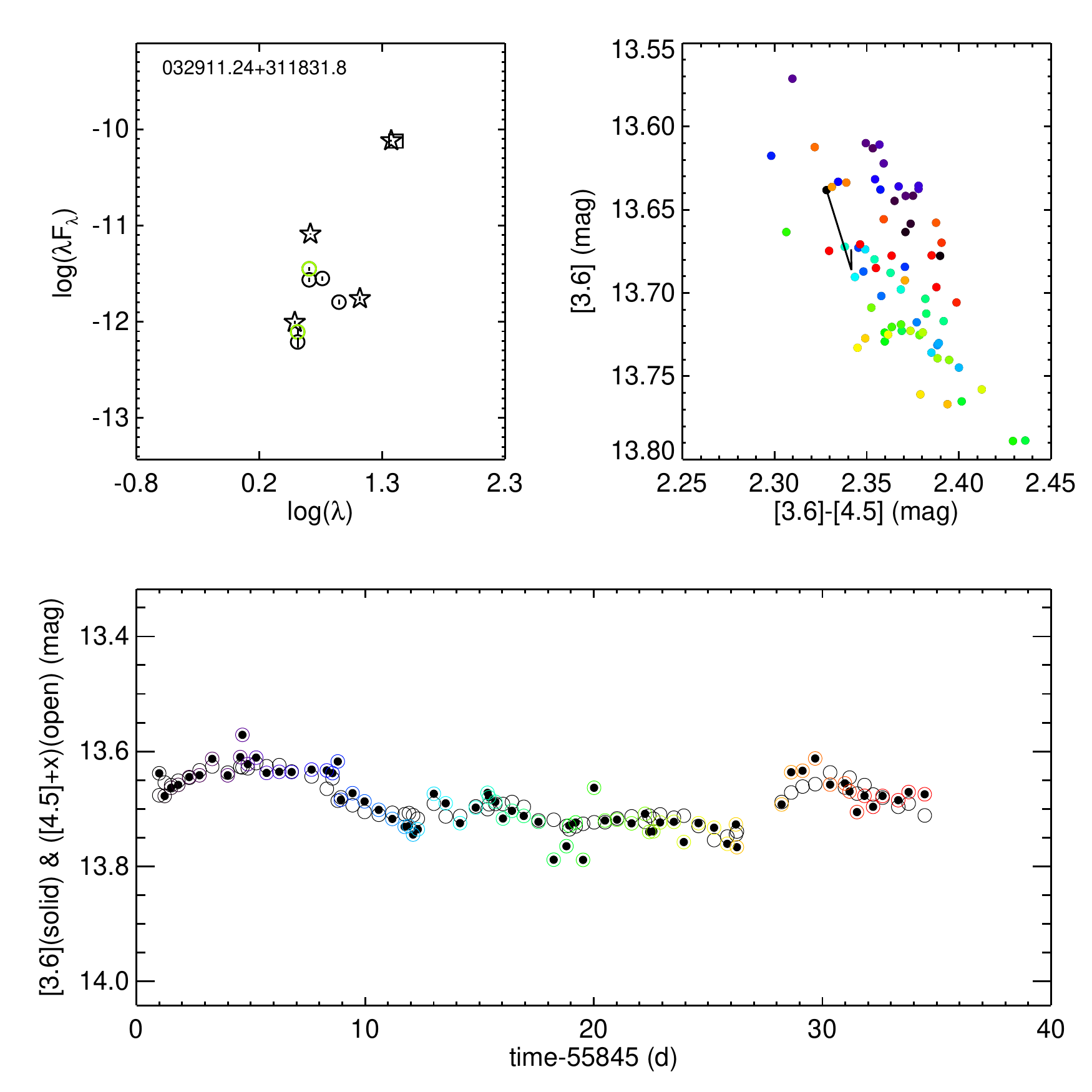}
\caption{Plots for the literature-identified Class 0 and jet-driving
source SSTYSV J032911.24+311831.8 (=IRAS7=West14=J07-24=VLA
27=Gutermuth 7, but {\em not} ASR 32; see R15). The SED in the upper left has the
same notation as prior SEDs. The upper right is a CMD, and the light
curve is at the bottom.}
\label{fig:032911.24+311831.8}
\end{figure}

Unfortunately, only 6 of the Class 0 or jet driving sources have
reasonable two-band coverage. Of those 6 sources, half of them are
variable over the YSOVAR campaign, and all three of those variables
have significant color variability consistent with reddening
variations in the [3.6] vs.\ [3.6]$-$[4.5] CMD. One of these objects,
SSTYSV J032910.70+311820.9, was called out above for having a very
large CY variation; it appears in Fig.~\ref{fig:032910.70+311820.9}.
The other two sources are SSTYSV J032856.11+311908.5
(Fig.~\ref{fig:032856.11+311908.5}) and SSTYSV J032911.24+311831.8
(Fig.~\ref{fig:032911.24+311831.8}). 

\begin{deluxetable}{ccccccccp{5cm}}
\tablecaption{Class 0 and Jet Drivers\label{tab:class0jets}}
\tabletypesize{\scriptsize}
\tablewidth{0pt}
\tablehead{\colhead{SSTYSV} & \colhead{Class 0}& \colhead{Jet driver}
& \colhead{Our class} & \colhead{CY var} & \colhead{Var}&
\colhead{[3.6]$_{\rm cryo}$} & \colhead{[4.5]$_{\rm cryo}$} &\colhead{Notes} } 
\startdata
032855.53+311436.3 & yes & yes & I  & yes & yes &  12.66&10.18& mostly [4.5] only \\
032856.11+311908.5 & yes & no  & I  & no  & yes &  11.22& 9.71& significant variability consistent with reddening variations in the [3.6] vs.\ [3.6]$-$[4.5] CMD\\
032857.37+311415.7 & no  & yes & I  & yes & yes &   9.87& 7.81& [4.5] only\\
032900.50+311200.7 & yes & yes & I  & yes & no  & \nodata &15.73& [4.5] only\\
032901.91+311541.4 & yes & no  & I  & no  & no  &  16.07&14.36& [4.5] only, only 8 points\\
032904.07+311446.5 & yes & yes & I  & no  & no  &  14.21&12.81& [4.5] only, but may be non-variable because of large errors; texture is in LC \\
032906.45+311534.4 & yes & no  & II & no  & no  &  14.38&13.38& \\
032910.70+311820.9 & no  & yes & I  & yes & yes &  12.57&10.82& significant variability consistent with reddening variations in the [3.6] vs.\ [3.6]$-$[4.5] CMD; large CY variation\\
032910.96+311825.6 & yes & no  & I  & yes & no  &  13.82&12.87& relatively few points\\
032911.24+311831.8 & yes & yes & I  & yes & yes &  13.95&11.61& significant variability consistent with reddening variations in the [3.6] vs.\ [3.6]$-$[4.5] CMD; potentially large variations on multi-year timescales\\
032912.05+311301.4 & yes & yes & I  & no  & no  &  13.95&10.47& [4.5] only\\
032912.05+311305.8 & yes & no  & I  & yes & yes &  17.44&13.16& [4.5] only\\
032913.60+311358.0 & yes & no  & I  & yes & no  &  15.69&13.53& [4.5] only\\
032914.96+312031.7 & yes & no  & II & no  & no  &  14.81&14.13& \\
032917.11+312745.5 & yes & no  & I  & no  & no  &  14.28&12.78& [3.6] only\\
032917.47+312748.2 & yes & no  & I  & yes & no  &  15.88&14.62& [3.6] only\\
032918.88+312313.0 & yes & no  & I  & no  & no  &  15.85&14.07& \\
\enddata
\end{deluxetable}

\subsection{Brown Dwarfs}
\label{sec:bds}

In NGC 1333, there are 55 objects with light curves and spectral types
of M4 or later, and 33 of those are type M6 or later (where M6 is
often taken as the stellar/sub-stellar boundary at $\sim$1 Myr). These
are among the youngest and most embedded brown dwarfs to ever be
monitored in the MIR. About a quarter of the brown dwarfs (8/33) are
variable. These variable objects include some that are relatively
faint, but they are not the majority; the mean [3.6] from the cryo era
is $\sim$12.6, and that for  [4.5] is $\sim$12.1; the mean magnitudes
for the ensemble of type M6 or later are $\sim$13 in both bands. 
These objects have widely ranging timescales and amplitudes, and do
not stand out from the rest of the population via those metrics. Of
the 8 variables, there are 1 Flat class, 5 Class IIs, and 2 Class
IIIs.

Five of the eight variable BDs are periodic, so the BDs stand out as
having a high fraction of periodic sources. At M7.3, SSTYSV
032857.11+311911.9 in Fig.~\ref{fig:032857.11+311911.9} above is one
of the periodic BDs.  Fig.~\ref{fig:032858.25+312202.0} above is a
burster, but is also a BD with a type of M6. Of the eight variable
brown dwarfs, three of them have significant color variabilty, and all
of those have significant redder-when-brighter (perpendicular to the
reddening vector) behavior, though admittedly just in the [4.5]
vs.~[3.6]$-$[4.5] CMD. This is significantly more common than in the
general population of variables.

\clearpage

\section{Summary}
\label{sec:concl}

We have presented YSOVAR data for NGC 1333.  There are 701 objects
with mid-IR light curves in one or both warm Spitzer bands. We find 92
objects that are mid-IR variable between the cryo-era observations and
the YSOVAR campaign, but only one (SSTYSV J032910.70+311820.9) with a
large color change between the cryo era and the YSOVAR campaign. There
are 78 objects that are variable over the YSOVAR campaign, nearly all
of which were identified as NGC 1333 members prior to this work. 
However, the fraction of NGC 1333 members that are mid-IR variable, at
$\sim$50\%, may be low compared to similar fractions from other
clusters (typically closer to 75\%-80\%).  We found evidence for a
larger variability fraction among more embedded sources. 

With amplitude defined as the difference between the 10th and 90th
percentile in the brightness distribution of points, typical
amplitudes are $\sim$0.1-0.15 and 20\% vary at 0.2 mags or more.
Amplitudes for disked sources are indistinguishable by class, but
Class III objects have lower amplitudes. 

In terms of color changes, the distribution of [3.6]$-$[4.5]
amplitudes is typically $\sim$0.05 mag. Larger color changes generally
(although not always) also translate to larger single-band changes. 
There are no discernible trends of amplitude of color change with
effective temperature. The most extreme color change is $>$0.2 mag
change (SSTYSV J032904.31+311906.3), implying $\Delta$\av$\sim$15-30
mag. For those variable objects with light curves in both IRAC bands,
there are 6 objects that have a significant redder-when-fainter
(`red') trend in CMDs, roughly consistent with the direction
expected for reddening variations. One object, SSTYSV
J032909.32+312104.1, has a significant bluer-when-fainter (`blue')
trend in both CMDs, roughly perpendicular to that expected for
reddening. There are 17 objects that have a significant red trend in
[3.6] vs.~[3.6]$-$[4.5] but not the other CMD, and 15 objects that
have a significant blue trend in [4.5] vs.~[3.6]$-$[4.5]. Sources with
all SEDs are found with the red trends, but no Class I sources are
found with blue trends.

Timescales for the variability seem to be $\lesssim$5d. We found only
weak evidence for  longer timescales among more embedded sources. Some
of the largest color changes are in objects that have small
timescales.

Finally, NGC 1333 provides a higher fraction of known or suspected
Class 0 sources, jet drivers, and brown dwarfs than the rest of the
YSOVAR clusters.  Class 0s and/or jet drivers are often faint in
[3.6], but the light curves we have for these objects span a wide
range of timescales and amplitudes, and they are more likely to be
variable on the 6-7 year timescale of the cryo-to-YSOVAR campaigns.
The brown dwarfs tend to be periodic, and more often found to have
significant color variability (and more often have `blue' color
variability) compared to stellar sources.

\acknowledgments

This work is based in part on observations made with the Spitzer Space
Telescope, which is operated by the Jet Propulsion Laboratory,
California Institute of Technology under a contract with NASA. Support
for this work was provided by NASA through an award issued by
JPL/Caltech. The research described in this paper was partially
carried out at the Jet Propulsion Laboratory, California Institute of
Technology, under contract with the National Aeronautics and Space
Administration.  The scientific results reported in this article are
based in part on data obtained from the Chandra Data Archive
including, observations made by the Chandra X-ray Observatory and
published previously in cited articles. 
This research has made use of NASA's Astrophysics Data System (ADS)
Abstract Service, and of the SIMBAD database, operated at CDS,
Strasbourg, France.  This research has made use of data products from
the Two Micron All-Sky Survey (2MASS), which is a joint project of the
University of Massachusetts and the Infrared Processing and Analysis
Center, funded by the National Aeronautics and Space Administration
and the National Science Foundation. The 2MASS data are served by the
NASA/IPAC Infrared Science Archive, which is operated by the Jet
Propulsion Laboratory, California Institute of Technology, under
contract with the National Aeronautics and Space Administration.
Brevis esse laboro osbcurus fio. (Horace)

A. Bayo acknowledges financial support from the Proyecto Fondecyt de
Iniciaci\'on 11140572.

H. Bouy is funded by the the Ram\'on y Cajal fellowship program number
RYC-2009-04497.  This research has been funded by Spanish grants
AYA2012-38897-C02-01, AYA2010-21161-C02-02, CDS2006-00070 and
PRICIT-S2009/ESP-1496. This work used Topcat (Taylor 2005) and Stilts
(Taylor 2006). Based in part on data collected at Subaru Telescope and
obtained from the SMOKA, which is operated by the Astronomy Data
Center, National Astronomical Observatory of Japan. This research used
the facilities of the Canadian Astronomy Data Centre operated by  the
National Research Council of Canada with the support of the Canadian
Space Agency. Based on observations obtained with MegaPrime/MegaCam, a
joint project of CFHT and CEA/DAPNIA, at the Canada-France-Hawaii
Telescope (CFHT) which is operated by the National Research Council
(NRC) of Canada, the Institute National des Sciences de l'Univers of
the Centre National de la Recherche  Scientifique of France, and the
University of Hawaii.

\clearpage
\appendix
\section{Appendix: Famous YSOs}

This section collects a list of the most famous YSOs in NGC 1333, and
adds figures for those objects not already presented above. 

\begin{deluxetable}{cp{3cm}cp{5cm}}
\tablecaption{Famous YSOs\label{tab:famousysos}}
\tabletypesize{\scriptsize}
\tablewidth{0pt}
\tablehead{\colhead{SSTYSV name} & \colhead{Synonyms}& \colhead{Figure}
& \colhead{Notes}  } 
\startdata
\sidehead{Variables over the YSOVAR campaign}
032847.82+311655.1 & SSV17 & Fig.~\ref{fig:032847.82+311655.1} & Long-term trends plus dip-like structure.\\
032851.01+311818.5 & SSV10, LkHa 352a & Fig.~\ref{fig:032851.01+311818.5} & Large ampl variable. CY var. `Step' up in light curve. 
Varies like reddening vector at start, then becomes
brighter, from which it moves fairly steadily bluer when brighter,
with smaller variations on top consistent with reddening variations.\\
032851.24+311739.3 & ASR41, LAL111 & Fig.~\ref{fig:032851.24+311739.3} & Known edge-on disk (EOD). Significant bluer-when-brighter trend in [3.6] vs.~[3.6]$-$[4.5].\\ 
032854.62+311651.2 & SSV18b & Fig.~\ref{fig:032854.62+311651.2} & Possible significant redder-when-brighter trend in [4.5] vs.~[3.6]$-$[4.5]. CY var. Long-term trend.\\
032855.53+311436.3 & IRAS 2A, SK8, J07-15 & Fig.~\ref{fig:032855.53+311436.3} &  Jet driver/class 0. Very little [3.6]; light curve is almost all [4.5]. CY var. Long-term trend.\\
032856.63+311835.6 & SSV11 & Fig.~\ref{fig:032856.63+311835.6} & Burster light curve. Long-term trend.  \\
032856.95+311622.3 & SSV15 & Fig.~\ref{fig:032856.95+311622.3} & Periodic source though not called out in text.  Significant redder-when-brighter trend in [4.5] vs.~[3.6]$-$[4.5]. CY var.\\
032857.37+311415.7 & IRAS 2b, J07-16 & Fig.~\ref{fig:032857.37+311415.7} &  Jet driver/class 0. All [4.5]. CY var. Long-term trends.\\
032859.32+311548.5 & SSV16, SVS16, SVS16ew & Fig.~\ref{fig:032859.32+311548.5} & Significant redder-when-brighter trend in [4.5] vs.~[3.6]$-$[4.5]. Long-term trend.\\
032901.53+312020.6 & SSV12, IRAS6 & Fig.~\ref{fig:032901.53+312020.6} & All [3.6]. Long-term trend.\\
032905.75+311639.7 & SSV14, ASR7 & Fig.~\ref{fig:032905.75+311639.7} & Significant redder-when-brighter trend in [4.5] vs.~[3.6]$-$[4.5]. Long-term trend.\\
032911.24+311831.8 & IRAS7, Sadavoy2014-West14, J07-24 & Fig.~\ref{fig:032911.24+311831.8} & Jet driver/class 0. CY var.\\
032917.66+312245.1 & SVS2 & Fig.~\ref{fig:032917.66+312245.1} & Periodic source though not called out in text.\\
032920.42+311834.3 & SSV5, HH17 & Fig.~\ref{fig:032920.42+311834.3} & Possible significant bluer-when-brighter trend in both CMDs. CY var. Long-term trend. \\
032921.87+311536.2 & SSV20, LkHalpha271 & Fig.~\ref{fig:032921.87+311536.2} & Periodic source though not called out in text. All [4.5]. CY var.\\
\tableline
032903.75+311603.9 & SVS13 & Fig.~\ref{fig:032903.75+311603.9} & Data
very close to saturation in [3.6] and saturated in [4.5]; the photometry
shown here is extracted separately from the pipeline, assuming it is
not quite saturated, and thus is not included in the rest
of the analysis. \\
\tableline
\sidehead{Objects not detected as variables over the YSOVAR campaign}
032912.91+311845.5 & HH6, IRAS7 & \nodata & CY var (see
\S\ref{sec:cyvar}).\\
032910.37+312159.1 & SVS3, IRAS8, X15 & \nodata & \\
032854.06+311654.3 & SSV18a & \nodata & \\
032846.18+311638.5 & SSV21, LkHalpha351 & \nodata & [3.6] only.\\
032857.20+311419.1 & SSV19, BD+30547 & \nodata & [4.5] only.\\
032909.64+312256.4 & SVS7 & \nodata & \\
032910.96+311825.6 & ASR 32?/33?, IRAS7 SM1/2, SK20/21, VLA2 & \nodata & Very few [3.6] points, mostly [4.5]. CY var.\\
032901.91+311541.4 & MMS3, Sadavoy2014-West19?, SVS13c & \nodata & Very few points, and only [4.5].\\
032912.05+311301.4 & IRAS4B, SK3, Sadavoy2014-West13, J07-25 & \nodata & Jet driver/class 0. [4.5] only.  \\
032913.60+311358.0 & IRAS 4C, SK5, VLA29, Sadavoy2014-West30, J07-26 & \nodata & [4.5] only. CY var. \\
032900.50+311200.7 & IRAS 4B1, SK1, Sadavoy2014-West33, J07-18 & \nodata & [4.5] only. CY var.\\
032904.07+311446.5 & IRAS5, Sadavoy2014-West40, SK14, J07-21 & \nodata & [4.5] only.\\
\enddata
\end{deluxetable}

\begin{figure}[h]
\epsscale{0.8}
\plotone{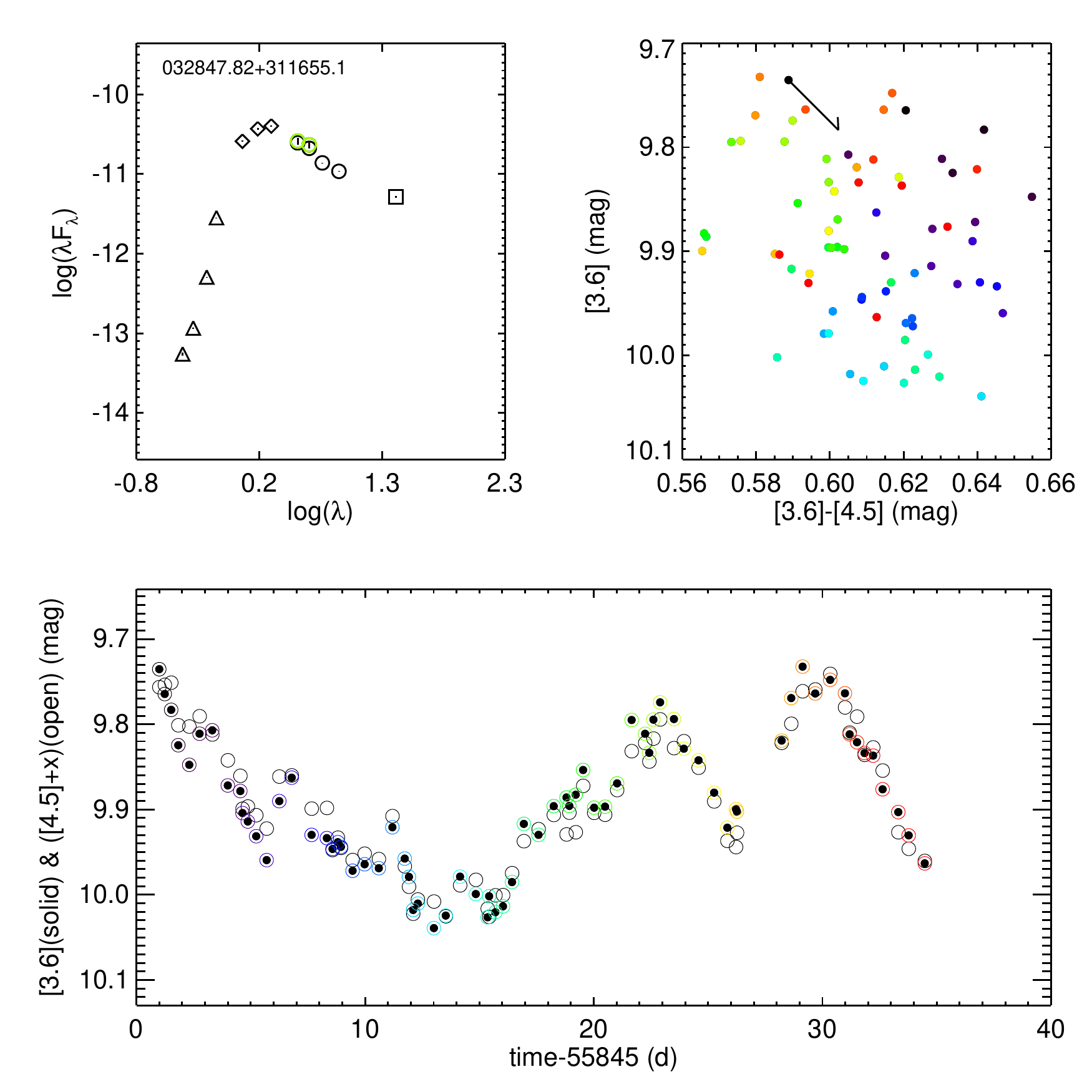}
\caption{Light curve for famous YSO SSTYSV J032847.82+311655.1 (SSV17);
 see Table~\ref{tab:famousysos}. }
\label{fig:032847.82+311655.1}
\end{figure}
\begin{figure}[h]
\epsscale{0.8}
\plotone{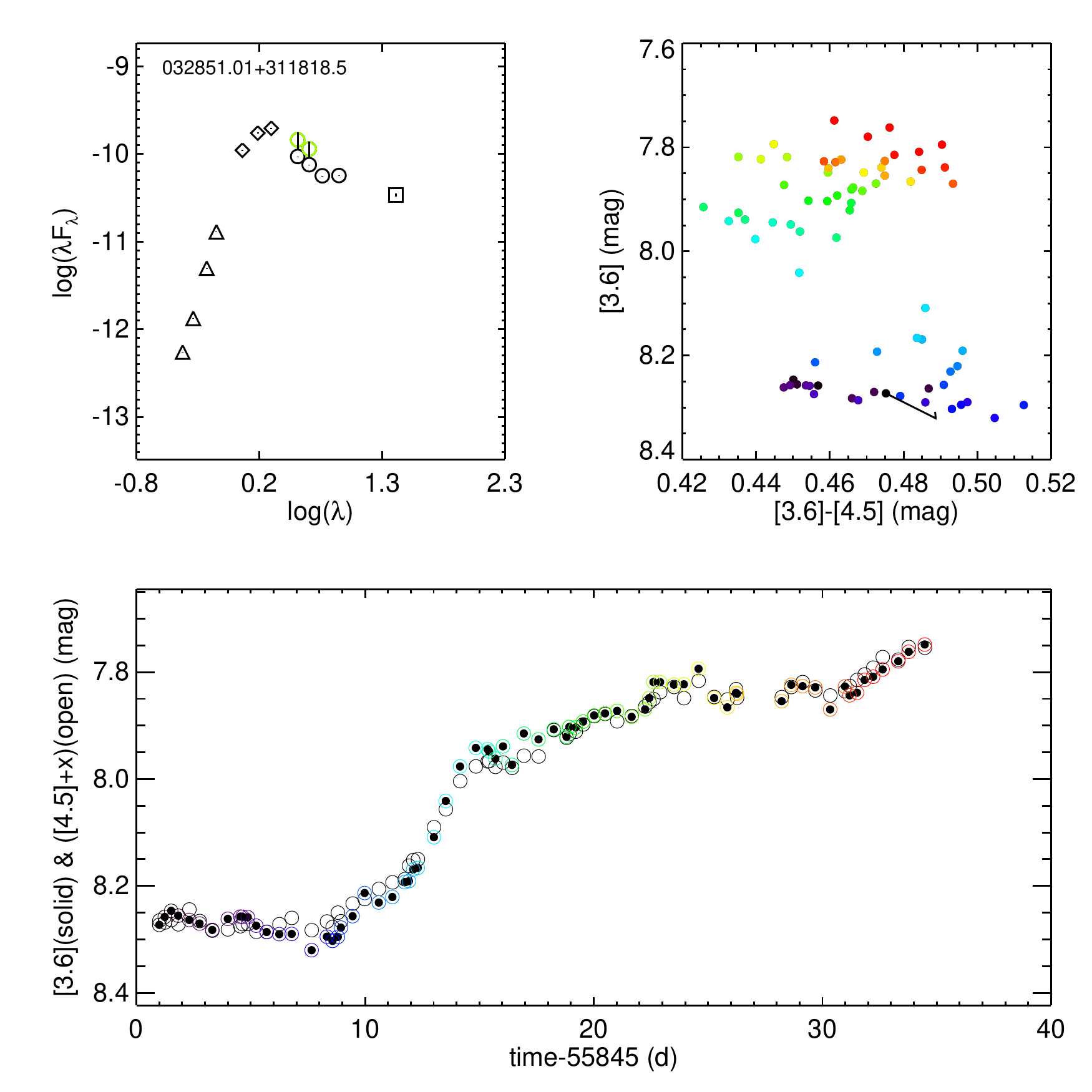}
\caption{Light curve for famous YSO SSTYSV J032851.01+311818.5 (SSV10,
LkHa 352a);  see Table~\ref{tab:famousysos}. }
\label{fig:032851.01+311818.5}
\end{figure}
\begin{figure}[h]
\epsscale{0.8}
\plotone{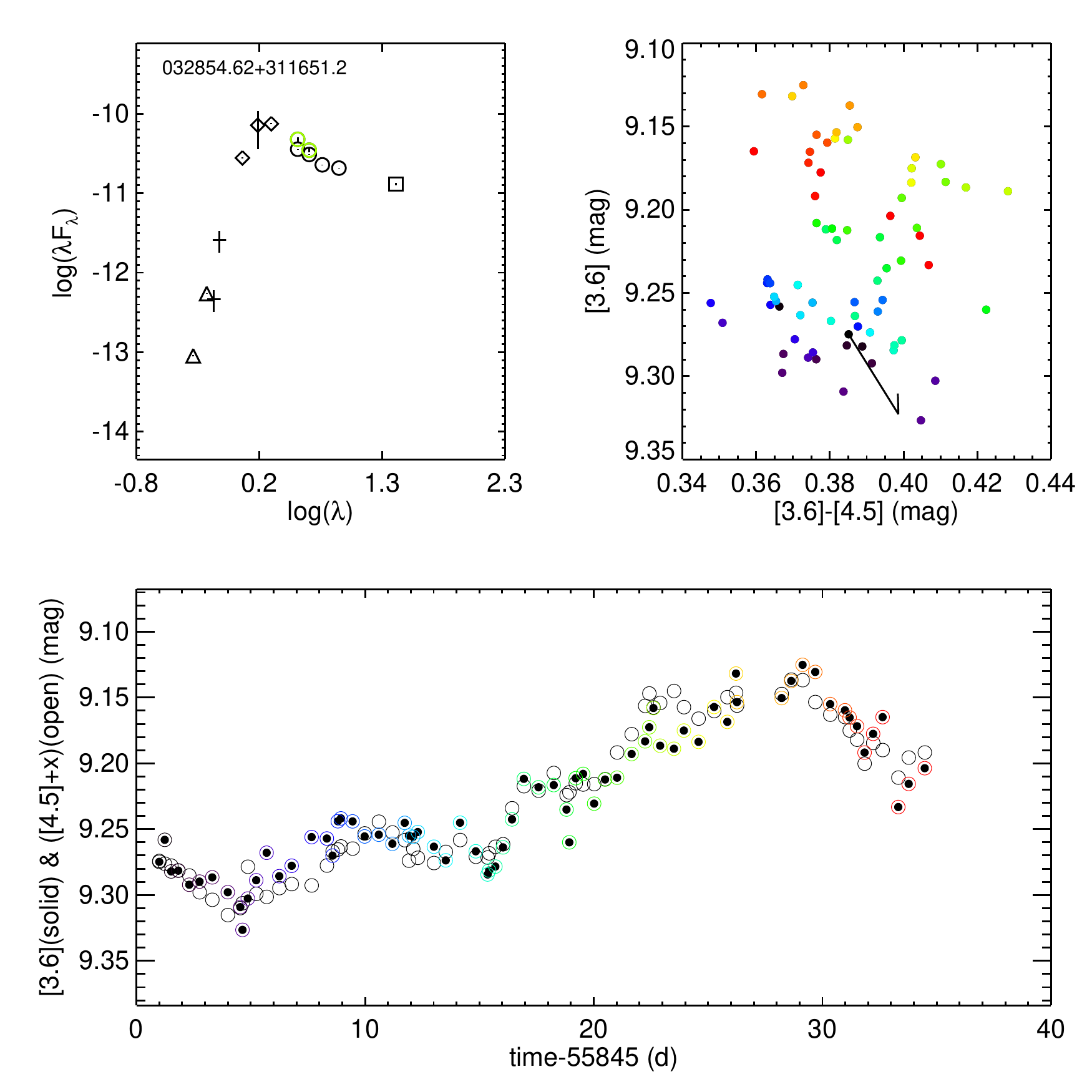}
\caption{Light curve for famous YSO SSTYSV J032854.62+311651.2
(SSV18b); see Table~\ref{tab:famousysos}. }
\label{fig:032854.62+311651.2}
\end{figure}
\begin{figure}[h]
\epsscale{0.8}
\plotone{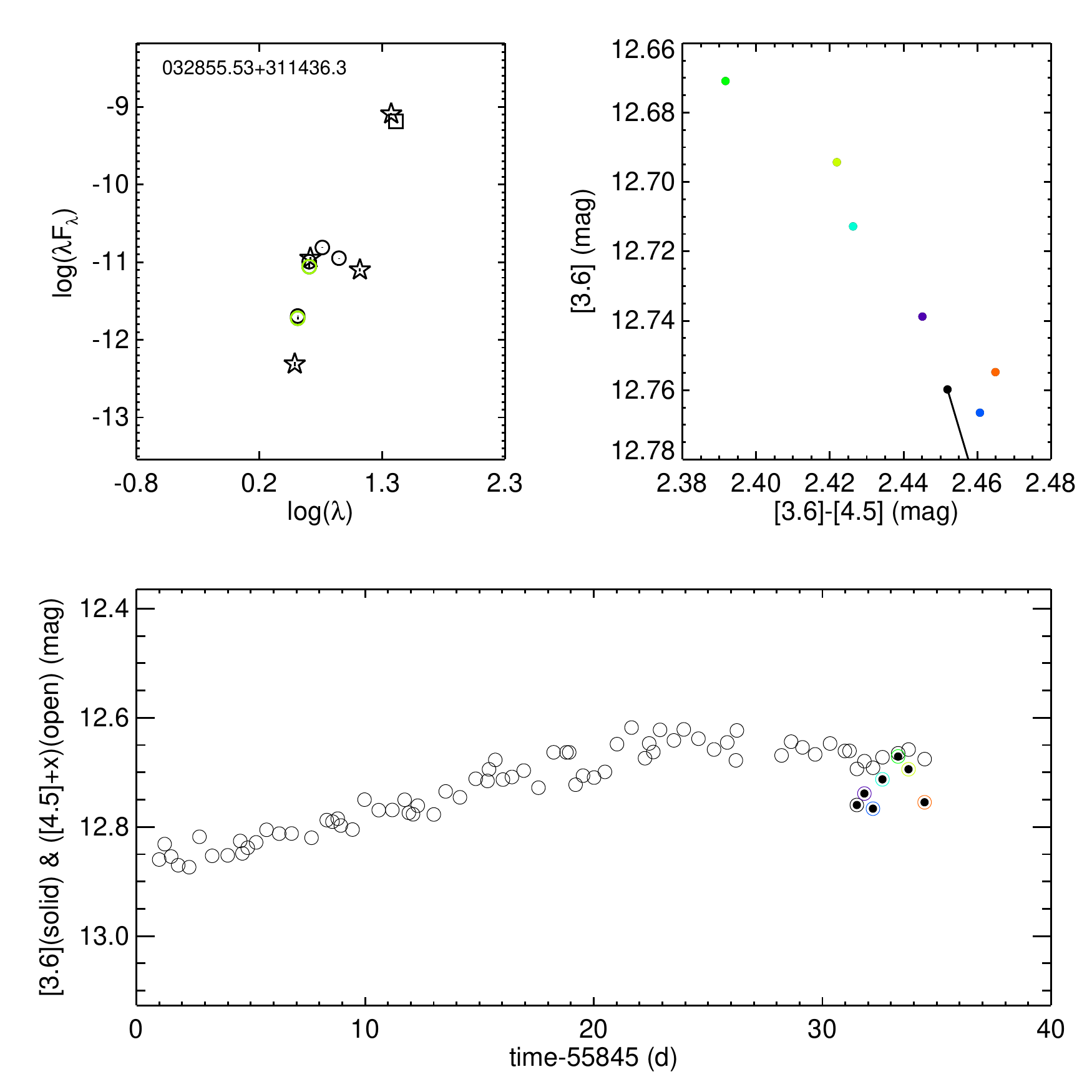}
\caption{Light curve for famous YSO SSTYSV J032855.53+311436.3 (IRAS
2A, SK8, J07-15);
 see Table~\ref{tab:famousysos}. }
\label{fig:032855.53+311436.3}
\end{figure}
\begin{figure}[h]
\epsscale{0.8}
\plotone{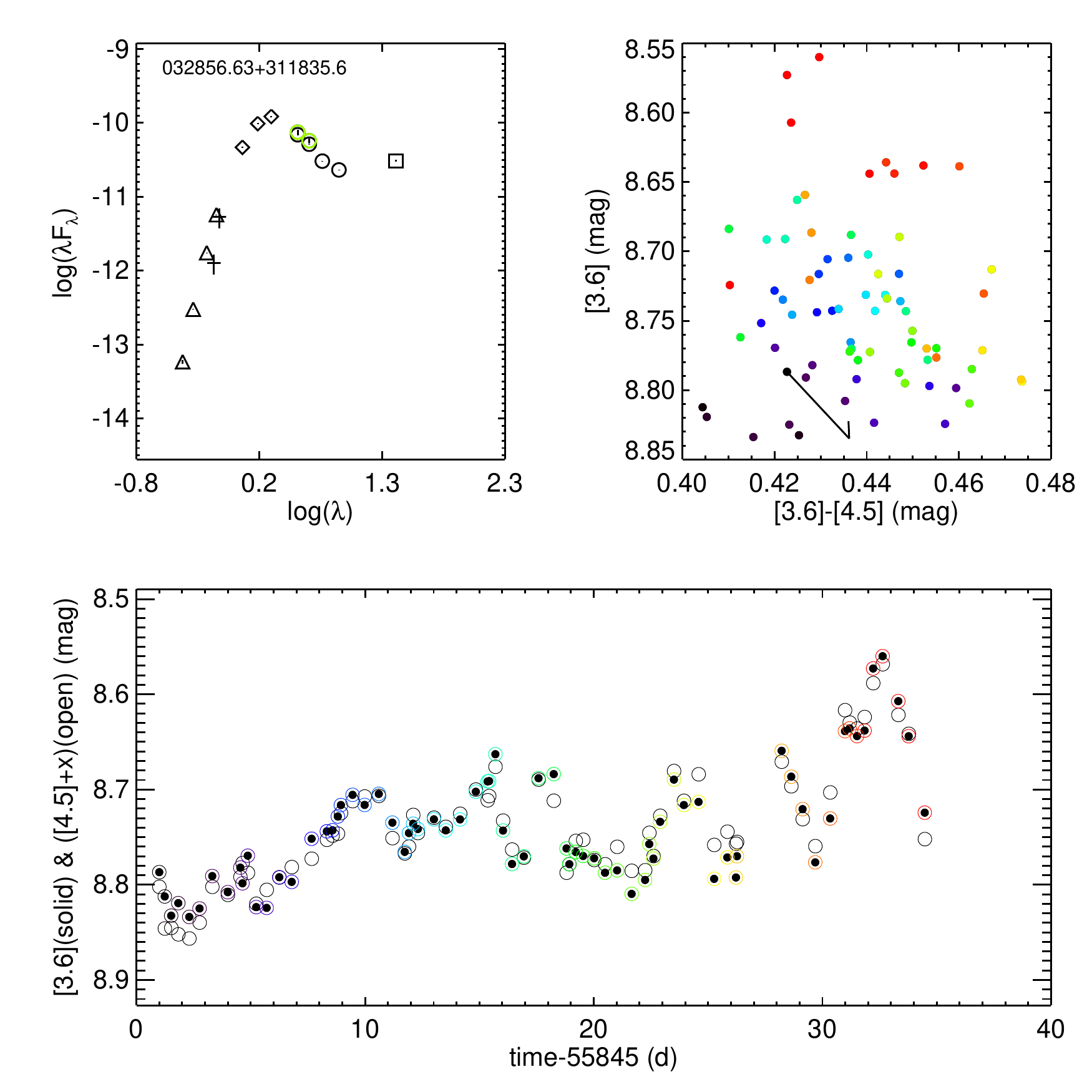}
\caption{Light curve for famous YSO SSTYSV J032856.63+311835.6 (SSV11);
 see Table~\ref{tab:famousysos}. }
\label{fig:032856.63+311835.6}
\end{figure}
\begin{figure}[h]
\epsscale{0.8}
\plotone{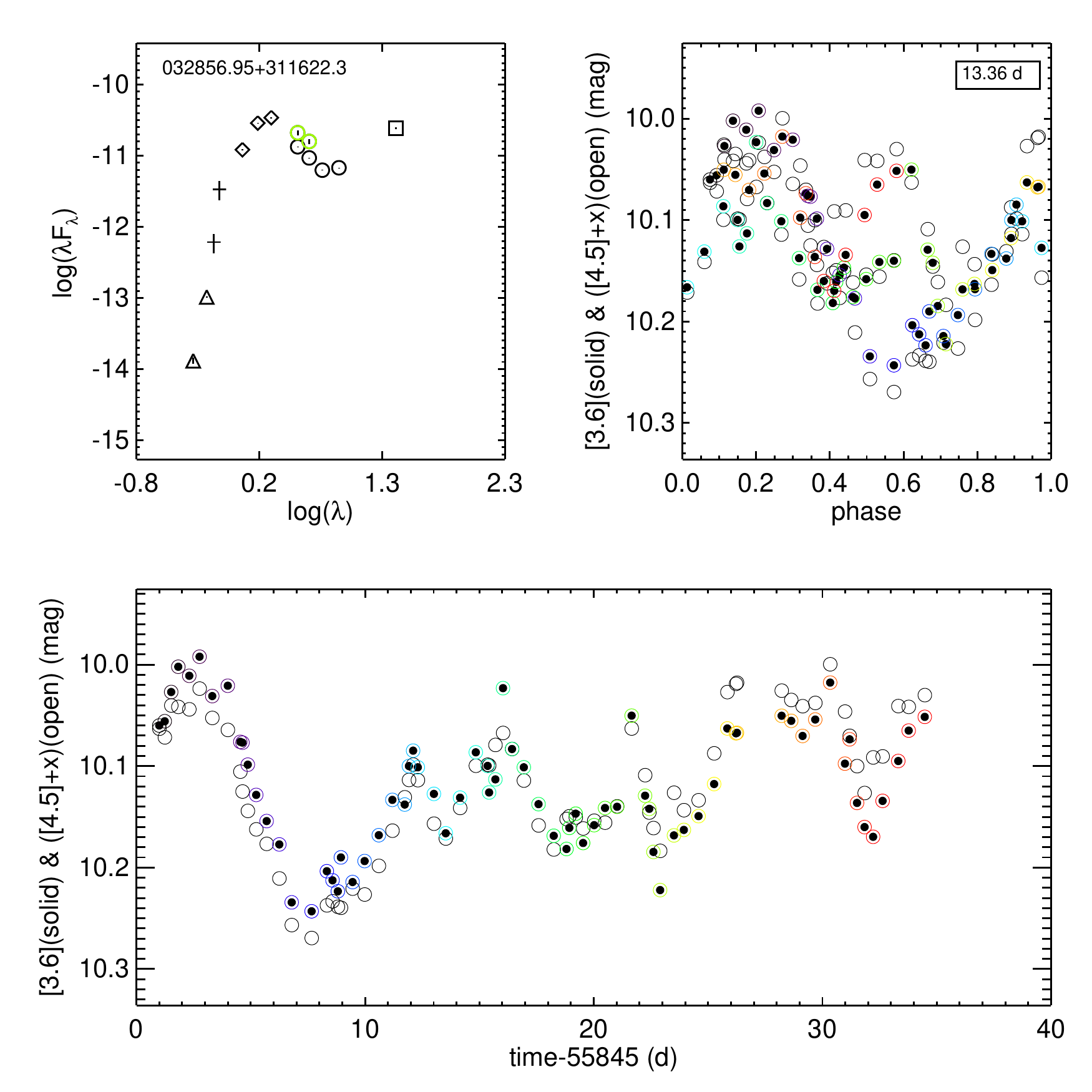}
\caption{Light curve for famous YSO SSTYSV J032856.95+311622.3 (SSV15);
 see Table~\ref{tab:famousysos}. }
\label{fig:032856.95+311622.3}
\end{figure}
\begin{figure}[h]
\epsscale{0.8}
\plotone{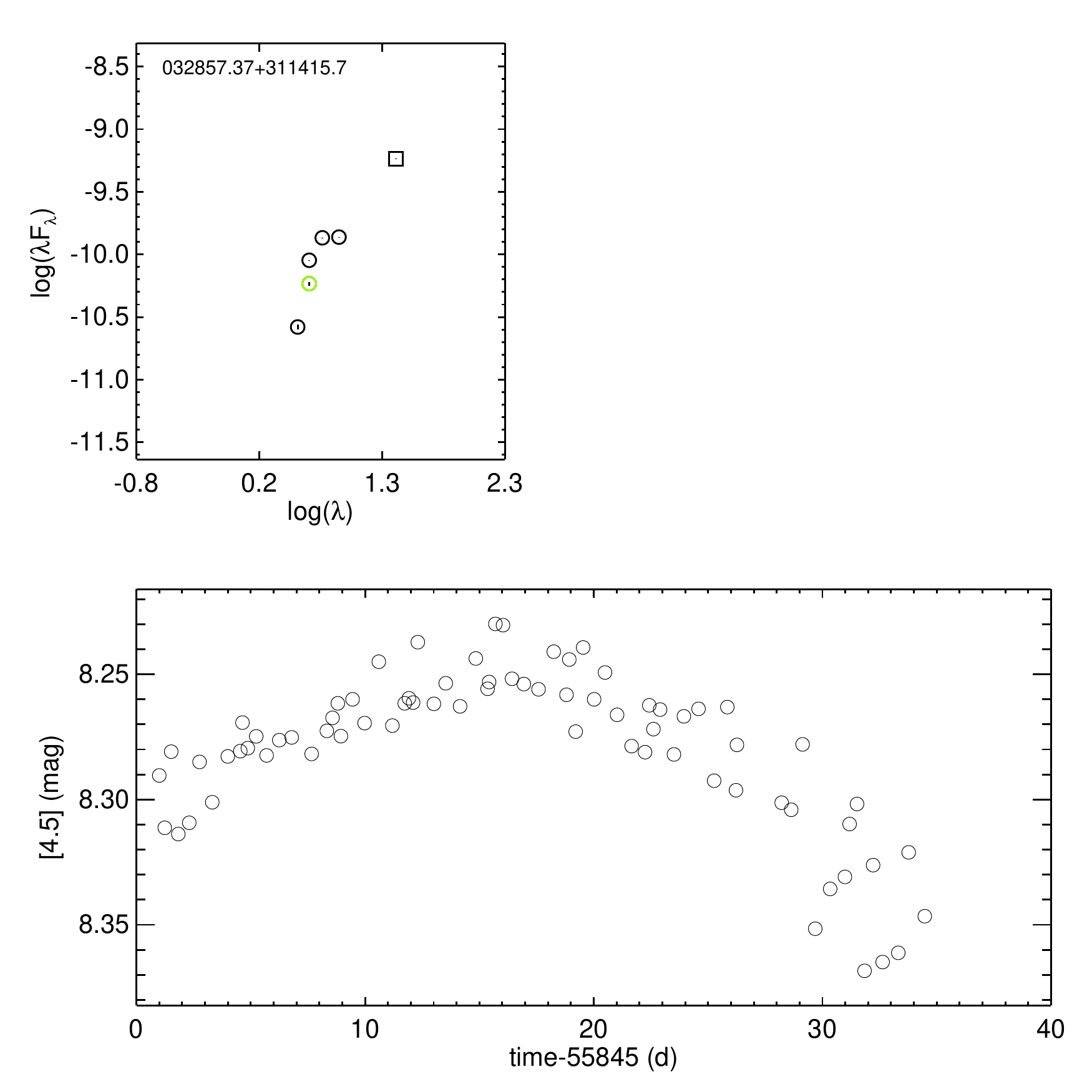}
\caption{Light curve for famous YSO SSTYSV J032857.37+311415.7 (IRAS
2b, J07-16);
 see Table~\ref{tab:famousysos}. }
\label{fig:032857.37+311415.7}
\end{figure}
\begin{figure}[h]
\epsscale{0.8}
\plotone{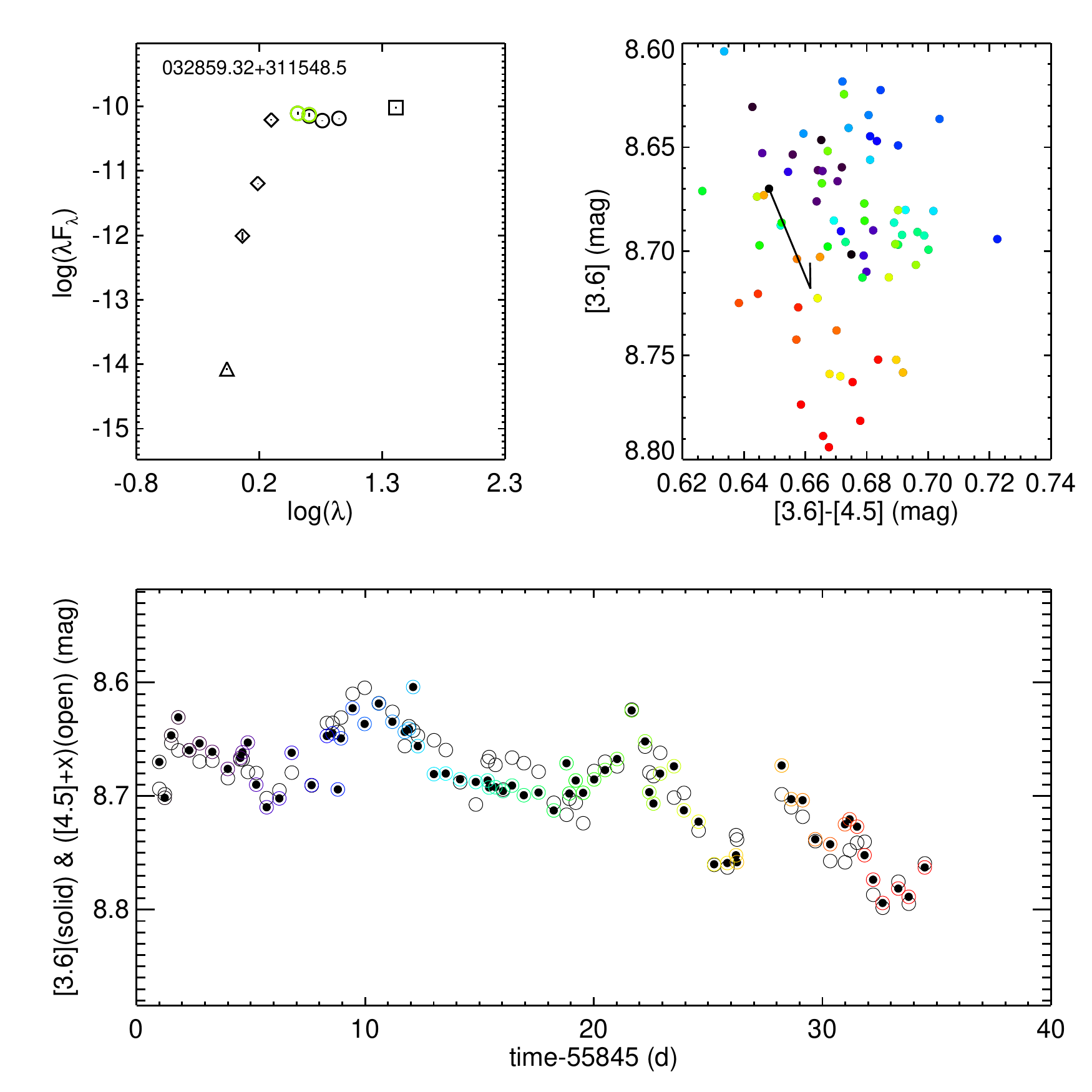}
\caption{Light curve for famous YSO SSTYSV J032859.32+311548.5 (SSV16,
SVS16, SVS16ew);
 see Table~\ref{tab:famousysos}. }
\label{fig:032859.32+311548.5}
\end{figure}
\begin{figure}[h]
\epsscale{0.8}
\plotone{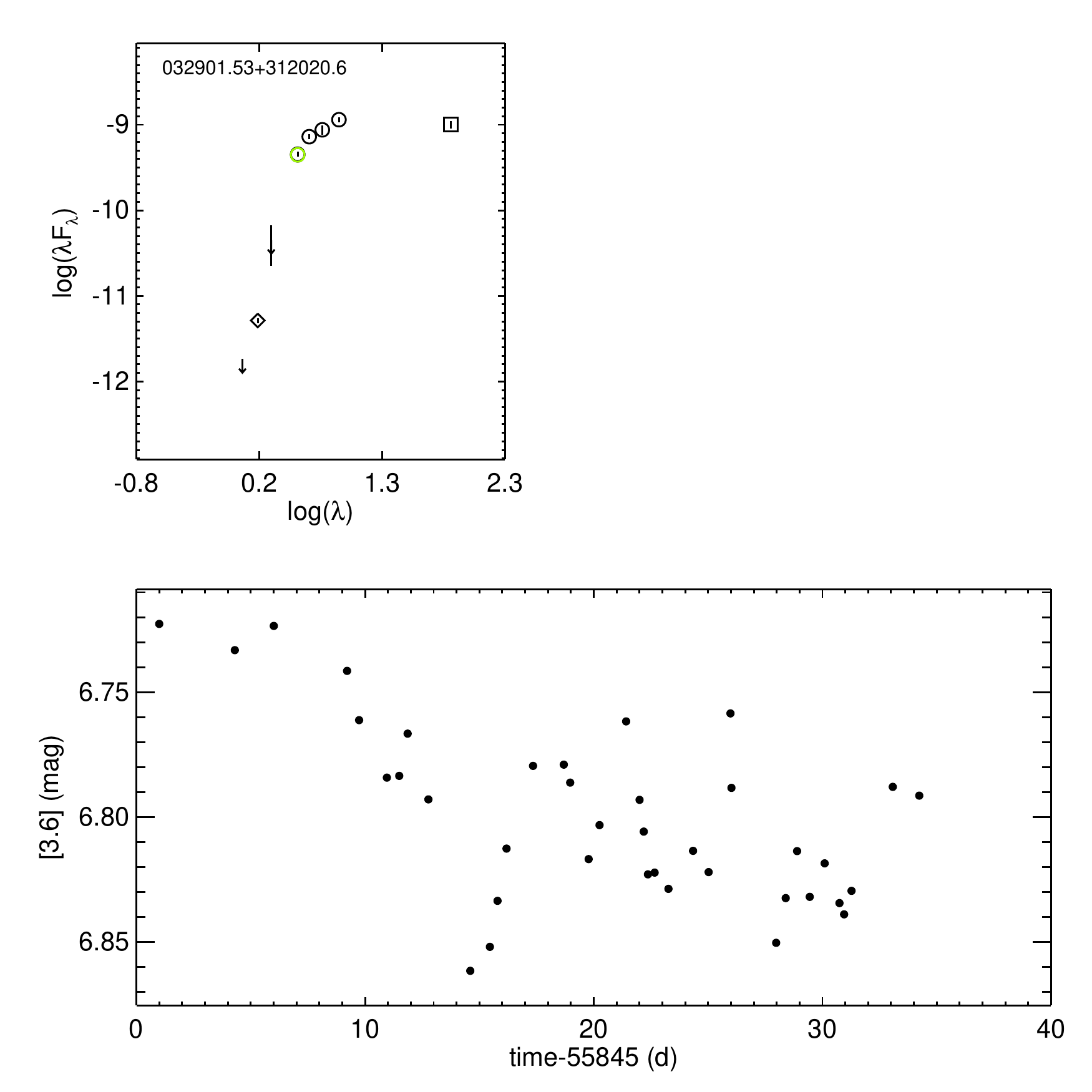}
\caption{Light curve for famous YSO SSTYSV J032901.53+312020.6 (SSV12,
IRAS6);
 see Table~\ref{tab:famousysos}. }
\label{fig:032901.53+312020.6}
\end{figure}
\begin{figure}[h]
\epsscale{0.8}
\plotone{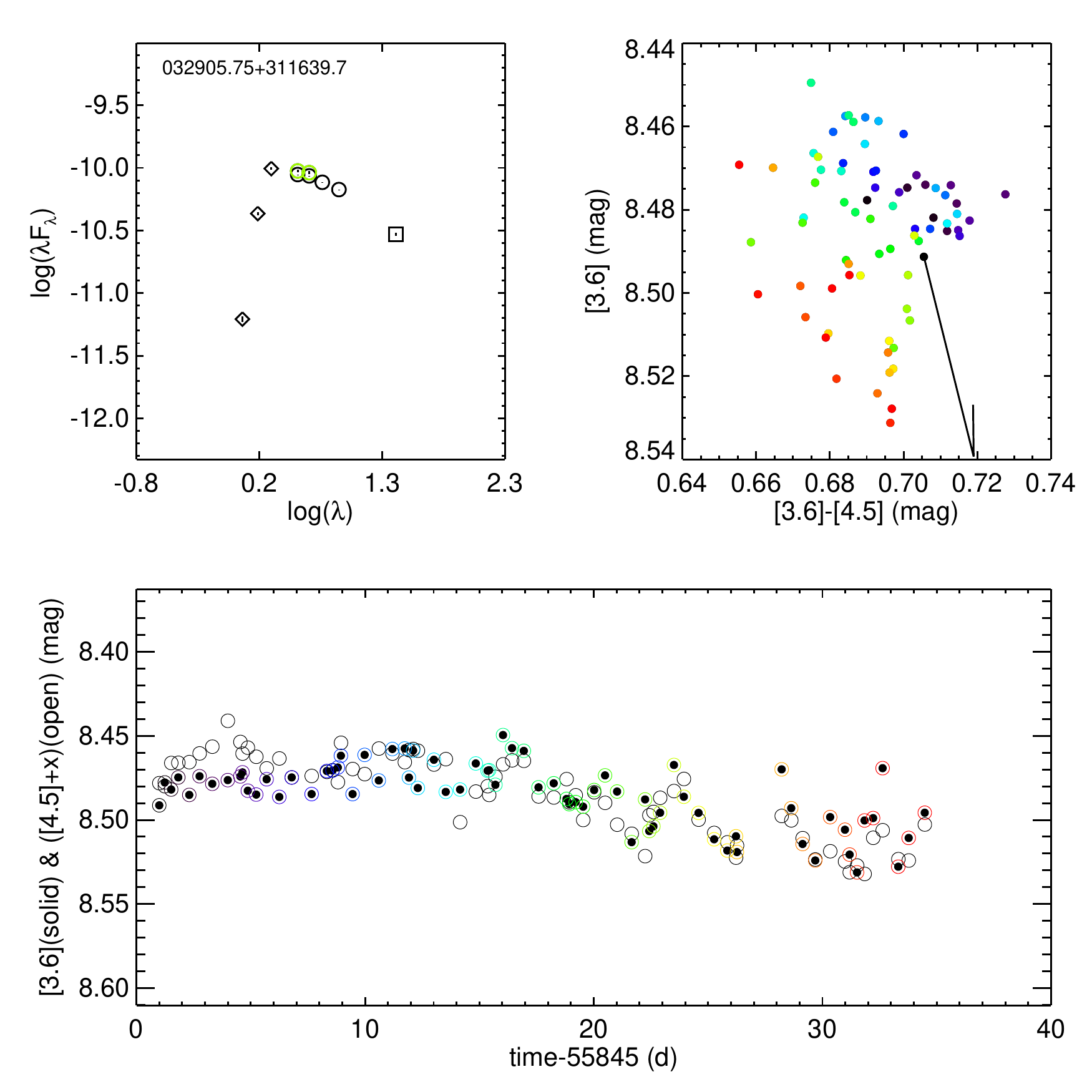}
\caption{Light curve for famous YSO SSTYSV J032905.75+311639.7 (SSV14,
ASR7);
 see Table~\ref{tab:famousysos}. }
\label{fig:032905.75+311639.7}
\end{figure}
\begin{figure}[h]
\epsscale{0.8}
\plotone{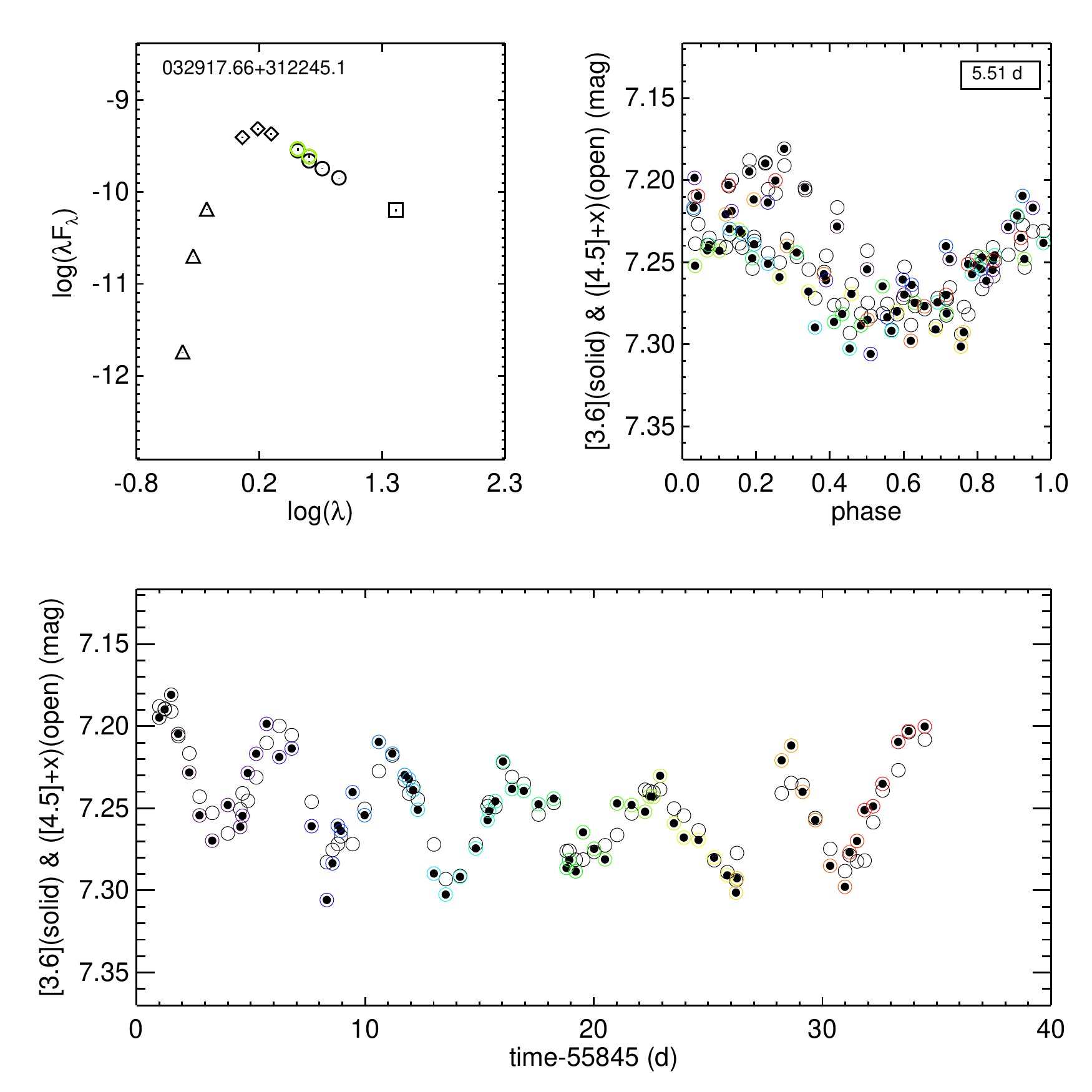}
\caption{Light curve for famous YSO SSTYSV J032917.66+312245.1 (SVS2);
 see Table~\ref{tab:famousysos}. }
\label{fig:032917.66+312245.1}
\end{figure}
\begin{figure}[h]
\epsscale{0.8}
\plotone{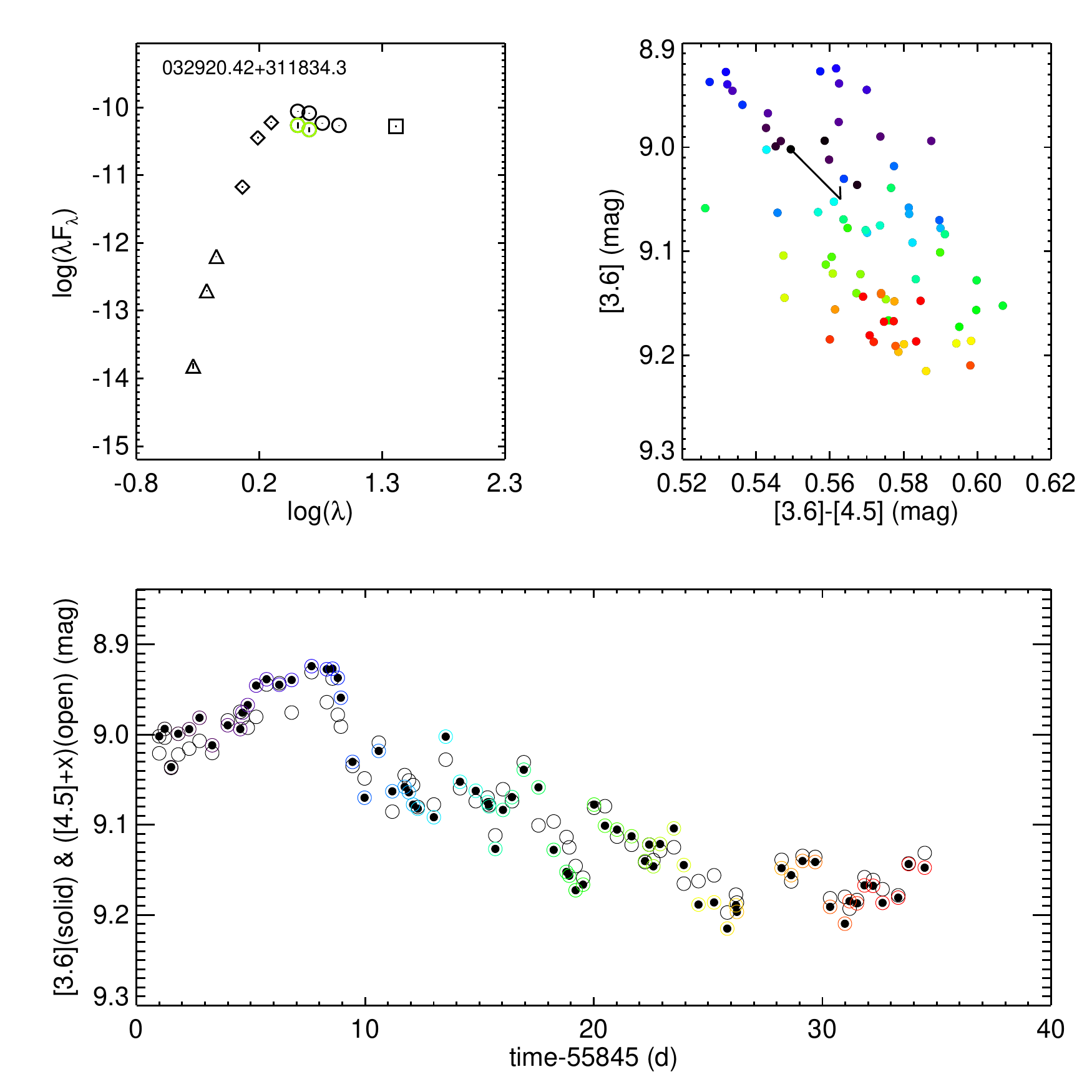}
\caption{Light curve for famous YSO SSTYSV J032920.42+311834.3 (SSV5,
HH17);
 see Table~\ref{tab:famousysos}. }
\label{fig:032920.42+311834.3}
\end{figure}
\begin{figure}[h]
\epsscale{0.8}
\plotone{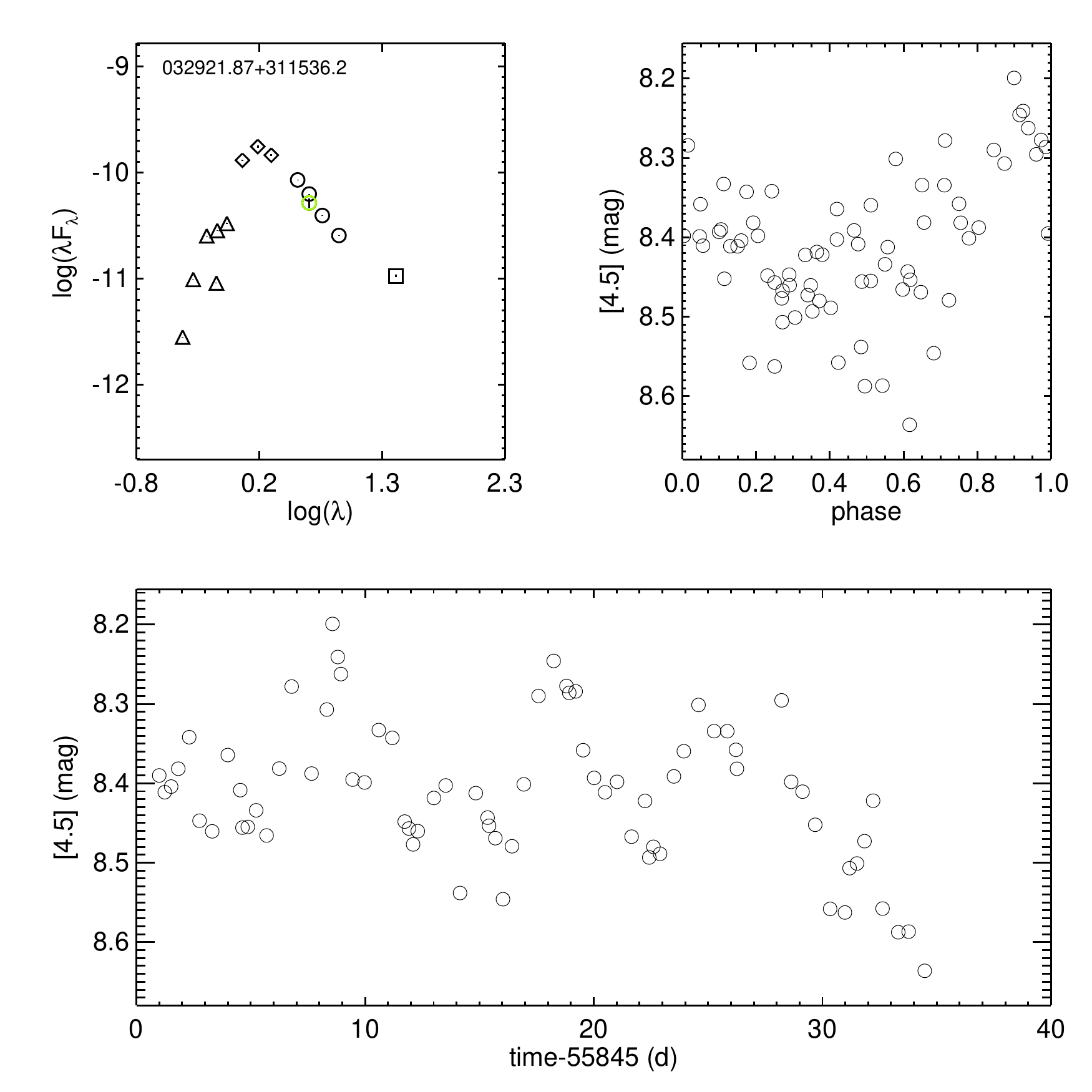}
\caption{Light curve for famous YSO SSTYSV J032921.87+311536.2 (SSV20,
LkHalpha271);
 see Table~\ref{tab:famousysos}. }
\label{fig:032921.87+311536.2}
\end{figure}

\begin{figure}[h]
\epsscale{0.8}
\plotone{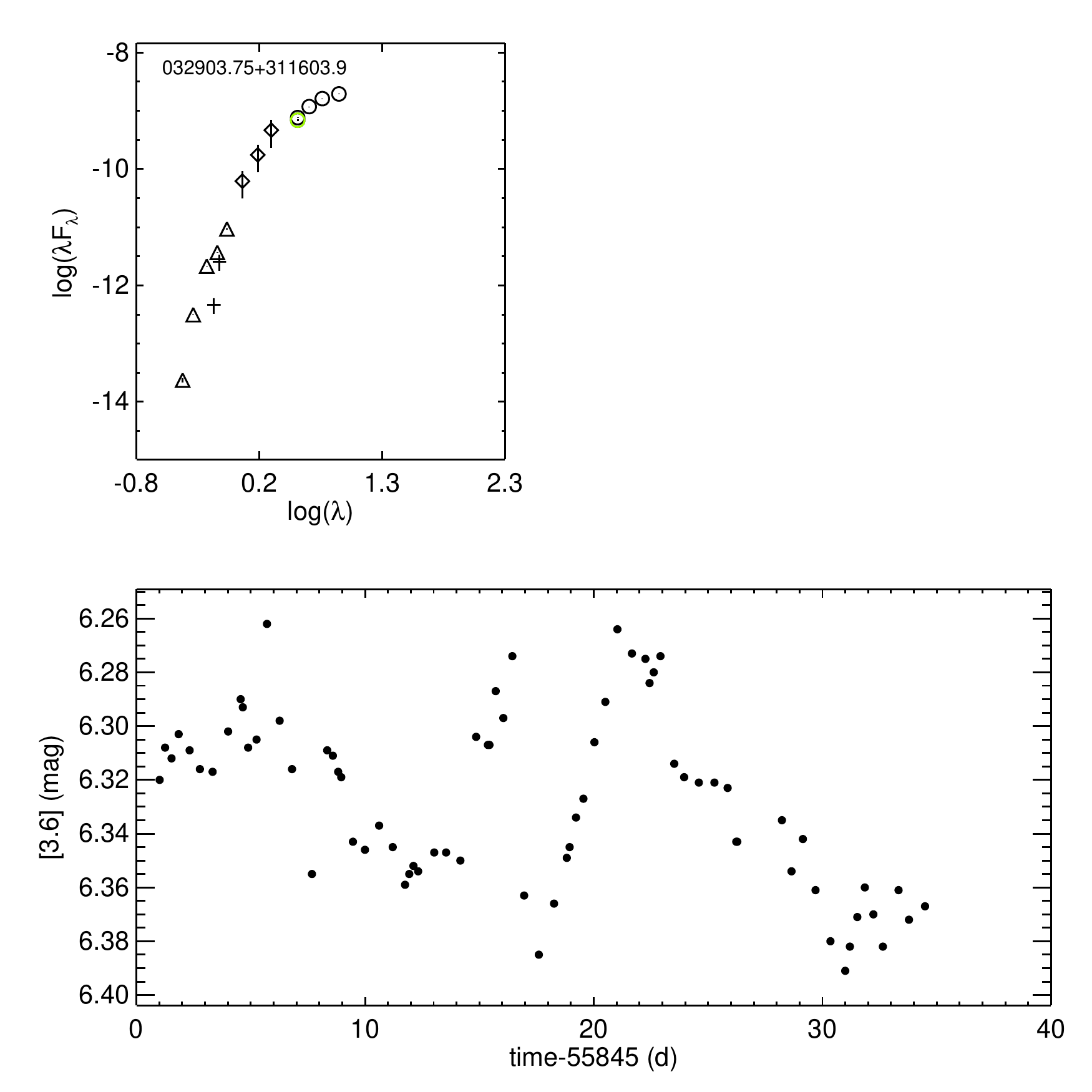}
\caption{Light curve for famous YSO SSTYSV J032903.75+311603.9, or SVS
13; see Table~\ref{tab:famousysos}. This light curve is extracted
separately from the pipeline, assuming that the [3.6] data are not
saturated, and thus is not included in the rest of the analysis.}
\label{fig:032903.75+311603.9}
\end{figure}

\end{document}